\documentclass[aps, pra, a4paper, showpacs, twocolumn, english, 10pt,nofootinbib]{revtex4-1}
\usepackage{bbm, amsthm, bm,textcomp, nicefrac,geometry,ragged2e}
\usepackage[dvipsnames]{xcolor}
\usepackage{float}
\usepackage[bbgreekl]{mathbbol}
\usepackage{graphicx,epstopdf,color,verbatim,enumitem,ulem}
\usepackage{pbox,array}
\usepackage{amsfonts}
\usepackage[english]{babel}
\usepackage{amssymb}
\usepackage{mathtools}
\usepackage{tikz}
\usepackage{mathrsfs}
\usepackage{amssymb}
\usepackage{array}
\usepackage{booktabs}
\usepackage{stackrel}
\usepackage{lipsum}
\usepackage[thinlines]{easytable}
\usepackage{amsmath}
\makeatletter
\usepackage[caption=false]{subfig}
\usepackage[T1]{fontenc}
\usepackage{hyperref}
\usepackage[euler]{textgreek}
\hypersetup{
    colorlinks = true,
    linkcolor = Black,
    citecolor = Black,
    filecolor = Black,
    urlcolor = Black,
}
\addto\captionsenglish{}

\newcommand\id{\mathbbm{1}}
\newcommand\ti{\text{i}}

\newcommand{\ket}[1]{\left| #1\right\rangle}

\definecolor{brickred}{rgb}{0.8, 0.0, 0.0}

\begin{document}

\title{A quantum simulator based on locally controlled logical systems}

\author{Ferran Riera-S\`{a}bat$^1$, Pavel Sekatski$^2$ and Wolfgang D\"ur$^1$}

\affiliation{$^1$Universit\"at Innsbruck, Institut f\"ur Theoretische Physik, Technikerstra{\ss}e 21a, 6020 Innsbruck, Austria \\ $^2$University of Geneva, Department of Applied Physics 1211 Geneva, Switzerland}

\date{\today}

\begin{abstract}
In a digital quantum simulator, basic two-qubit interactions are manipulated by means of fast local control operations to establish a desired target Hamiltonian. Here we consider a quantum simulator based on logical systems, i.e. where several physical qubits are used to represent a single logical two-level system to obtain enhanced and simple control over effective interactions between logical systems. Fixed, distance-dependent pairwise interactions between the physical qubits lead to effective interactions between the logical systems, which can be fully controlled solely by the choice of their internal state. This allows one to directly manipulate the topology and strength of effective interactions between logical systems. We show how to choose and generate the required states of logical systems for any desired interaction pattern and topology, how to perform arbitrary logical measurements, and how to obtain full control over single logical systems using only the intrinsic two-body interactions and control of individual physical qubits. This leads to a universal quantum simulator based on logical systems. We discuss the advantages of such a logical quantum simulator over standard ones, including the possibility to reach target topologies that are only accessible with large overheads otherwise. We provide several examples of how to obtain different target interaction patterns and topologies from initial long-ranged or short-ranged qubit-qubit interactions with a specific distance dependence.
\end{abstract}

\maketitle

\section{Introduction}

Quantum simulation is one of the pillars of quantum technologies, where a well-controlled quantum system is used to simulate another quantum system by reproducing its Hamiltonian \cite{Feynman1982,abrams1997simulation,georgescu2014quantum,Schaetz_2013,blatt2012quantum,bloch2012quantum,gross2017quantum,monroe2021programmable}. This opens exciting possibilities in the study of condensed matter systems, but also in quantum chemistry or high-energy physics \cite{georgescu2014quantum}. One can study parameter regimes in the simulator that are not accessible in the original system, and obtain access to internal states and features that cannot be measured directly otherwise. Significant experimental progress has been reported in recent years on both analogue and digital quantum simulators, where e.g. individual atoms or ions, photons, or cold gases are used to simulate other quantum systems \cite{lanyon2011universal,zhang2017observation,scholl2021quantum,semeghini2021probing,ebadi2021quantum,espinoza2021engineering}. 

In a digital quantum simulator, fast intermediate control pulses are used to manipulate the interaction between qubits. In this way, the strength and type of given bipartite interactions can be manipulated \cite{dodd2002universal,jane2002simulation,bennett2002optimal,abrams1997simulation}. For any fully connected interaction pattern, also additional pairwise and multi-qubit interactions not present in the original system can be generated, leading to a universal quantum simulator \cite{dodd2002universal,jane2002simulation,bennett2002optimal}. However, the latter process only works in higher order, leading to significant overhead in simulation time. 

Here we introduce a quantum simulator based on logical systems, where multiple physical qubits are used to represent each logical two-level system. In such a scenario, multiple pairwise \textit{ZZ} interactions among the physical qubits contribute to the effective interaction between the logical systems. The main advantage of such an approach is enhanced and direct control of effective interactions between such logical systems. By restricting the states of each logical system to two orthogonal $m$-qubit states $\big\{\big|0^L\big\rangle, \big|1^L\big\rangle\big\}$, one obtains a system of interacting logical qubits where the strength of interaction can be controlled and manipulated by the choice of the internal states, without the need to manipulate the basic two-body interaction among physical qubits. We utilize the distance dependence of the physical pairwise interactions to tailor internal states in such a way that desired effective coupling among logical systems is generated. This allows one not only to manipulate interaction patterns and topologies but also to obtain interaction patterns among logical systems that are not accessible directly within the original system. For instance, a system with solely nearest-neighbour (n.n.) couplings in a rectangular lattice can be grouped in such a way that the logical systems have long-ranged couplings whose interaction strength can be controlled by the choice of internal states.

The main results of this paper are as follows:
\begin{itemize}
    \item We show how to establish arbitrary interaction patterns among logical systems for generic commuting short- and long-ranged physical interactions.
    \item We provide efficient methods to maximize interaction strength for specific target topologies.
    \item We show how to add full control to logical systems using solely control of individual physical qubits together with the intrinsic two-body interactions, leading to a universal quantum simulator based on logical systems. 
\end{itemize}

For commuting interactions, we show how to choose and manipulate internal states to obtain arbitrary effective interaction patterns among the logical systems. For specific target topologies and patterns, we provide explicit optimized solutions with large effective coupling strength. Importantly, this requires only an initial preparation in some entangled state, which can however be done using the intrinsic physical interaction in the system as we demonstrate. All further manipulations only require control of individual physical qubits, without the need to manipulate physical qubit-qubit interactions -- not even to turn them on or off at will.

Similarly, we show that single-qubit measurements suffice to perform arbitrary measurements on each logical system.

Arbitrary rotations among the logical qubits (or equivalently effective logical single-qubit terms) can be obtained by utilizing the intrinsic pairwise physical interactions, together with control of individual qubits. This leads then to a universal quantum simulator based on logical systems, where standard techniques from Hamiltonian simulation \cite{dodd2002universal,jane2002simulation,bennett2002optimal,abrams1997simulation} are used to transform commuting interactions and local control operations to arbitrary target Hamiltonians. Notice that in contrast to the manipulation of interaction patterns, here fast local control is required.

We remark that in our approach logical systems or encodings are not used to increase noise resilience as in quantum error correction \cite{nielsenchuang2010}, but to enhance the accessibility and control of effective interactions.

This article is organized as follows. In Sec.~\ref{sec:setting} we describe the setting. We introduce the underlying physical many-body system and how the logical systems are implemented. In Sec.~\ref{sec:control:logical:system} we demonstrate how logical qubits and their interactions are controlled by means of local operations on the constituting physical system. In Sec.~\ref{sec:applications} we analyze several particular cases given by different interaction ranges and target systems. In Sec.~\ref{sec:extensions} we consider general interaction types for the physical system, and we show how any kind of interaction can be implemented between the logical systems by using known techniques of Hamiltonian simulation. We also give the scheme for implementing logical qudits by using extra degrees of freedom on the physical system. In Sec.~\ref{sec:comparison} we compare the efficiency of our approach with some standard Hamiltonian simulation techniques. Finally, in Sec.~\ref{sec:summary} we conclude with a summary and point out further extensions.

\section{Setting}
\label{sec:setting}

\subsection{Physical layer}

\begin{figure}
    \centering
    \subfloat[\centering]{\includegraphics[width=0.5\columnwidth]{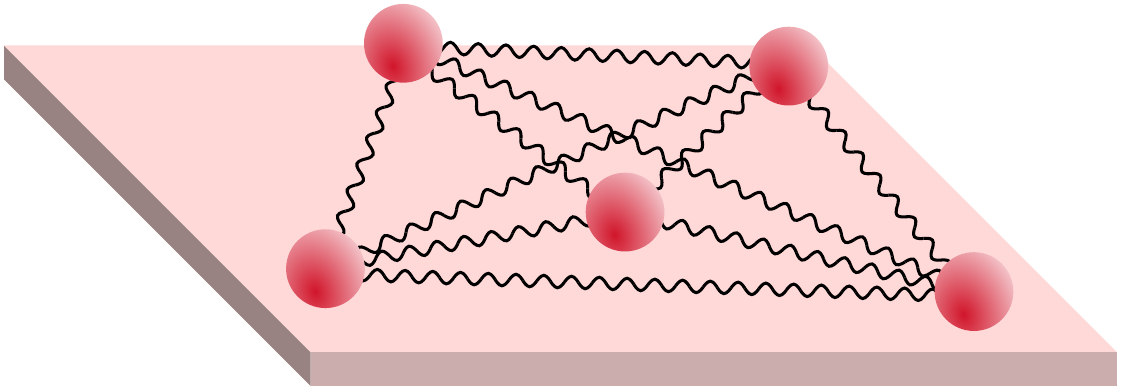} } \\
    \subfloat[\centering]{\includegraphics[width=0.9\columnwidth]{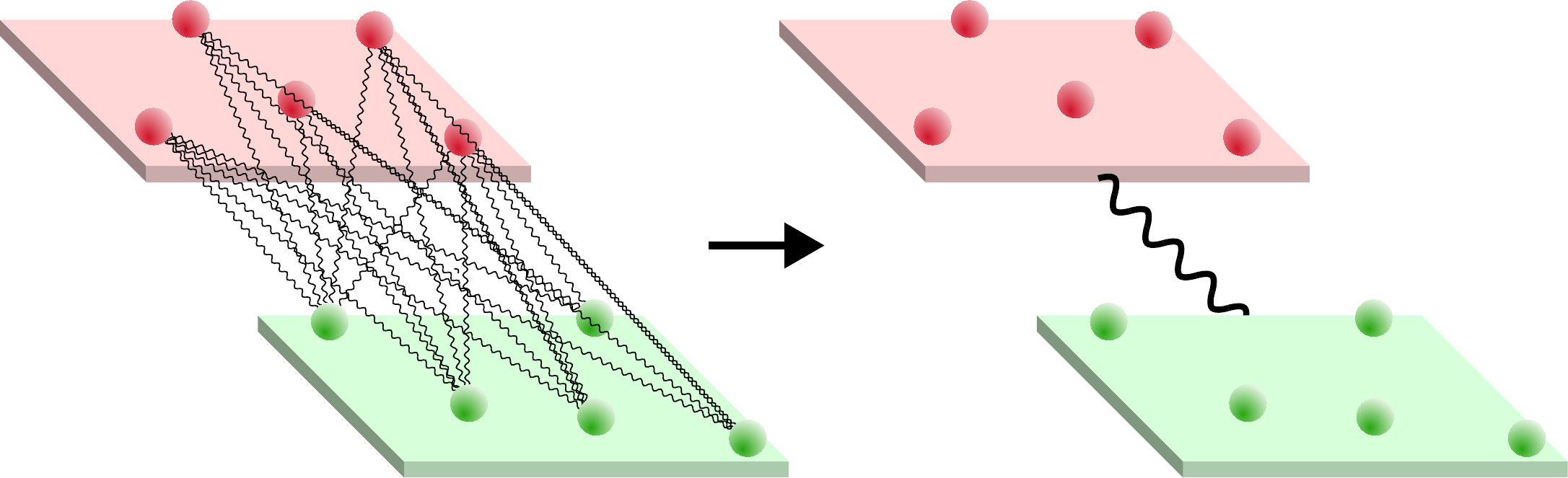} } \\
    \subfloat[\centering]{\includegraphics[width=0.9\columnwidth]{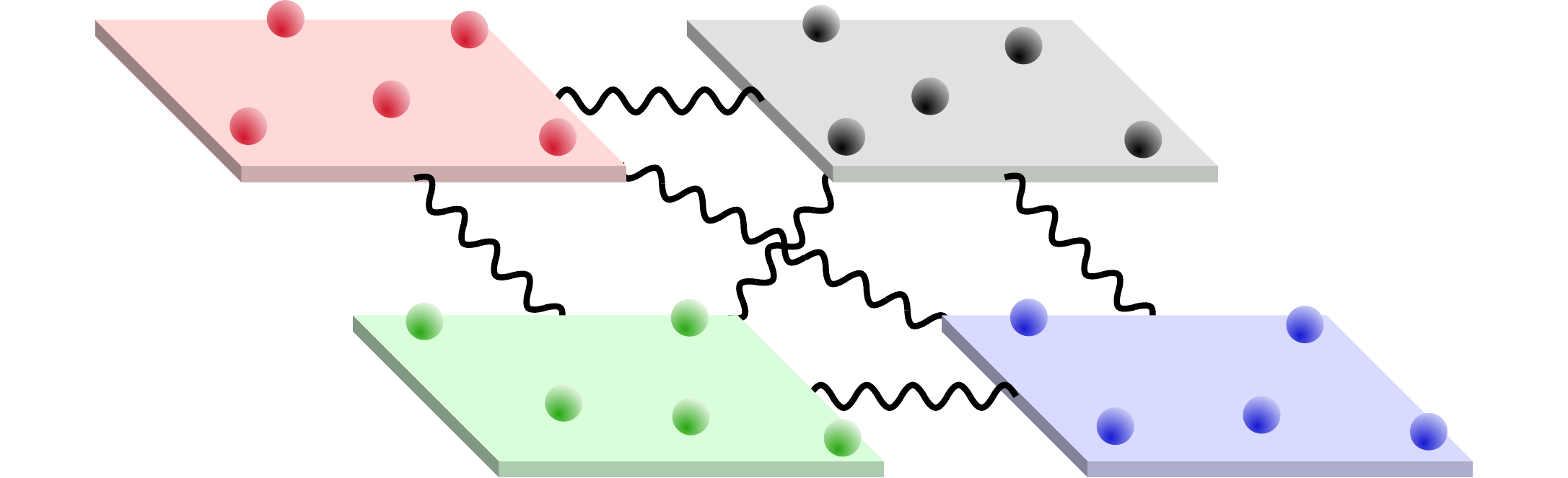} } \caption{\label{fig:1} (a) Representation of a set of physical qubits and its inner interactions, given by Eq.~\eqref{eq:H:i}. (b) Representation of the interaction between two sets, see Eq.~\eqref{eq:H:ij}. The interaction between two sets corresponds to the sum of all interactions between pairs of physical qubits from the two different sets. (c) Representation of four sets of five qubits each in the logical layer. When a logical qubit is encoded in each set only the interactions between sets are relevant, see Eq.~\eqref{eq:Ham2}.}
\end{figure}

Consider a spatially distributed many-body system of qubits. We assume an intrinsic always on pairwise \textit{ZZ} distance-dependent interaction, i.e., if qubit-$1$ and qubit-$2$ are at positions $\boldsymbol{r}_1$ and $\boldsymbol{r}_2$ respectively, they interact via $J \, \text{f} \, (\left| \boldsymbol{r}_1 - \boldsymbol{r}_2 \right|)Z_1 Z_2$ where $J \, \text{f} \, (\left| \boldsymbol{r}_1 - \boldsymbol{r}_2 \right|)$ is the coupling strength which depends on the coupling constant $J$ and on the distance between the two qubits. Precisely, we consider qubit-qubit couplings inversely proportional to a power, $\alpha$, of the distance and in some cases up to some interaction range $r$, i.e., the function $\text{f} \, (x)$ is given by
\begin{equation}\label{eq:interaction:range}
    \text{f} \, ( x ) = \left\{
    \begin{matrix}
        x^{-\alpha} & \text{ if } \left| x \right| \leq r \\[0.3em]
        0 & \text{ if } \left| x \right| > r 
    \end{matrix}\right. .
\end{equation}
We group the qubits in $N$ sets (or groups) $S_i$ of $n_i=|S_i|$ qubits each for $i=1,\dots, N$ (see Fig.~\ref{fig:1}), where we denote as $S^{(k)}_i$ the qubit $k = 1,\dots,n_i$ of set $i$, and $s^{(k)}_i$ labels its state in the $Z$-basis. In principle, $s^{(k)}_i = \pm 1$, but we consider $s^{(k)}_i \in [-1,1]$ as we demonstrate in Sec.~\ref{sec:eff:spin} that we can effectively obtain any intermediate non-integer value by flipping the qubits at specific times of the evolution. While physical interactions do not have a cutoff distance, for interactions that quickly decay with the distance it is natural to simplify the description with a cutoff approximation.

Having grouped the qubits, the Hamiltonian describing the dynamics of the whole system can be written as
\begin{equation}\label{eq:H:tot}
    H = \sum_{ i = 1 }^N H_i + \sum_{1\leq i<j\leq N} H_{ ij }
\end{equation}
where
\begin{equation}\label{eq:H:i}
    H_i = \sum_{1\leq k<l \leq n_i} f_{ii}^{(kl)} Z_i^{(k)} Z_i^{(l)}
\end{equation}
describes the inner interactions of qubits within $S_i$, and
\begin{equation}\label{eq:H:ij}
    H_{ ij } = \sum_{ \substack{ 1\leq k \leq n_i\\ 1\leq l \leq n_j}} f_{i j}^{(kl)} Z_i^{(k)} Z_j^{(l)}
\end{equation}
describes the interactions between qubits in $S_i$ with qubits in $S_j$ for $i\neq j$, as $Z^{(k)}_i$ acts on qubit $S^{(k)}_i$ and $f^{(kl)}_{ij}$ is the coupling strength between $S_i^{(k)}$ and $S_j^{(l)}$.

\subsection{Logical layer}
\label{Sec:Logical:layer}

For each set, $S_i$, we choose a vector, $\boldsymbol{s}_i$, and we implement a logical qubit by defining the logical computational basis as $\ket{0^L}_i \equiv \ket{\boldsymbol{s}_i}$ and $\ket{1^L}_i \equiv \ket{-\boldsymbol{s}_i}$ which spans the logical subspace, where $\ket{\boldsymbol{s}_i} \equiv \big|s_i^{(1)}\big\rangle \big| s_i^{(2)}\big\rangle\cdots \big|s_i^{(n_i)}\big\rangle$ fulfils $Z_i^{(k)} \ket{\boldsymbol{s}_i} = s_i^{(k)} \ket{\boldsymbol{s}_i}$. From now on, we refer to ``physical qubits'' as the qubits at the physical level and ``logical qubits'' as the effective two-level systems implemented in each set of physical qubits. 

Reducing the Hilbert space of the whole system in this particular way allows us to simplify the Hamiltonian of Eq.~\eqref{eq:H:tot}. First, notice that the logical qubits are not affected by the self-interactions of physical qubits within the same set. Any pair of states $\ket{\boldsymbol{s}_i}$ and $\ket{-\boldsymbol{s}_i}$ are degenerate with respect to $H_i$, i.e., they have the same eigenvalue
\begin{equation}
    \label{eq:DFS}
    H_i \ket{\pm\boldsymbol{s}_i} = \sum_{ 1\leq k<l\leq n_i } f_{ii}^{(kl)} s_i^{(k)} s_i^{(l)} \ket{\pm \boldsymbol{s}_i}.
\end{equation}
Therefore, within the logical subspace, the evolution generated by $H_i$ only yields a global phase that can be ignored.

The second part of the Hamiltonian, Eq.~\eqref{eq:H:ij}, is also diagonal in the computational basis, and its eigenvalues are given by
\begin{equation*}
    H_{ ij } \ket{\boldsymbol{s}_i, \boldsymbol{s}_j} = \sum_{\substack{1 \leq k \leq n_i \\ 1 \leq l \leq n_j }} f_{ i j }^{(kl)} s_i^{(k)} s_j^{(l)} \ket{ \boldsymbol{s}_i, \, \boldsymbol{s}_j },
\end{equation*}
that we conveniently denote 
\begin{equation}\label{eq:eigenvalues}
    \lambda_{ij} \equiv \sum_{\substack{1 \leq k \leq n_i \\ 1 \leq l \leq n_j }} f_{ i j }^{(kl)} s_i^{(k)} s_j^{(l)} = \boldsymbol{s}_i^T \boldsymbol{F}_{ij} \boldsymbol{s}_j,
\end{equation}
where $\boldsymbol{F}_{ij}$ is a $n_i \times n_j$ non-negative matrix with components given by $( \boldsymbol{F}_{ij} )_{kl} = f_{i j}^{(kl)}$. We call $\boldsymbol{F}_{ij}$ the \textit{interaction matrix} of the $ij$-pair.

When the state of each set of qubits is restricted to the logical subspace, the eigenvalues of $H_{ij}$ are doubly degenerate and they are given by $\pm \, \boldsymbol{s}_i^T \boldsymbol{F}_{ij} \boldsymbol{s}_j$. In the logical basis, the action of $H_{ij}$ is given by
\begin{equation*}
    H_{ ij } \ket{ k^L \, l^L }_{ij} = (-1)^{ k + l } \, \lambda_{ij} \ket{ k^L \, l^L }_{ij}
\end{equation*}
for $k$ and $l \in \{ 0, 1 \}$. Therefore, up to a constant, the whole Hamiltonian of Eq.~\eqref{eq:H:tot} can be written as
\begin{equation}\label{eq:Ham2}
    H = \sum_{1\leq i<j \leq N} \lambda_{ij} \, Z^L_i Z^L_j
\end{equation}
where $Z^L_i$ is the Pauli-$Z$ operator acting on the logical subspace of $S_i$, i.e., $Z^L_i \ket{ \pm \boldsymbol{s}_i } = \pm \ket{ \pm \boldsymbol{s}_i }$, and $\lambda_{ij}$ is the effective coupling strength.

In summary, restricting each set into a logical subspace, we obtain an ensemble of logical qubits that interact pair-wise according to \textit{ZZ} interactions with a specific interaction pattern. The respective interaction strength, $\lambda_{ij}$, only depends on the coupling $\text{f}(x)$ and the spatial distribution of the physical qubits, via $\{ \boldsymbol{F}_{ij} \}$, and on the choice of the logical subspaces for each set, via $\{ \boldsymbol{s}_i \}$. In the following, we will show how this can be used to control the effective interactions between logical systems by proper choice of logical states.

\section{Control of the logical systems}
\label{sec:control:logical:system}

In this section, we discuss how local manipulations of logical qubits (initialization, unitaries and measurements) can be implemented. Then we explain how an arbitrary effective spin value for each physical qubit can be realized by flipping it at specific times during the evolution. Finally, we show how arbitrary logical interaction patterns $\lambda_{ij}$ can be simulated.

\subsection{Interactions inside logical sets}

The control of each logical qubit encoded on the set $S_i$ relies on the intrinsic interaction between the physical qubits in the set. 
For that reason, it will be useful to distinguish between three cases depending on the interaction graph of a set. The interaction graph of a set is given by vertices representing the qubits and edges between interacting pairs of qubits.

First, we say a set is \textit{fully-connected} if in its interaction graph every pair of vertices is connected by an edge. In the case of long-range interactions, $r\to \infty$, any set is fully-connected. Second, a set is said to be \textit{connected} if its interaction graph is connected, i.e. there is a path between any pair of vertices. A direct connection is not required. Finally, if this is not the case we say that a set is \textit{disconnected}.

In disconnected sets, local control of physical qubits does not suffice to fully control the corresponding logical qubit, while for connected and fully connected sets such control can be achieved solely by single-qubit operations. Notice that one may also achieve full control within a logical set by other means, e.g. by utilizing controllable gates as in a small-scale quantum processor. In this case, our protocols become much easier, as one only needs to consider the manipulation of interactions between different sets.


\subsection{State preparation: connected sets}
\label{sec:state:preparation:connected:sets}

The very first thing to consider is how to initialize each set in the logical subspace by means of local operators on physical qubits. Since the logical states, $\ket{0^L}_i= \ket{\boldsymbol{s}_i}$ and $\ket{1^L}_i = \ket{-\boldsymbol{s}_i}$ are product, they can be prepared just by measuring each physical qubit in the $Z$-basis followed by a correction operation on each physical qubit. Nevertheless, as the logical systems are coupled via a pairwise \textit{ZZ} interaction, Eq.~\eqref{eq:Ham2}, the computational states are eigenstates of the interaction Hamiltonian and preparing them is uninteresting.

In order for the interactions to generate entanglement between logical systems, they need to be prepared on a different basis. A good example is the logical \textit{X}-basis, which corresponds to entangled states $\ket{+^L}_i =  (\ket{\boldsymbol{s}_i} + \ket{-\boldsymbol{s}_i})/\sqrt{2}$ (GHZ states). This can be done by first preparing each set in the $\ket{\boldsymbol{1}}_i = \ket{(1,\dots,1)}_i$ state. Then, we apply the Hadamard gate, H, to one of the physical qubits and we transfer the qubit state to the logical subspace by performing a sequence of control gates between the physical qubits, i.e.,
\begin{equation}\label{eq:initialization:GHZ}
    \text{CX}_i^{(1,\text{all})} \, \text{H}_i^{(1)} \ket{\boldsymbol{1}}_i = \frac{1}{\sqrt{2}} \big( \ket{\boldsymbol{1}}_i + \ket{-\boldsymbol{1}}_i\big),
\end{equation}
where $\text{CX}_i^{(1,\text{all})}$ is a sequence of $n-1$ control-\textit{X} gates of the from $\prod_{k=2}^{n_i} \text{CX}_i^{(l_k k)}$ for any $\{ l_k \}_{k=2}^{n_i}$ fulfilling $l_k < k$, e.g., $l_k = 1$ or $l_k = k-1$. The Hadamard gate and the control-$X$ gate are given in the computational basis by $\text{H} \ket{k} = (\ket{0} + (-1)^k \ket{1})/\sqrt{2}$ and $\text{CX} \ket{i}\ket{j}= \ket{i}\ket{j\oplus i}$ respectibely.

A control-\textit{X} gate between any pair of qubits can be obtained by letting them evolve under their intrinsic interaction $Z_k^{(i)} Z_k^{(j)}$ for a time $\tau = \pi/\big(4 f_{kk}^{(ij)}\big)$ with some extra single-qubit operations, see Fig.~\ref{fig:circuit:2a}. We can isolate the interaction of any pair $f_{kk}^{(ij)} Z_k^{(i)} Z_k^{(j)}$ by setting the effective spin value of all the other qubits to zero $s_k^{(l)}=0$ for $l \neq i,j$, as we explain in Sec.~\ref{sec:eff:spin}. This allows us to implement any sequence of control gates between the qubits of the set. In particular, in a fully-connected set, the sequence $\prod_{i=2}^{n_k} \text{CX}_k^{(1,i)} \mapsto \text{CX}_k^{(1,\text{all})}$ can be performed by consuming a time $\eta_0 = \sum_{j=2}^{n_k} \pi / \big(4 f_{kk}^{(1j)} \big)$.

In connected sets, it is possible that there is no qubit that couples to all the others, which prevents us from implementing $\prod_{k=2}^{n_k} \text{CX}_i^{(1k)}$. However, this is not an obstacle as there are other control-\textit{X} sequences that can be used, e.g., if there is a $\prod_{k=2}^{n_k} \text{CX}_i^{(k-1,k)} \mapsto \text{CX}_i^{(1,\text{all})}$. In particular, from the definition of a connected set, we always can find a sequence of control operations between coupling qubits to implement the logical Hadamard gate.

Note that in some cases, several control gates can be performed simultaneously, which allows us to reduce the implementation time. For instance, consider a connected set with the interaction given by $H_i = \sum_{k=2}^{n_i} f Z_i^{(1)} Z_i^{(k)}$. In this case, the sequence of Fig.~\ref{fig:circuit:2a} can be implemented in a time $\tau = \pi/(4f)$, as all control gates can be implemented simultaneously. This is a particular case where all qubits only couple to one particular qubit with the same coupling strength. This allows us to implement all control gates simultaneously. In other situations, only some of the control gates can be implemented simultaneously.

Notably, for finite range interaction between the physical qubits, a set can be disconnected, which in general prevents us from directly implementing the logical Hadamard gate. In this case, an alternative procedure to initialize the sets in the $\ket{+^L}_i$ is required, which is discussed in Sec.~\ref{sec:state:preparation:disconnected}.

\subsection{Logical unitary operations}
\label{sec:LU:control}

\begin{figure}
    \subfloat[\centering]{
    \includegraphics[width=0.98\columnwidth]{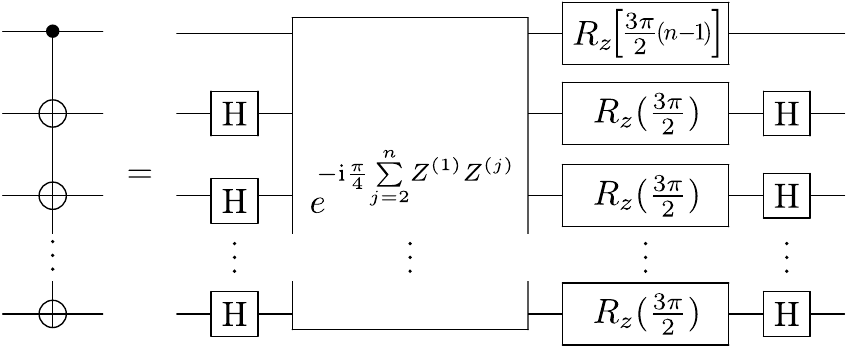}\label{fig:circuit:2a}} \hfill \\
    \subfloat[\centering]{
    \includegraphics[width=0.98\columnwidth]{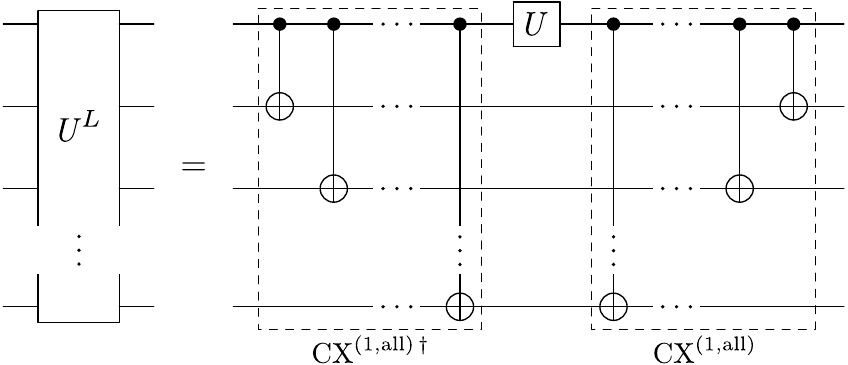}\label{fig:circuit:2b}} \hfill
    \caption{\label{fig:circuit22}In (a) quantum circuit implement a $\prod_{k=2}^n \text{CX}^{(1,k)}$ between several physical qubits by using the intrinsic interactions. In (b), a quantum circuit to implement an arbitrary single-qubit gate $U$ on a logical qubit.}
\end{figure}

An arbitrary logical single qubits gate, $U$, can be implemented on an arbitrary logical qubit by first localizing its state into one of the physical qubits (decoding), applying the gate on this physical qubit, and then delocalizing (encoding) the information again. This is done by applying the decoding/encoding operation $\text{CX}_i^{(1,\text{all})\,\dagger}$, i.e., given an arbitrary logical qubit state $\ket{\psi^L} = \alpha \ket{0^L} + \beta \ket{1^L}$ we obtain
\begin{equation}\label{eq:localizing}
\begin{aligned}
    \text{CX}_i^{(1,\text{all}) \, \dagger} \Big( \alpha \ket{s^{(1)}_i,\dots, s^{(n_i)}_i} + \beta \ket{-s_i^{(1)},\dots, -s_i^{(n_i)} } & \Big) \\
    = \left(\alpha \ket{s^{(1)}_i\Big\rangle + \beta \Big|-s^{(1)}_i}\right) \ket{s^{(2)}_i s^{(1)}_i, \dots, s^{(n_i)}_i s^{(l_{n_i})}_i } &
\end{aligned}
\end{equation}
where $\{ l_k \}_{k=2}^{n_i}$ determines the sequence of control gates used in $\text{CX}_i^{(1,\text{all})}$. Then, $U$ is applied to qubit $S_i^{(1)}$, i.e., if $U\ket{\psi} = \alpha' \ket{0} + \beta' \ket{1}$ then
\begin{equation*}
    U_i^{(1)} \left(\alpha \ket{s^{(1)}_i}+ \beta \ket{-s^{(1)}_i}\right) = \alpha' \ket{s^{(1)}_i} + \beta' \ket{-s^{(1)}_i}.
\end{equation*}
Finally, the state of qubit $S^{(1)}_i$ is delocalized again into the logical subspace by inverting Eq.~\eqref{eq:localizing}, i.e., applying $\text{CX}_i^{(1,\text{all})}$, and hence
\begin{equation}\label{eq:UL}
    U^L_i = \text{CX}_i^{(1,\text{all})} \, U_i^{(1)} \, \text{CX}_i^{(1,\text{all}) \, \dagger}.
\end{equation}
For instance, if one of the qubits couple to all the rest $U_i^L$ can be implemented as shown in Fig.~\ref{fig:circuit:2b}, otherwise we need to use a different sequence of control-$X$ gates to implement $\text{CX}_i^{(1,\text{all})}$.

Note that Eq.~\eqref{eq:UL} requires $2n-2$ two-physical-qubit gates. However, that is a general method and there are particular cases of gates that do not require any entangling gate between the physical qubits. For instance, an arbitrary \textit{Z}-rotation, i.e., $R_{\text{z}} (\phi) = \exp \{ -\ti \frac{\phi}{2} Z \}$ where $\phi \in [0,2\pi)$, of a logical qubit, can be obtained by individually rotating the physical qubits (or only one of them) of the underlying set in the \textit{Z} direction, i.e. the rotation of the logical qubit with the desired angle $\varphi$ is given by
\begin{equation*}
    R_{\text{z},i}^L (\varphi) = R_{\text{z},i}^{(1)} (\varphi_1) \otimes \cdots \otimes R_{\text{z},i}^{(n_i)} (\varphi_{n_i})
\end{equation*}
such that $\sum_k \varphi_k / s^{(k)} = \varphi$. Another example is the logical $X$ gate, which can be implemented by flipping each of the qubits in the set, i.e.,
\begin{equation*}
    X_i^L = X^{(1)}_i \otimes \cdots \otimes X^{(n_i)}_i.
\end{equation*}


Note that in the case of connected sets, the above method makes single qubit control sufficient to implement any single qubit operation in the logical subspace. However, to make use of the self-interaction term, the state of the set must leave the logical subspace as such subspaces are inherently insensitive to the self-interactions, see Eq.~\eqref{eq:DFS}. Therefore, to implement a logical unitary we need to ``turn off'' all interaction of a set with all external systems, while the constituting physical qubits interact. For a time $\tau$, this is achieved by flipping all the qubits of that set at time $\tau/2$, i.e., 
\begin{equation*}
\begin{gathered}
    \left( \bigotimes_{k=1}^{n_1} X_1^{(k)} \right) e^{-\ti H \tau/2} \left( \bigotimes_{l=1}^{n_1} X_1^{(l)} \right) e^{-\ti H \tau/2} \\
    = e^{-\ti \, \left( \sum_{k=1}^N H_k + \sum_{2\leq i<j\leq N} H_{ij} \right) \, \tau},
\end{gathered}
\end{equation*}
where we have used that $X e^{-\ti Z t} X = e^{\ti Z t}$, and $H$, $H_{ij}$, $H_i$ are given in Equations~\eqref{eq:H:tot}, \eqref{eq:H:i} and \eqref{eq:H:ij}. Iterating this step, we can disconnect all sets for a certain time $\tau$ which allows us to perform unitary operations in different logical sets simultaneously.

In contrast, in disconnected sets the control of a logical qubit is limited. While arbitrary \textit{Z}-rotations and the logical \textit{X}-gate are non-entangling gates between the physical qubits and can be performed straightforwardly, entangled operations such as the logical Hadamard gate can not be directly implemented.

\subsection{Logical measurements}

The possibility to implement general unitary operations on the logical qubit as outlined above, allows one to perform any single qubit projective measurement by applying the desired basis change followed by the measurement in the logical $Z$-basis. The logical $Z$-measurement is accomplished by measuring any of the physical qubits in the $Z$-basis. Nevertheless, this is a rather costly procedure, requiring entangling operations between the physical qubits.

An alternative possibility follows from the results of Ref. \cite{walgate2000local}, where it is shown that any two orthogonal $K$-qubit states can be deterministically distinguished by means of local operations and classical communication alone. This implies that any logical observable can be measured by a sequence of local operations on physical qubits, provided that the state of the system is initially restricted to a two-dimensional subspace as is the case for our logical qubits. Thus, for both connected and disconnected sets, the state of the logical qubit can be read out on any basis by only performing a local operation on the physical qubits.

We now briefly illustrate the procedure following \cite{walgate2000local}. The main observation is that any two $K$-qubit orthogonal states $\big|\psi\big\rangle$ and $\big|\psi^\perp\big\rangle$ can always be written as
\begin{eqnarray*}
    \big|\psi\big\rangle_{A\dots K} &=& \big|a_1\big\rangle_A \, \big|b\big\rangle_{B \dots K} + \big|a_2\big\rangle_A \, \big|c\big\rangle_{B \dots K} \vspace{0.015in} \\
    \big|\psi^\perp\big\rangle_{A\dots K} &=& \big|a_1\big\rangle_A \ket{b^\perp}_{B \dots K} + \big|a_2\big\rangle_A \ket{c^\perp}_{B \dots K}
\end{eqnarray*}
where $\big\{ \ket{a_1}, \ket{a_2} \big\}$ is an orthonormal basis, $\big\{ \ket{b}, \big|b^\perp\big\rangle \big\}$ and $\big\{ \ket{c}, \big|c^\perp\big\rangle \big\}$ are two pairs of non-normalized but orthogonal states. After measuring the first qubit in the basis $\big\{\ket{a_1}, \ket{a_2} \big\}$, the problem reduces to distinguishing between the orthogonal states $\ket{b}$ and $\ket{b^\perp}$ (or $\ket{c}$ and $\ket{c^\perp}$) of the remaining $K-1$ qubits. By recursively repeating the procedure for $K$ steps one is able to distinguish between the original states by combining the results of all the local measurements. This procedure in general requires classical communication and adaptive measurements. Note that these measurements do not project the state of the system into $\ket{\psi}$ or $\big|\psi^\perp\big\rangle$ but in a known random product state.

If we now apply this procedure to distinguish between two orthogonal single logical qubit states, $\ket{\psi^L}$ and $\big|\psi^{\perp L}\big\rangle$, to one part of an arbitrary $N$ logical qubits state, $\ket{\Phi^L}_{1\dots N}$, the probabilities for the corresponding outcomes are given by
\begin{eqnarray*}
    \text{Pr}(\psi|\Phi) &=& \big\langle \Phi^L \big| \left( \big|\psi^L \big\rangle\big\langle\psi^L \big|_1 \otimes \id_{2\dots N} \right) \ket{\Phi^L} \\
    \text{Pr}(\psi^\perp|\Phi) &=& \big\langle \Phi^L \big| \left( \big|\psi^{\perp L} \big\rangle\big\langle\psi^{\perp L}\big|_1 \otimes \id_{2\dots N} \right) \ket{\Phi^L},
\end{eqnarray*}
and the respective post-measurement states by
\begin{equation*}
\begin{gathered}
    \frac{1}{\sqrt{\text{Pr}(\psi|\Phi)}} \left( \big|\tilde{\psi}\big\rangle\big\langle\psi^L\big|_1 \otimes \id_{2 \dots N} \right) \ket{\Phi^L}_{1\dots N} \vspace{0.01in} \\
    \frac{1}{\sqrt{\text{Pr}(\psi^\perp|\Phi)}} \left( \big|\tilde{\psi}^\perp \big\rangle\big\langle \psi^{\perp L} \big|_1 \otimes \id_{2 \dots N} \right) \ket{\Phi^L}_{1 \dots N},
\end{gathered}
\end{equation*}
where $\big|\tilde{\psi}\big\rangle$ and $\big|\tilde{\psi}^\perp \big\rangle$ are two $n_1$-physical-qubit product states that depend on the outcome of the $n_1$ single-qubit measurements performed in the process. Therefore, after performing a correction operation on $S_1$ to set its state to $\ket{\psi^L}$ or $\ket{\psi^{\perp L}}$, the whole procedure is equivalent to performing the projective logical measurement given by $\{ \big|\psi^L\big\rangle\big\langle\psi^L\big|,\big|\psi^{\perp L}\big\rangle\big\langle\psi^{\perp L}\big| \}$.

\begin{figure*}
    \centering
    \includegraphics[width=2.01\columnwidth]{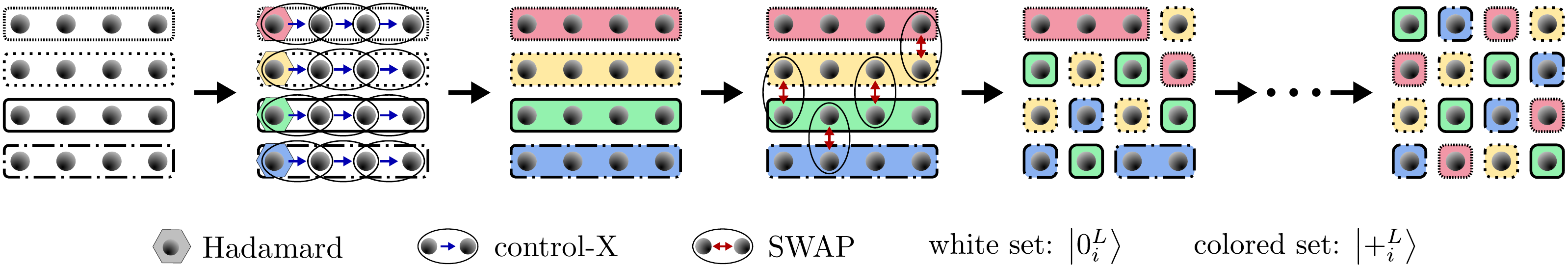}
    \caption{\label{fig:NN:FR0} We show the initialization procedure for the logical systems in a square lattice with nearest neighbour interactions. First the qubits are initialized in the state $\ket{0^L}_i$ and grouped in four connected sets. Next, we prepare each set in state $\ket{+^L}_i$ by applying a Hadamard gate to the first qubit in each row, followed by three control-\textit{X} gates between n.n. from left to right within each set. Finally, we distribute the logical qubits in the lattice by means of SWAP operations between physical qubits of different sets.}
\end{figure*}

\subsection{State preparation: disconnected sets}
\label{sec:state:preparation:disconnected}

Let us now discuss how to initialize all logical qubits in the $\ket{+^L}_i$ state in the case where they are encoded in disconnected sets. Given that at least two subsets of a disconnected set do not interact, it is impossible to prepare the genuinely multipartite entangled state $\ket{+^L}_i$ by considering the physical qubit of the set alone. Nevertheless, as we now show it is made possible by going back to the global picture where the physical qubits are not yet divided into logical sets. In other words, the preparation of the desired states of the logical qubits relies on their interaction with physical qubits from other sets.

First, we group the qubits in $N$ connected sets. We prepare each set in the logical 0 states, i.e., set $S_i$ is prepared in the state $\ket{0^L}_i = \ket{\boldsymbol{s}_i}$ for an arbitrary $\boldsymbol{s}_i$, by projecting the state of each physical qubit on the \textit{Z}-basis with an extra single physical qubit correction operation. Next, we prepare a GHZ state as detailed in Sec.~\ref{sec:state:preparation:connected:sets}. Then, note that the state of two physical qubits within their interaction range can be interchanged by performing a SWAP gate, as $\text{SWAP}^{(kl)} \ket{\psi \, \phi}^{(kl)} = \ket{\phi \, \psi}^{(kl)} \; \forall \, \psi \, \phi$, where it can be implemented as a sequence of three control-$X$ gates, i.e., $\text{SWAP}^{(kl)} = \text{CX}^{(kl)} \text{CX}^{(lk)} \text{CX}^{(kl)}$. The implementation of a CX gate between any two connected physical qubits is described in Sec.~\ref{sec:state:preparation:connected:sets}. Therefore, by performing several SWAP gates between physical qubits of different sets, we can arbitrarily distribute (or delocalize) the logical qubits and obtain an arbitrary grouping while keeping them in an entangled state, see Fig.~\ref{fig:NN:FR0} for an illustrative example. Note that with \textit{k} qubits in a 1D lattice with nearest neighbour interactions at most $(k^2 - k)/2$ SWAP operations are necessary to implement any permutation of the qubits. Therefore, in a 2D lattice where there are more physical qubit-qubit interactions, the upper bound on the number of required operations is significantly reduced. Observe, that by means of SWAP gates we could also implement any logical operation on the logical sets, but it is costly.

\subsection{Effective spin values}
\label{sec:eff:spin}

In Sec.~\ref{sec:setting}, we have shown that the interaction between any pair of logical qubits depends on the corresponding logical subspaces, i.e., the interaction strength between qubits encoded on the sets $S_i$ and $S_j$ depends on vectors $\boldsymbol{s}_i$ and $\boldsymbol{s}_j$. The vectors $\{ \boldsymbol{s}_i \}$ are restricted to have integer components of the form $s_i^{(k)} \in \{-1, 1\}$, as each component corresponds to (twice) the spin value of the physical qubits. In this section, we show how we can overcome this restriction, and obtain an arbitrary effective non-integer spin value for each physical qubit. This provides us with the possibility of implementing a logical qubit in any logical subspace given by a vector with non-integer components, i.e., $s_i^{(k)} \in \left[ -1, 1 \right]$, and leads to more freedom in the control of the interactions \cite{sekatski2020optimal,W_lk_2020}. 

In an ensemble of multiple physical qubits with commuting interactions, such effective non-integer spin values can be achieved by inducing spin flips at some specific times during the evolution. For that, the flipping time has to be negligible compared to the speed of phase accumulation. To illustrate this, we consider $m$ physical qubits that interact with a pair-wise \textit{ZZ} interaction. The evolution of the qubits is then given by
\begin{equation*}
    O_0(t) = e^{-\ti \sum_{1\leq i<j \leq m} f^{(ij)} Z^{(i)} Z^{(j)} t}.
\end{equation*}
If we consider the evolution for a fixed but arbitrary time $\tau$, we can establish an effective non-integer spin value for one of the qubits by flipping it at an intermediate time of the evolution, i.e., if we flip qubit-1 at $\tau_1$ we obtain
\begin{equation*}
\begin{gathered}
    O_1(\tau) = X^{(1)} \, O_0(\tau - \tau_1) \, X^{(1)} \, O_0(\tau_1) \\
    = e^{-\ti \, \left( \sum_{k=2}^m f^{(1k)} s^{(1)} Z^{(1)} Z^{(k)} + \sum_{2\leq i < j \leq m} f^{(ij)} Z^{(i)} Z^{(j)} \right) \tau}
\end{gathered}
\end{equation*}
where $s^{(1)} = 1 - (2\tau_1 / \tau) \in[-1,1]$. Note that all interactions associated with qubit-1 are multiplied by $s^{(1)}$ while the others remain untouched. This reduction in interaction strength can be interpreted as an effective spin value of the corresponding qubit, see also Eq. \ref{eq:eigenvalues}.

Next, we concatenate the evolution given by $O_1(\tau-\tau_2)$ and $O_1(\tau_2)$ by flipping qubit-2 in between, where $\tau_2 = \big(1+s^{(2)}\big)/2$. This results in $O_2(\tau)$, an evolution where interactions associated to qubit-2 are affected by a factor of $s^{(2)}$ while other interactions remain as in $O_1(\tau)$, i.e.,
\begin{equation*}
\begin{gathered}
    O_2(\tau) = X^{(2)} \, O_1 (\tau - \tau_2) \, X^{(2)} \, O_1(\tau_2) \\ 
    = e^{-\ti \, \left( f^{(12)} s^{(1)} s^{(2)} Z^{(1)} Z^{(2)} + \sum_{k=3}^m f^{(1k)} s^{(1)} Z^{(1)} Z^{(k)} \right. } \hspace{0.6in}
    \\ \hspace{0.6in} ^{\left. + \sum_{k=3}^m f^{(2l)} s^{(2)} Z^{(2)} Z^{(l)} + \sum_{3\leq i < j \leq m} f^{(ij)} Z^{(i)} Z^{(j)} \right) \tau}
\end{gathered}
\end{equation*}
Therefore, we can iterate this step until we establish an effective spin value for each of the qubits, i.e., if we define
\begin{equation}\label{eq:Oj}
\begin{gathered}
    O_j(\tau) \equiv X^{(j)} \, \widetilde{O}_{j-1} (\tau - \tau_j) \, X^{(j)} \, O_{j-1}(\tau_j),
\end{gathered}
\end{equation}
for $2\leq j \leq m$ with
\begin{equation*}
    \widetilde{O}_{j-1}(\tau) \equiv O_{j-2}(\tau_{j-1}) X^{(j-1)} O_{j-2}(\tau- t_{j-1}) X^{(j-1)},
\end{equation*}
then we obtain
\begin{equation*}
    O_m(\tau) = e^{-\ti \sum_{1\leq i< j\leq m} f^{(ij)} s^{(i)} s^{(j)} Z^{(i)} Z^{(j)} \tau}.
\end{equation*}
Observe that $\widetilde{O}_j(\tau) = O_j(\tau)$, however, it makes a difference in the resulting gate sequence. In particular, if we use $\widetilde{O}_k(\tau)$, the number of \textit{X}-gates is reduced by half as it contains some terms of the form $(X^{(i)})^2 = \id$ that we can ignore. See Appendix~\ref{app:sec:flipping} for a particular example.

Note that the total number of fast flips required, $\chi$, is given by $\chi \leq 2^m$. The exponential scaling applies to the general setting of long-ranged interactions, where all qubits interact pairwise. However, in many relevant cases one  needs a  considerably reduced  number of flips. In particular, if the interaction graph is not fully-connected, some physical qubits do not couple with each other. This allows us to flip them  ``in parallel'' which reduces the number of iterations of the procedure explained before, see Appendix~\ref{app:sec:flipping2} for the detailed analysis. For instance, in the case of the physical qubits arranged in a square lattice with n.n. interactions, the number of flips is given by $\chi_{\text{n.n.}} \leq (m/2)^2 + 2m$, and if the lattice also contains diagonal interactions $\chi_{\text{diag}} \sim (m/4)^4$. In the case of fully-connected interaction graphs, the physical qubits that support the same logical qubit are effectively decoupled. Again we can flip these qubits ``in parallel'', which allows us to establish the effective spin values with a number of flips given by $\mathcal{O}\left(n^N\right)$ where $n$ is the size of the sets. 


Notice that other ways to manipulate effective spin values are conceivable. First, one can in principle place $r$ physical qubits at the same position (or very close to each other), which leads to an effective spin value of $s_i^{(k)} = \pm r$ for the system when using only states $\ket{0}^{\otimes r}$ and $\ket{1}^{\otimes r}$. Intermediate integer values are possible by using states of the form $\ket{0}^{\otimes l} \ket{1}^{\otimes (r-l)}$. Second, effective spin values essentially represent the coupling strength of the system. Depending on how the coupling is induced -e.g. via external laser fields, or e.g. by dipole-dipole interactions of Rydberg ions, direct manipulation of the coupling strength via change of some of the parameters might be possible. For example by changing (spectrally adjusting) the dipole moments of individual atoms via locally induced Stark shifts as proposed in~\cite{Sekatskii_2003}, or by choosing the used Rydberg level $n$.

\subsection{Control of interactions}
\label{sec:int:control}

In Sec.~\ref{sec:setting}, we have shown that the logical qubits couple with each other via a pairwise \textit{ZZ} interaction, where the coupling strength of the $ij$-pair is given by $\lambda_{ij} = \boldsymbol{s}_i^T \boldsymbol{F}_{ij} \boldsymbol{s}_j$ in Eq.~\eqref{eq:Ham2}. From this expression, one can see that we can modify the interaction strengths between the logical qubits by changing either the spatial distribution of the physical qubits, i.e., the interaction matrices $\boldsymbol{F}_{ij}$, Eq.~\eqref{eq:eigenvalues}, or the logical subspaces, i.e., vectors $\boldsymbol{s}_i$. Given a target interaction pattern described by a set of interaction strengths $\{ \lambda_{ij} \}$, we aim to find a spatial distribution for the physical qubits and a logical subspace for each set that reproduces that interaction pattern. Formally, this corresponds to find a set of matrices $\{ \boldsymbol{F}_{ij} \}$ and a set of vectors $\{ \boldsymbol{s}_i \}_{i=1}^N$ which fulfill the equations
\begin{equation}\label{eq:hells:equation}
    \left\{ \boldsymbol{s}_i^T \boldsymbol{F}_{ij} \boldsymbol{s}_j = \lambda_{ij} \right\}_{1 \leq i < j \leq N}.
\end{equation}
In this article, we consider the positions of the physical qubits fixed. We thus aim to reproduce target interaction pattern $\lambda_{ij}$ by a clever choice of the logical vectors $\{\boldsymbol{s}_i\}$. As discussed above, to manipulate these vectors it suffices to have local control of the internal degrees of freedom of the individual subsystems (the physical qubits). Formally, we want to find a solution of a system of $N(N-1)/2$ non-linear Eqs.~\eqref{eq:hells:equation} for the variables $\boldsymbol{s}_1,\dots, \boldsymbol{s}_N $ and fixed sets of interaction matrices $\{ \boldsymbol{F}_{ij} \}$ and the target interaction pattern $\{ \lambda_{ij} \}$.

Our goal is to derive a sufficient condition for the interaction matrices $\{ \boldsymbol{F}_{ij} \}$ that guarantees that the corresponding system of equations can be solved for any target interaction pattern $\{ \lambda_{ij} \}$. We conclude that generically, i.e., assuming $\boldsymbol{F}_{ij}$ being a general random matrix, $dim(\boldsymbol{s}_i) \geq i-1$ suffices to ensure the existence of such set of vectors. In other words, if set $S_i$ contains $n_i = i - 1$ physical qubits (with the exception of $S_1$ containing $n_1 = 1$), given any target interaction pattern we can always find a set of logical subspaces that generates it. This implies the total number of physical qubits required to simulate $N$ logical qubits is given by $[N (N-1) / 2 ] +1$. We show this statement by providing an algorithm to solve such kind of non-linear systems of equations:

\begin{center}
\noindent\makebox[\linewidth]{\rule{\columnwidth}{0.4pt}}
\textbf{Algorithm-1}
\end{center}
\vspace{-16pt}
\noindent\makebox[\linewidth]{\rule{\columnwidth}{0.4pt}}

\begin{enumerate}
    \item[]\textit{Input}: A set of $N(N-1)/2$ matrices $\{ \boldsymbol{F}_{ij}\}_{i<j}$ and a real value for each element $\{ \lambda_{ij} \}_{i<j}$.
    
    \item Construct the following non-linear system of $N(N-1)/2$ equations
    \begin{equation}\label{eq:algortihm:1}
        \left\{ \boldsymbol{x}_i^T \boldsymbol{F}_{ij} \boldsymbol{x}_j = \lambda_{ij} \right\}_{1 \leq i < j \leq N},
    \end{equation}
    where vectors $\{ \boldsymbol{x}_i \}_{i=1}^N$ are the variables.
    
    \item Take a random real vector $\boldsymbol{s}_1$ and assign it to $\boldsymbol{x}_1$, i.e., $\boldsymbol{x}_1 := \boldsymbol{s}_1$.
    
    \item Consider the equation containing $\boldsymbol{F}_{12}$, i.e.,
    \begin{equation}\label{eq:19}
        \boldsymbol{s}^T_1 \boldsymbol{F}_{12} \boldsymbol{x}_2 = \lambda_{12}.
    \end{equation}
    As $\boldsymbol{x}_1$ is already fixed, this corresponds to a linear equation for $\boldsymbol{x}_2$. Find the solution set of Eq.~\eqref{eq:19}, and choose a random solution $\boldsymbol{s}_2$. Assign that solution to vector $\boldsymbol{x}_2$ i.e., $\boldsymbol{x}_2 := \boldsymbol{s}_2$.

    \item Consider the two equations involving $\boldsymbol{F}_{13}$ and $\boldsymbol{F}_{23}$, i.e., $\boldsymbol{s}_1 \boldsymbol{F}_{13} \boldsymbol{x}_3 = \lambda_{13}$ and $\boldsymbol{s}_2 \boldsymbol{F}_{23} \boldsymbol{x}_3 = \lambda_{23}$. The equations form a linear system of two equations for $\boldsymbol{x}_3$ and hence they can be written as
    \begin{equation*}
        \begin{pmatrix}
            \boldsymbol{s}_1^T \boldsymbol{F}_{13} \\
            \boldsymbol{s}_2^T \boldsymbol{F}_{23}
        \end{pmatrix}
        \boldsymbol{x}_3
        = 
        \begin{pmatrix}
            \lambda_{13} \\ \lambda_{23}
        \end{pmatrix} .
    \end{equation*}
    Find the solution set of the system, choose a random solution $\boldsymbol{s}_3$ and assign it to vector $\boldsymbol{x}_3$, i.e., $\boldsymbol{x}_3 := \boldsymbol{s}_3$.

    \item Iterate step 3 for the rest of vectors, i.e., consider equations involving $\{ \boldsymbol{F}_{i,k} \}_{i=1}^{k-1}$ and solve the linear system of $k-1$ equations, i.e.,
    \begin{equation*}
        \begin{pmatrix}
            \boldsymbol{s}_1^T \boldsymbol{F}_{1k} \\ \vdots \\
            \boldsymbol{s}_{k-1}^T \boldsymbol{F}_{k-1,k}
        \end{pmatrix}
        \boldsymbol{x}_k
        = 
        \begin{pmatrix}
            \lambda_{1k} \\ \vdots \\ \lambda_{k-1,k}
        \end{pmatrix},
    \end{equation*}
    pick a solution $\boldsymbol{s}_k$ and assign it to vector $\boldsymbol{x}_k := \boldsymbol{s}_k$.
    
    Note that system $k$ can be solved for any values of $\{\lambda_{ij}\}$ if and only if the vectors
    \begin{equation*}
        \left\{ \boldsymbol{s}_1^T \boldsymbol{F}_{1k}, \dots, \boldsymbol{s}_{k-1}^T \boldsymbol{F}_{k-1,k} \right\}
    \end{equation*}
    are linearly independent. We call this condition the \textit{LI condition}.
    
    \item \textit{Extra.} If vector components are bounded as $x_i^{(k)} \in [-1,1]$, divide all vectors $\{ \boldsymbol{s}_i \}$ by $\big(\max_{i,k} |s_i^{(k)}|\big)^2$. This scales all interaction strength by
    \begin{equation*}
        \lambda_{ij} \to \frac{\lambda_{ij}}{\Big(\underset{k,l}{\max} \big|s_l^{(k)}\big|\Big)^2}.
    \end{equation*}
    
    \item[] \textit{Output}: A set of vectors $\{ \boldsymbol{s}_i \}$ fulfilling Eq.~\eqref{eq:algortihm:1}. If variables are bounded, this algorithm provides a set of vectors that fulfil Eq.~\eqref{eq:algortihm:1} up to some scaling.
    
    \item[*] Find in Appendix~\ref{app:sec:eqsystem.example} a detailed application of this algorithm for a particular qubit distribution and interaction pattern.
\end{enumerate}
\vspace{-0.5cm}
\noindent\makebox[\linewidth]{\rule{\columnwidth}{0.4pt}}
\vspace{0.15cm}

\vspace{10 pt}
Observe that given a set of matrices $\{ \boldsymbol{F}_{ij} \}$, algorithm-1 provides a particular solution of the system of equations for any set $\{ \lambda_{ij} \}$ if in each step the LI condition is fulfilled. A necessary condition for fulfilling the LI condition is given by $n_k \geq k-1$ if $\boldsymbol{F}_{ij}$ is a $n_i \times n_j$ matrix (and hence $dim(\boldsymbol{s}_i)\geq i-1$ except for $sim(\boldsymbol{s}_1) \geq 1$). The LI condition is not always fulfilled, but if we assume generic matrices the probability of non-being fulfilled is vanishing, i.e., only for particular configurations of measure zero we do not obtain a solution. If in some step the LI condition is not fulfilled, we can restart the procedure with a different choice of the vectors in the previous steps. However, it may be the case that the system has no solution and hence the algorithm would always be stuck at some point. In that case, we could rearrange the qubit positions or choose a different grouping. We have not encountered such a problem in any of the examples we considered.

Due to physical limitations, i.e., $s_i^{(k)} \in [-1,1]$, clearly there is a maximum effective coupling strength that can be established between any pair of logical qubits. For that reason, for a given interaction pattern, we are not just interested in finding a particular solution that generates it. Moreover, we want to obtain the set of vectors $\{ \boldsymbol{s}_i \}$ that generate the target interaction pattern with the maximal coupling strength. As shown in the next section, while algorithm 1 demonstrates the existence of a solution in a constructive way, it does not provide an optimal solution with respect to the coupling strength. We, therefore, use other optimization techniques.

To simulate a generic two-body interaction pattern $\lambda_{ij}$ for $N$ logical qubits we need to solve the Eq.~\eqref{eq:hells:equation}. There, the total number of physical qubits corresponds to the number of variables, while the number of tunable parameters specifying the interaction pattern, i.e. $N(N-1)/2$, gives the number of equations. Thus, for a generic solution to exist, the total number of physical qubits must be larger or equal to $N(N-1)/2$. We have seen that considering completely general spatially distributed qubits, $n_i = i-1$ qubits in set $S_i$ guarantees that any interaction pattern can be simulated (up to some scaling). This means we need at least $1 + \sum_{i=1}^{N-1} i = [ N (N-1) / 2] + 1$ physical qubits in total (which is roughly minimal). However, if we want to have the same number of qubits in each set in order to have systems of the same size, we need $N(N-1)$ physical qubits, as following algorithm-1 one always needs $N-1$ qubits in $S_N$. 

We stress that this method is fully general, and allows one to produce arbitrary interaction patterns. This includes in particular the possibility to set certain interaction strengths to zero, thereby inducing a specific topology. In addition, the strengths of interactions can be varied at will, allowing one to mimic any desired geometry and also simulate random interaction strengths corresponding to some disordered model. 

\begin{figure*}
    \centering
    \subfloat[\centering]{\includegraphics[width=0.20\columnwidth]{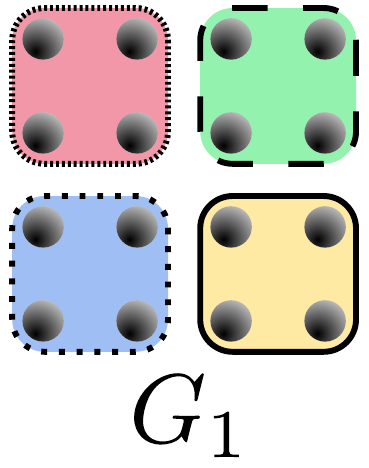} \label{fig:toy:modelg1}} \hspace{0.133in}
    \subfloat[\centering]{\includegraphics[width=0.20\columnwidth]{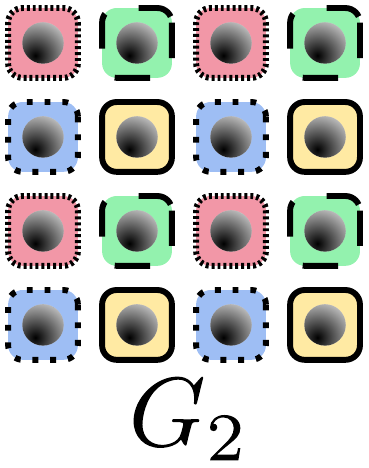} \label{fig:toy:modelg2}} \hspace{0.133in}
    \subfloat[\centering]{\includegraphics[width=0.20\columnwidth]{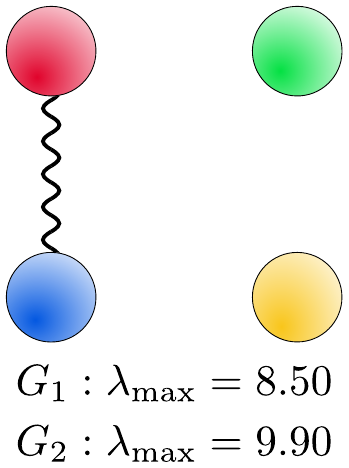} \label{fig:toy:model1} } \hspace{0.133in}
    \subfloat[\centering]{\includegraphics[width=0.20\columnwidth]{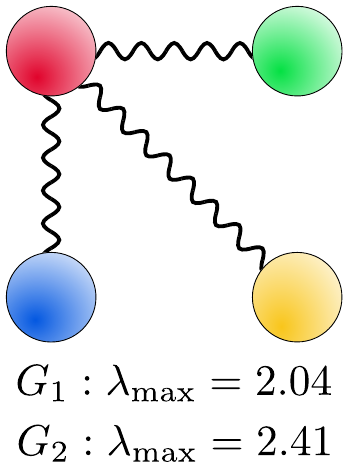} \label{fig:toy:model2} } \hspace{0.133in}
    \subfloat[\centering]{\includegraphics[width=0.20\columnwidth]{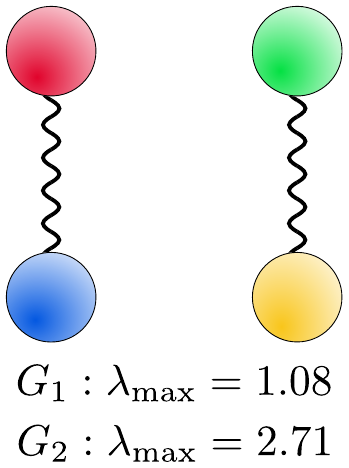} \label{fig:toy:model3} } \hspace{0.133in}
    \subfloat[\centering]{\includegraphics[width=0.20\columnwidth]{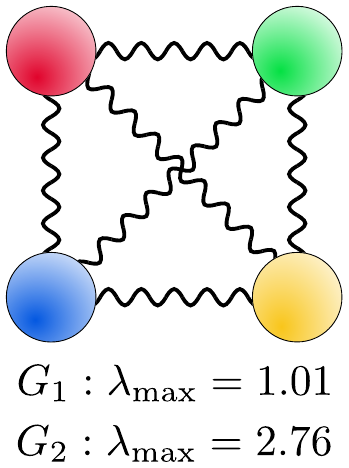} \label{fig:toy:model4} } \hspace{0.133in}
    \subfloat[\centering]{\includegraphics[width=0.20\columnwidth]{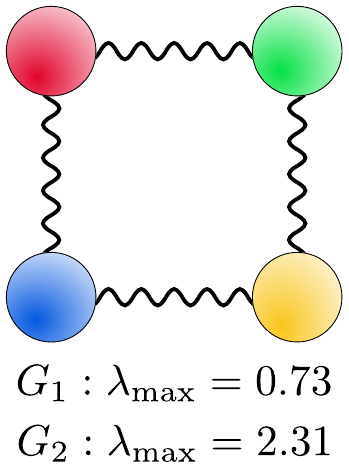} \label{fig:toy:model5} } \hspace{0.133in}
    \subfloat[\centering]{\includegraphics[width=0.20\columnwidth]{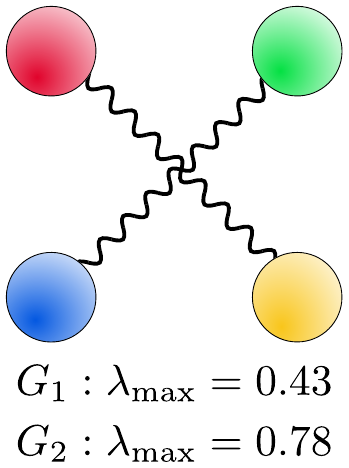} \label{fig:toy:model6} } \hfill
   
    \caption{\label{fig:toy:model} In (a) and (b) we show a $4 \times 4$ square lattice of physical qubits where the qubits are grouped in four sets where a wavy line links the logical qubit are implemented. We assume all qubits interact with each other with a distance dependence coupling, given in Eq.~\eqref{eq:interaction:range} with $\alpha=1$ and $r \to \infty$. In (c)-(h), we show different interaction patterns where two logical qubits if their interaction strength is given by $\lambda_{ij} = \lambda_{\max} J /\delta$, and a pair is not connected if they interaction strength is $\lambda_{ij} = 0$.}
\end{figure*}

\begin{table*}
\begin{tabular}{|c|c|c|c|c|}
    \hline
    $\;$ Fig.~\ref{fig:toy:model} $[G_2]\;$ & $\boldsymbol{s}_1$ & $\boldsymbol{s}_2$ & $\boldsymbol{s}_3$ & $\boldsymbol{s}_4$ \\ \hline
    c & $(1,1,1,1)$         & $(0,0,0,0)$        & $(1,1,1,1)$         & $(0,0,0,0)$         \\
    d & $(1, 1, 1, 1) $     & $(1, -0.64, 1, -1)$& $(0.29,1,-1,0.35)$  & $(-0.29,-0.19,1,1)$ \\
    e & $(1,-0.06,1,-0.44)$ & $(-0.06,1,-0.44,1)$& $(1,-0.44,1,-0.06)$ & $(-0.44,1,-0.06,1)$ \\
    f & $(1,0.09,0.09,1)$   & $(0.09,1,1,0.09)$  & $(0.09,1,1,0.09)$   & $(1,0.09,0.09,1)$   \\
    g & $(1,1,1,-1)$        & $(1,1,-1,1)$       & $(1,-1,1,1)$        & $(-1,1,1,1)$        \\
    h & $(1,-0.52,-0.52,1)$ & $(-0.52,1,1,-0.52)$& $(-0.52,1,1,-0.52)$ & $(1,-0.52,-0.52,1)$ \\ \hline
\end{tabular}
\caption{\label{tab:g2} Logical subspace for each logical qubit to generate the different interaction patterns of Fig.~\ref{fig:toy:model} with grouping $G_2$.}
\end{table*}

\section{Applications}
\label{sec:applications}

In this section, we analyse in detail several cases of particular interest with different physical interaction ranges, target systems and interaction patterns. 

First, in Sec.~\ref{sec:Frange}, we consider cases with a full interaction range where the intrinsic qubit-qubit interaction strength decreases as the inverse of the distance, i.e., $r\to \infty$ and $\alpha = 1$ in Eq.~\eqref{eq:interaction:range}. This is the most general situation and due to a large number of physical interactions, a quadratic scaling in the number of physical qubits in each logical set is required to simulate  general interaction patterns. 

Then, in Sec.~\ref{Sec:nn}, we consider settings with a finite interaction range, e.g., nearest neighbour (n.n.) interactions. This is a model of particular interest, given that it is commonly used to describe various physical systems. There, we show that our setting can still be used to reproduce fully interacting many body systems. Moreover, a finite interaction range of the physical qubits allows us to achieve linear scaling in the number of physical qubits when one is only interested in generating finite-range interaction patterns.

\begin{figure*}    
    \subfloat[\centering]{\includegraphics[width=0.7\columnwidth]{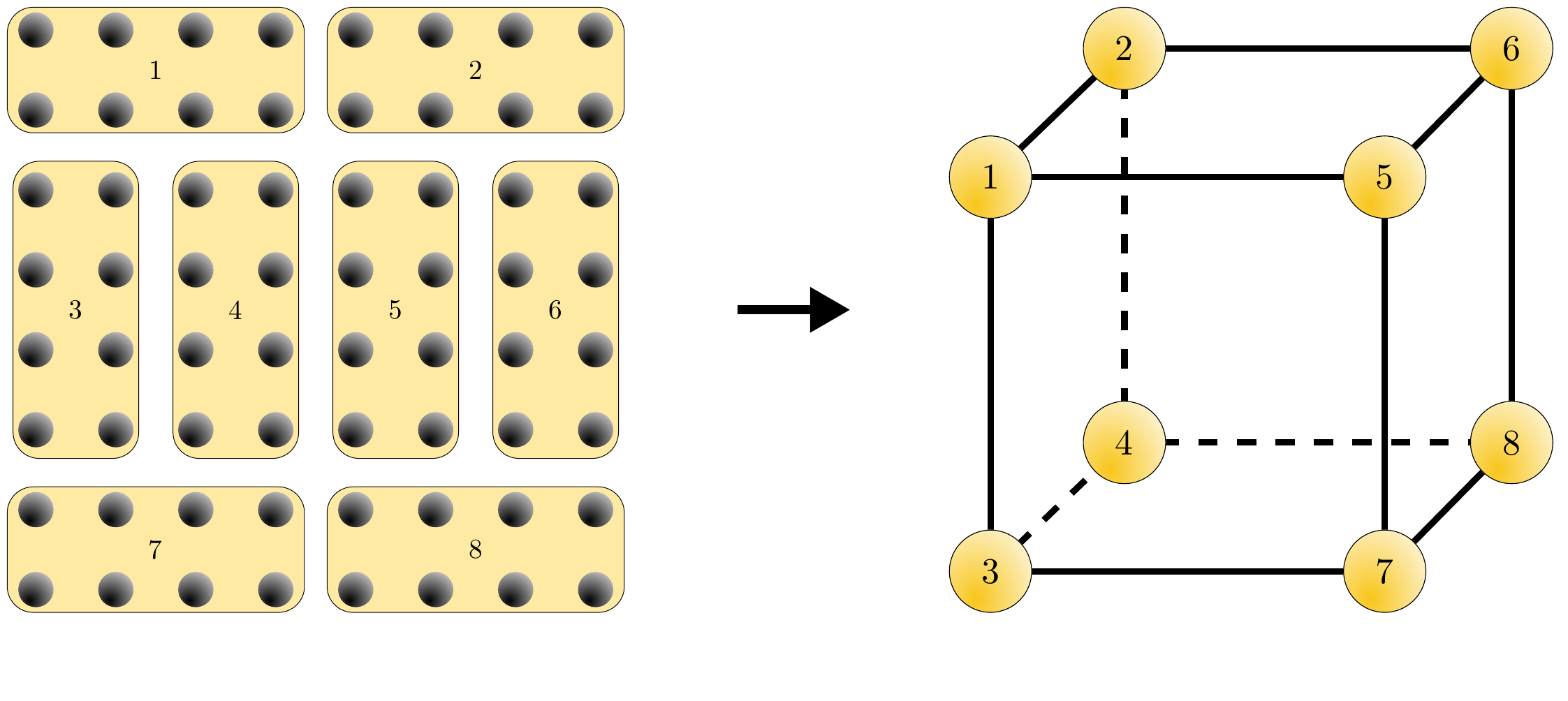} \label{fig:cube}} \hspace{0.6in}
    \subfloat[\centering]{\includegraphics[width=0.95\columnwidth]{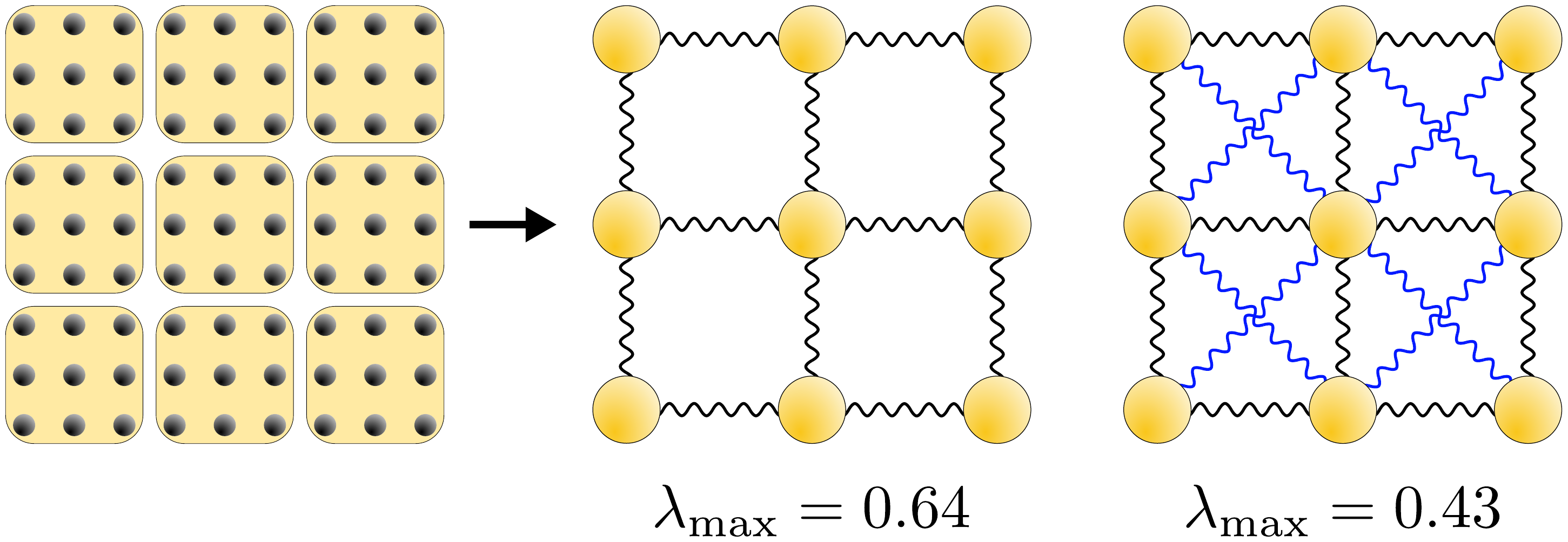} \label{fig:FR-NN}}
    \caption{\label{fig:5} (a) A $8 \times 8$ square lattice of physical qubits grouped in eight blocks of eight qubits each. By properly choosing internal states, this leads to a system of 8 logical qubits arranged on a cube that interact according to this 3d geometry. (b) A $9 \times 9$ square lattice of physical qubits grouped in nine blocks of nine qubits each. By properly choosing internal states, an n.n. interaction pattern and an n.n. and diagonal interaction pattern can be implemented. Black wavy lines represent interactions with an interaction strength of $\lambda_{ij} = \lambda_{\max} J / \delta$, and blue wavy lines represent interactions with an interaction strength of $\lambda_{ij} = (\lambda_{\max}/\sqrt{2}) (J / \delta)$.}
\end{figure*}

\subsection{Full-range intrinsic qubit-qubit interaction}
\label{sec:Frange}

\subsubsection{Arbitrary interaction patterns in a 4 \texttimes 4 square lattice}
\label{sec:toymodel}

First, we consider a simple but illustrative example of 16 physical qubits in a $4 \times 4$ square lattice. In this setting, we group the qubits into sets of four qubits each. Restricting the state of each set into a logical subspace, the whole system is made to behave as four logical qubits with the interaction pattern depending on the chosen logical subspaces. Given a target interaction pattern we construct the logical subspaces that simulate it by finding the set of vectors that fulfils Eq.~\eqref{eq:hells:equation}. Then, from any interaction pattern, we can switch to another pattern just by changing the logical subspace of each set without perturbing the state of the whole system. We consider two different ways to group the qubits. In the grouping $G_1$, we group the four neighbouring qubits of each corner, see Fig.~\ref{fig:toy:modelg1}. In the grouping $G_2$, we group the qubits in a way that the sets are spread through the lattice, as we show in Fig.~\ref{fig:toy:modelg2}.

In Figures~\ref{fig:toy:model1}-\ref{fig:toy:model6}, we show some different interaction patterns that can be generated on the logical level. For the considered patterns, some interactions are turned off, while the others are set to the same strength, i.e., $\lambda_{ij} = 0$ (no line connecting the qubits in Figs.~\ref{fig:toy:model1}-\ref{fig:toy:model6}) or $\lambda J/\delta$ where $\delta$ is the separation between two n.n. physical qubits. In Fig.~\ref{fig:toy:model}, we also provide the maximum value of $\lambda$ we could find for each interaction pattern. To obtain the larger coupling strength for each pattern we used numerical optimization algorithms\footnote{For each interaction pattern, we used the Wolfram Mathematica function ``NMaximize'' to find the set of vectors that leads to a maximum of the coupling strength. For some patterns, the function did not converge and we used ``FindMaximum''. Both functions may output a local maximum as the functions are non-concave.}, as algorithm-1 introduced in Sec.~\ref{sec:int:control} does not provide optimal solutions. In Tables~\ref{tab:g2} and~\ref{tab:g1} we show the logical subspaces for each logical qubit in order to generate the interaction patterns shown in Fig.~\ref{fig:toy:model}.

Observe that in this particular case, for both groupings the maximum interaction strength depends on the target interaction pattern. We point out the two qubits interaction pattern, Fig.~\ref{fig:toy:model1}, which reaches the maximum interaction strength of $\lambda_{\text{max}} = 8.50$ for $G_1$ and $9.90$ for $G_2$. Since only two qubits interact, we can set each of the two interaction qubits to the logical subspace given by $\boldsymbol{s}_{1,3} = (1,1,1,1)^T$ what maximizes the interaction strength, and decouple the rest of qubits by setting them to $\boldsymbol{s}_{2,4} = (0,0,0,0)^T$. On the other hand, note that if we also want to establish an interaction of the same strength with the other pair of qubits, i.e., generate the interaction pattern of Fig.~\ref{fig:toy:model4}, the maximum interaction strength is considerably reduced. This is because in this interaction pattern we have to impose the interaction pairs to be insensitive to the other pair, unlike in the first pattern where the other qubits were just decoupled. Each of the rest of the interaction patterns shows a different maximum interaction strength, and for some of them, the interaction strength is smaller than 1. This means the effective interaction strength is weaker than the direct physical qubit-qubit interaction given by $J/\delta$.

Notice as well that grouping $G_2$ provides a larger maximum interaction strength between the logical qubits. This is because, in $G_2$, distances between qubits of different sets are on average smaller than in $G_1$. This leads to higher effective coupling strengths. However, single logical gates which cannot be implemented by local operation on the physical qubits are faster implemented in grouping $G_1$ as the distances between qubits of each set are smaller, e.g., the time required to implement a logical Hadamard gate is twice larger in grouping $G_2$. Generally speaking, one can choose the best way of grouping the qubits, depending on the set of interaction patterns that one is interested in generating and the amount of required unitary operations on the logical qubits.

\subsubsection{Simulation of three-dimensional geometries from a two-dimensional lattice}

With a two-dimensional square lattice, it is also possible to reproduce interaction patterns that correspond to three-dimensional qubit arrangements. We have already encountered such an example in Fig.~\ref{fig:toy:model4}, where all four qubits interact with the same coupling strength. This reproduces the interaction pattern of four qubits located at the vertices of a regular tetrahedron.

We also can consider more complex target topologies. Grouping the qubits of an $8 \times 8$ square lattice in eight sets of eight qubits each as we show in Fig.~\ref{fig:cube}, we obtain a network of eight logical qubits. By generating the corresponding interaction pattern the logical qubits reproduce a lattice where each qubit is located in a vertex of a cube. Assuming that qubits in the cube interact with a coupling strength decreasing with the inverse of the distance, the interaction pattern between the logical qubits is given by an interaction strength of $\lambda_{ij} = \lambda J /\delta$ for adjacent vertices, $\lambda_{ij} = (\lambda/ \sqrt{2}) ( J /\delta )$ for non-adjacent vertices sharing a face, and $\lambda_{ij} = ( \lambda/\sqrt{3} ) ( J /\delta )$ for vertices diametrically opposed in the cube. Numerical optimization leads to a maximum achievable coupling strength of $\lambda_{\max} = 0.57$. Alternatively, one can simulate eight qubits in a cubic lattice with n.n. interactions, i.e., the qubits in the cube only interact with the qubits on adjacent vertices. In this case, the interaction pattern can be generated with a maximum effective interaction strength of $\lambda_{\max} = 0.31$. Find in Appendix \ref{app:sec:data}, Table \ref{tab:fig:5a}, the logical subspaces that generate the interaction pattern of Fig.~\ref{fig:cube}. Similarly as in the example in Sec.~\ref{sec:toymodel}, one can find other ways of grouping the qubits that lead to a higher maximum coupling.

\subsubsection{Reducing interactions range}

A case of particular interest consists of simulating a logical finite range interaction pattern, where each qubit only interacts with the qubits within its own neighbour, with an ensemble of physical qubits subject to full range interaction.

In a $9 \times 9$ square lattice we can implement 9 logical qubits, by grouping the qubits in groups of 9 as shown in Fig.~\ref{fig:FR-NN}. Assuming that the logical qubits are located in a $3 \times 3$ square lattice, we can implement an n.n. pattern and an n.n. and diagonal interaction pattern and compute the corresponding maximum coupling (see Fig.~\ref{fig:FR-NN}). The coupling between n.n. is given by $\lambda_{ij} = \lambda J /\delta$ and between diagonal neighbours by $\lambda_{ij} = (\lambda / \sqrt{2}) ( J /\delta )$ as they are at distance $\delta\sqrt{2}$. Find in Appendix \ref{app:sec:data}, Table \ref{tab:fig:5b}, the logical subspaces that generate the interaction patterns of Fig.~\ref{fig:FR-NN}.

\subsection{Finite range physical qubit-qubit interactions} \label{Sec:nn}

\subsubsection{Generating arbitrary interaction patterns from n.n. interaction}
\label{Sec:nn:1}

\begin{figure*}
    \subfloat[\centering]{\includegraphics[width=\columnwidth]{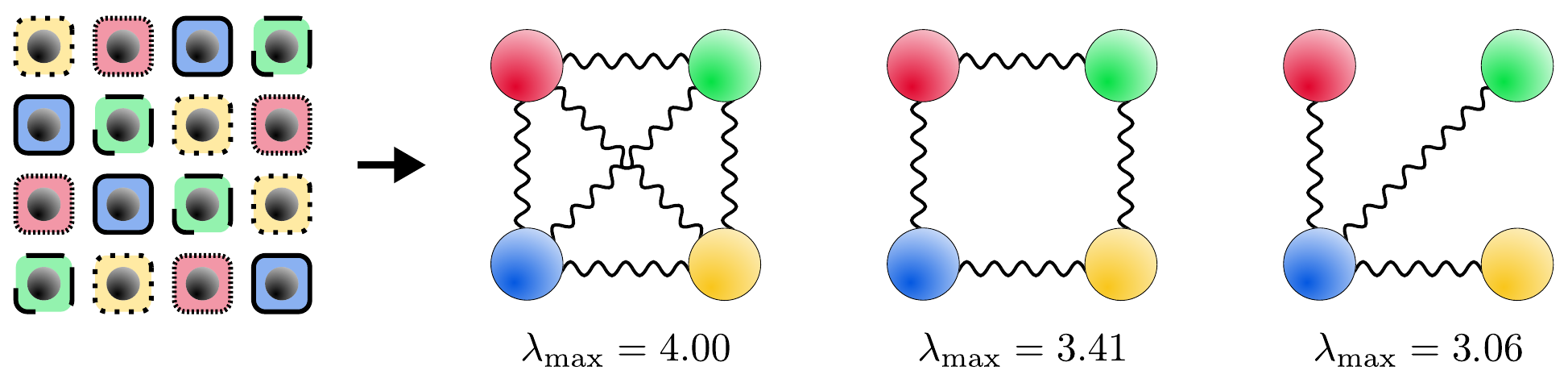} \label{fig:NN4:FR} } \hspace{0.3cm}
    \subfloat[\centering]{\includegraphics[width=\columnwidth]{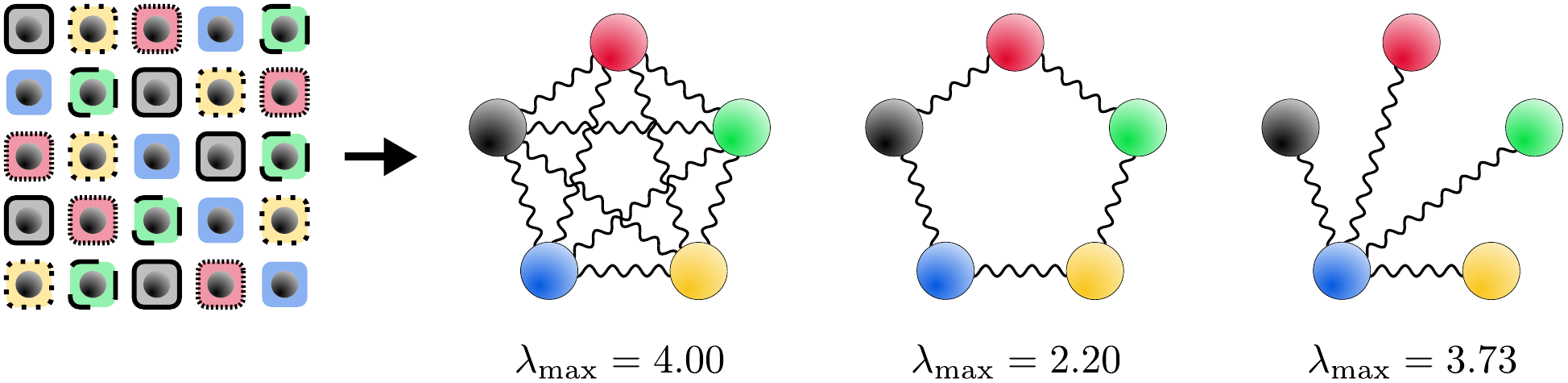} \label{fig:NN5:FR} }
    \caption{\label{fig:NN:FR} We consider square physical lattices with n.n. interactions where we implement full interacting logical qubits. In (a) [(b)] we show a $4 \times 4$ [$5 \times 5$] qubit lattice with n.n. interaction. We group the qubits in four [five] sets of four [five] qubits each, in such a way there are four physical interactions between any pair of sets. By properly choosing internal states, we can implement different interaction patterns with the maximal coupling strength $\lambda_{ij} = \lambda_{\max} J /\delta$.}
\end{figure*}

Now we consider a setting where the physical qubits are arranged in a square lattice with nearest neighbours interactions. Due to the finite range of the physical interactions, by locally grouping the physical qubits we obtain an ensemble where most of the interactions between logical qubits are zero, independently of the chosen logical subspaces. Note that in this case, our sufficient condition derived in Sec.~\ref{sec:int:control} does not apply because the interaction matrices $\boldsymbol{F}_{ij}$ correspond to singular cases where most of their entries are zero. However, there are particular ways of grouping the qubits that allow us to establish arbitrary interaction between the logical qubits. In particular, as each physical qubit only interacts with its surroundings, we need to delocalize (or spread out) the logical sets in the lattice, to obtain interactions between any pair of logical qubits.

In a $\ell \times \ell$ square lattice with $n=\ell^2$ physical qubits, there are $2\ell(\ell-1)$ physical qubit-qubit interactions. Once we group the physical qubits, the physical interactions are combined, resulting in the interactions between the logical qubits. Therefore to implement a full-connected ensemble of $N = \ell$ logical qubits we need to group the physical qubits in sets such that all pairs of sets couple to each other through at least one physical qubit-qubit interaction. In particular, to give the same weight to all logical qubit-qubit interactions, we can group the qubits in the lattice in a way that there is the same number of physical interactions (four in this case) between each pair of logical sets. 

On the left of Figures~\ref{fig:NN4:FR} and~\ref{fig:NN5:FR}, we show how to group the qubits in a way that all groups have four physical interactions with each of the other groups with 16 and 25 physical qubits. On the right of these figures, we show different interaction patterns using these groupings. All interactions that are represented by a wavy line have a coupling strength given by $\lambda_{ij} = \lambda J /\delta$, where $\delta$ is the distance between two n.n. physical qubits. Observe that by delocalizing the logical qubits we can obtain higher couplings strengths compared with the general case analyzed in Sec.~\ref{sec:toymodel}, as the distances between logical qubits are reduced.

For larger lattices, one can try to find a similar grouping to implement more logical qubits. However, we could not prove that such grouping always exists when implementing $N$ logical qubits in a $N \times N$ square lattice. Nevertheless, in Appendix~\ref{app:nn:scale}, we derive an alternative way of implementing $N$ logical qubits with $n = (5 / 9) N^{\log_2 6} - 4 \approx (5/9) N^{2.58} - 4$ physical qubits guaranteeing that between any pair of sets there are at least four physical qubit-qubit interactions in the same way as grouping showed in Fig.~\ref{fig:NN4:FR}. The idea is to divide the lattice into square blocks (sub-lattices) of 16 physical qubits. For each block one then chooses four logical qubits and assigns them to the physical qubits as depicted in Fig.~\ref{fig:NN4:FR}. This way any two logical qubits share at least four physical interactions if they appear in the same block once.

\subsubsection{Simulation of programmable chiral lattices with a linear scaling}
\label{sec:scalable:lattices}

In Sec.~\ref{sec:int:control}, we showed that to simulate an ensemble of $N$ logical qubits where all $N(N-1)/2$ interactions can be tuned, one needs a physical system of quadratically many physical qubits. Nevertheless, this scaling can be reduced if we only aim to obtain interaction patterns with some degree of locality, i.e., to obtain finite range interaction between the logical qubits. In this case, we can encode the logical qubits in sets of size independent of $N$.

\begin{figure}
    \centering
    \subfloat[\centering]{\includegraphics[width=0.9\columnwidth]{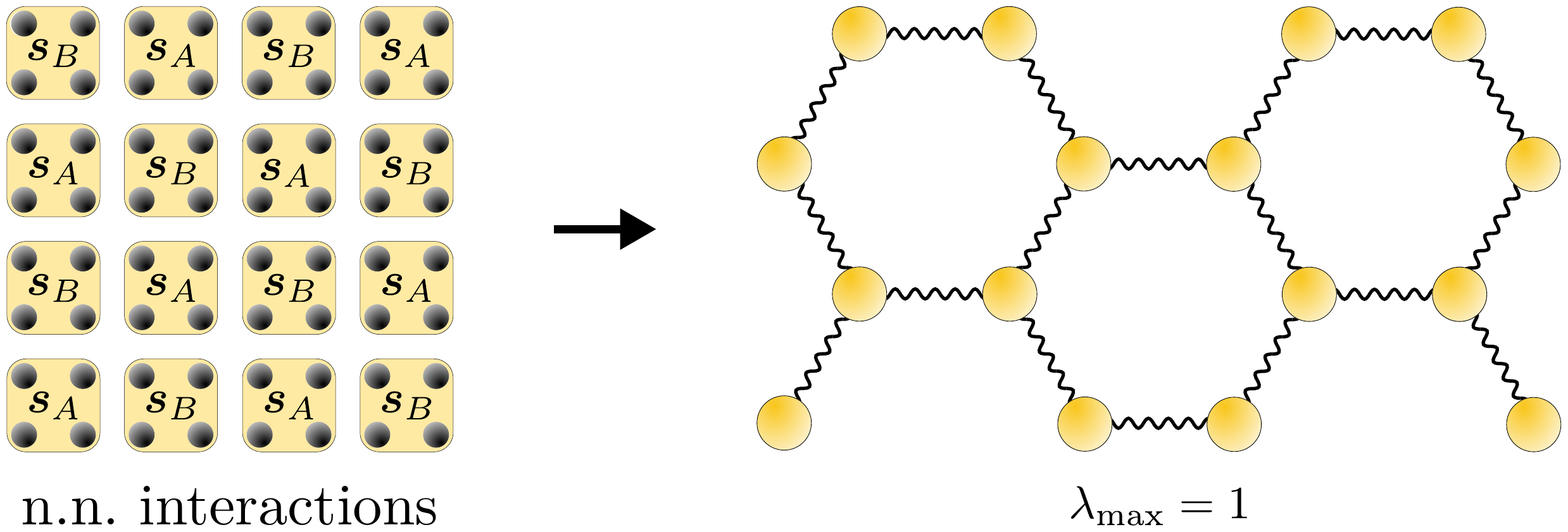} \label{fig:scalable:patterns:a} } \hfill
    \subfloat[\centering]{\includegraphics[width=0.9\columnwidth]{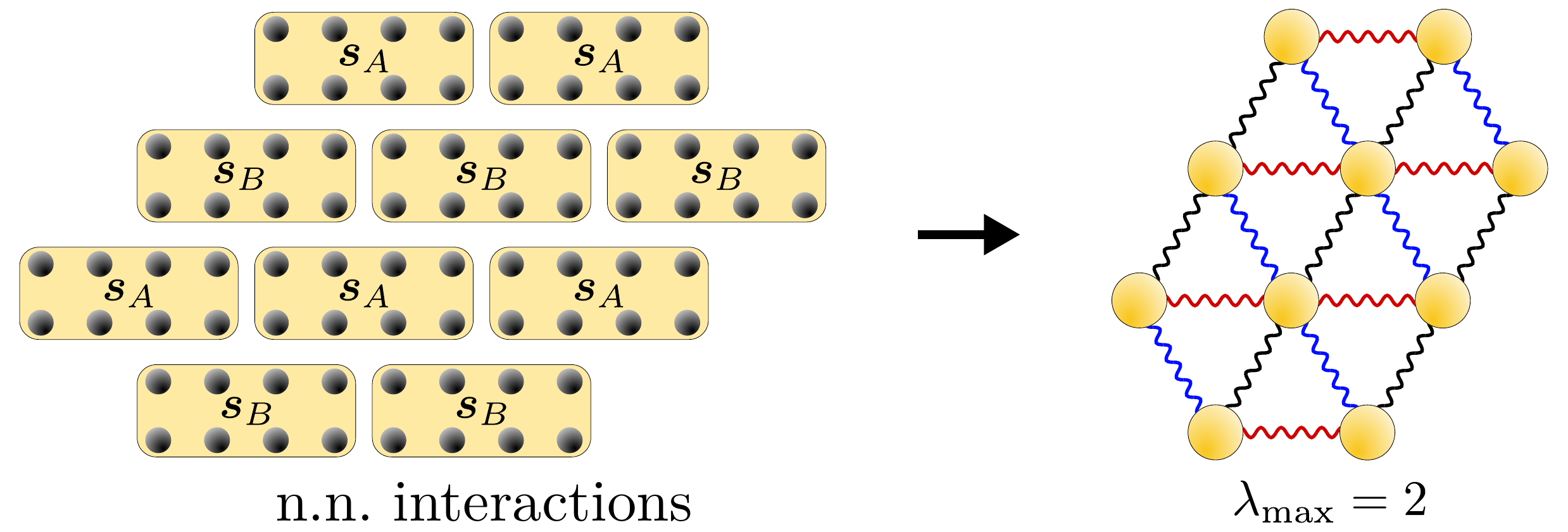} \label{fig:scalable:patterns:b} } \hfill
    \subfloat[\centering]{\includegraphics[width=0.9\columnwidth]{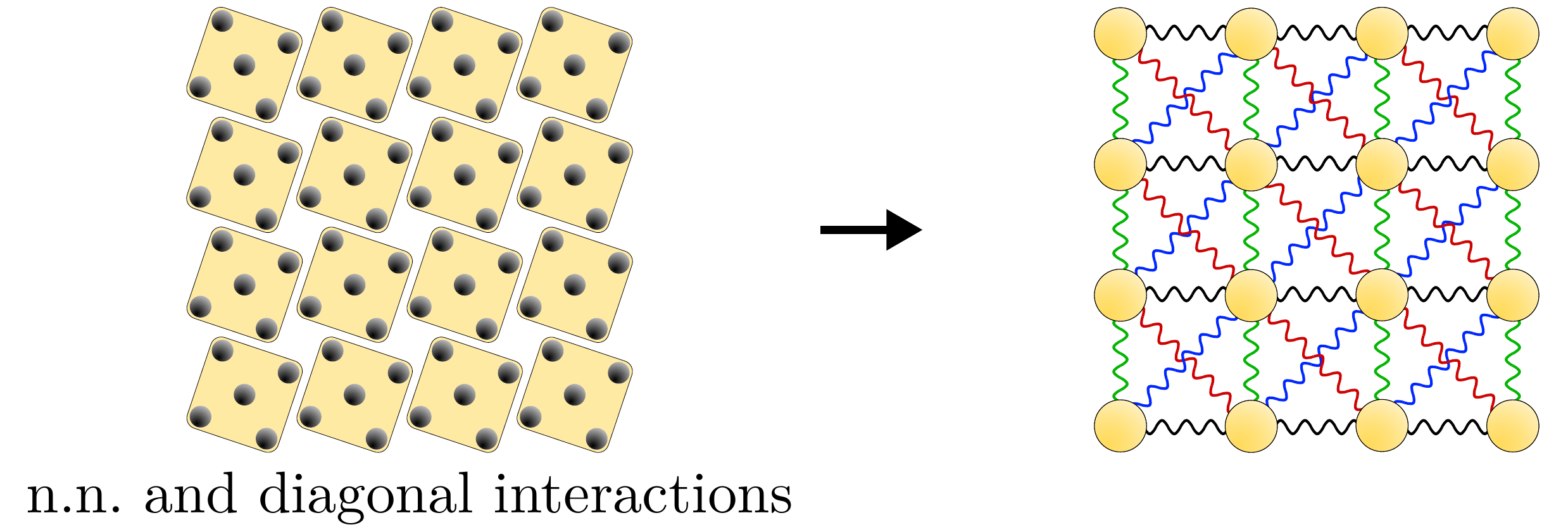} \label{fig:scalable:patterns:c} } \caption{\label{fig:scalable:patterns} In (a) and (b) we consider square latices of physical qubits with n.n. interactions. In (a) we reproduce an hexagonal interaction pattern. The maximum coupling is $\lambda_{\max} = 1$. In (b) we reproduce a triangular interaction pattern. The coupling strength in each direction (black, blue and red) can be independently tuned up to $\lambda_{\text{max}} = 2$. In (c) we consider n.n. and diagonal interaction in the physical lattice. There we reproduce a regular lattice with the same range of interaction where we can tune the coupling strength in each direction (black, blue, red and green). In all lattices, we impose periodic constraints to make the solution directly scalable. In (a) and (b) we use two logical subspaces, $\boldsymbol{s}_A$ and $\boldsymbol{s}_B$, which we alternate in the lattice. In (c) we use the same logical subspace for each set.}
\end{figure}

Consider a regular qubit lattice with a finite interaction range, e.g., n.n. interactions. Grouping the qubits in a localized way, e.g., as shown in Fig.~\ref{fig:scalable:patterns}, we do not obtain a fully interacting effective Hamiltonian but a finite interaction range, and we can not make directly couple any pair of sets. In this case, each logical qubit only couples to the $M$ sets within its interaction range and thus the total number of non-zero tunable interactions is given by $MN/2$. Notice, the number of tunable interactions for each logical qubit only depends on $M$ and it is independent of $N$. This implies the size of the logical qubits is independent of $N$, and hence, to simulate $N$ logical qubits we just need linearly many physical qubits.

We support this statement by considering different settings and finding solutions for an arbitrary number of logical qubits. To do so, we consider periodic patterns in the logical ensemble. Given a periodic target pattern and a specific grouping, we take a big enough number of sets such that we can find a set of vectors $\{ \boldsymbol{s}_i \}$ that generates the pattern while fulfilling a periodic repetition of logical subspaces in the sets. These restrictions on the logical qubits allow us to iterate the solution to extend the pattern for an arbitrary number of logical qubits.

In the first example, we reproduce $N$ logical qubits in a hexagonal lattice with n.n. interactions from a square physical lattice with n.n. interactions. We group the physical qubits in sets of four as shown in Fig.~\ref{fig:scalable:patterns:a}. To find a periodic solution we only use two different logical subspaces $\boldsymbol{s}_A$ and $\boldsymbol{s}_B$ which are alternated in each row and column. With this restriction we generate the pattern by setting $\boldsymbol{s}_A = (0, 1, 0, 1)^T$ and $\boldsymbol{s}_B = (0.83, 1, 0.17, 1)^T$ leading to the maximum coupling of $\lambda_{ij} = J /\delta$ with $\lambda_{\text{max}} = 1$.

The second example consists in obtaining $N$ logical qubits in a triangular lattice with n.n. interactions from a square physical lattice with n.n. interactions. This is a situation of particular interest as the number of interactions for each qubit is increased (from four to six neighbours). In Fig.~\ref{fig:scalable:patterns:b} we show how we group the physical qubits in sets of eight. Note that each set couples to its six neighbours with two physical qubit-qubit interactions and just by setting each logical qubit in the logical space given by $\boldsymbol{s} = (1, 1, 1, 1)^T$ we reproduce a triangular lattice with a coupling strength of $\lambda_{ij} = 2J /\delta$. In order to obtain more complex patterns, we impose the two following periodic conditions: (1) all sets from the same horizontal row are in the same logical sub-space, and (2) in each diagonal column, we alternate between two logical sub-spaces $\boldsymbol{s}_A$ and $\boldsymbol{s}_B$. With these constraints, we obtain that we can fully tune independently the interaction coupling in each direction of the lattice up to a maximum coupling of $\lambda_{ij} = \lambda_{\text{max}}J /\delta$ with $\lambda_{\text{max}} = 2$, see Appendix~\ref{app:sec:data} for particular solutions.

The last example we consider is shown in Fig.~\ref{fig:scalable:patterns:c}. In a physical square lattice with n.n. and diagonal interactions we group the qubits in sets of five to obtain a logical square lattice where we can tune the n.n. and diagonal interactions. If we impose the periodic condition that requires all the sets to be in the same logical sub-space, $\boldsymbol{s}_i = \boldsymbol{s}_j \; \forall \, ij$, we can tune the coupling strength in each of the four directions of the lattice. In this case, the coupling in each direction cannot be independently tuned, see Appendix~\ref{app:sec:data} for particular solutions.

\section{Extensions}
\label{sec:extensions}

In this section, we present several extensions of our approach. We show that:
\begin{itemize}
    \item A restriction to commuting physical interactions is not necessary, and in fact, any initial (non-commuting) two-body interaction can be used to generate effective, commuting interactions among the logical systems.
    \item One can use techniques from Hamiltonian simulation, i.e., intermediate fast control operations, to manipulate the logical \textit{ZZ}-interactions such that any target interaction Hamiltonian can be generated \cite{dodd2002universal}. Here we explicitly show how to reproduce an \textit{XYZ}-interaction and a multiqubit $Z$ interaction.
    \item One can also use multiple levels of the logical system to obtain a quantum simulator for \textit{d}-level systems.
\end{itemize}
In all cases, additional fast local control, but only on individual physical qubits, is required. Here control gates have to be repeatedly performed with frequencies that are much fast than the physical interaction timescale. This is in contrast to the manipulation of interaction patterns presented in Sec.~\ref{sec:control:logical:system}, which only requires a finite number of single physical operations depending on the size of the groups, as shown in Sec.~\ref{sec:eff:spin}. 

\subsection*{General physical interactions }

\subsection{Physical layer: bringing general two-body interactions to \textbf{ZZ}-type}
\label{sec:general:physical:interactions}

So far our analysis requires a commuting \textit{ZZ}-interaction between physical qubits to start with. We will now show that it can be extended to settings with general two-body interactions between physical qubits. This can be done because any two-qubit interaction can be reduced to \textit{ZZ}-coupling by means of local control.

In \cite{dodd2002universal}, it has been shown that given any two-body entangling Hamiltonians $\tilde H$ and $H_0$ on $m$ qubits, there is a decomposition of the form
\begin{equation}\label{eq:sumH}
    \tilde{H} = \sum_{j=0}^{4^m-1} \gamma_j \, H_j
\end{equation}
where $H_j = V_j H_0 V_j^\dagger$ is a local unitary transformation of $H_0$. Do not confuse $H_j$ with the self-interaction Hamiltonian of a logical set, Eq.~\eqref{eq:H:i}. In this Sec.~\ref{sec:general:physical:interactions}, we only consider physical qubits. Then by the Trotter formula,
\begin{equation*}
    e^{-\ti \tilde{H} t} = \lim_{k\to \infty} \left( \prod_{j=0}^{4^m-1} e^{- \ti H_j t_j/k} \right)^k
\end{equation*}
where $t_j = \gamma_j t$, we can approximate the evolution generated by $\tilde{H}$, by fast alternating between the evolution generated by $\{ H_j \}$, where these are achieved as
\begin{equation*}
    e^{-\ti H_j t} = V_j \, e^{-\ti H_0 t} \, V_j^{\dagger}.
\end{equation*}

Let us now show how to reproduce a two-body \textit{ZZ} Hamiltonian from an \textit{XYZ}-type interaction we need to find a set of local unitary operations $\{ V_k \}$ such that
\begin{equation}\label{eq:VHV:ZZ}
    \sum_{k=0}^{4^m-1} \gamma_k V_k \, H_{\textit{XYZ}} \, V_k^\dagger = \sum_{1\leq i<j \leq n} f^{(ij)}_{zz} Z^{(i)} Z^{(j)} 
\end{equation}
where
\begin{equation*}
    H_{\textit{XYZ}} = \!\!\! \sum_{1\leq i<j \leq m} \!\!\! f^{(ij)}_{xx} X^{(i)} X^{(j)} + f^{(ij)}_{yy} Y^{(i)} Y^{(j)} + f^{(ij)}_{zz} Z^{(i)} Z^{(j)}.
\end{equation*}

To find the set of unitary operations, we can use the procedure introduced in \cite{dodd2002universal}. First, we divide the ensemble into two subsets of the same size $\Omega_0$ and $\Omega_1$, applying a $Z$ flip to the qubits in $\Omega_0$ we obtain a local transformation of $H_{\textit{XYZ}}$ what combined with $H_{\textit{XYZ}}$ allows us to generate
\begin{equation}\label{eq:H'first}
    H' = \frac{1}{2} \left( H_{\textit{XYZ}} + W_1 \, H_{\textit{XYZ}} \, W_1 \right)
\end{equation}
where $W_1 = \bigotimes_{i \in \Omega_0} Z^{(i)}$. Note that $H'$ leaves invariant the interaction within $\Omega_0$ and $\Omega_1$, but projects into the \textit{ZZ} term the interaction between pairs of qubits in $\Omega_0$ with $\Omega_1$. Next we divide each subset $\Omega_i$ in $\Omega_{i0}$ and $\Omega_{i1}$ and we apply a $Z$ flip to all qubits in $\Omega_{00} \cup \Omega_{10}$ to obtain a local transformation of $H'$ that combined with $H'$ generates
\begin{equation*}
    H'' = \frac{1}{2} \left(H' + W_2 H' \, W_2 \right)
\end{equation*}
where $W_2 = \bigotimes_{i \in \Omega_{00}\cup\Omega_{10}} Z^{(i)}$, what leads to a \textit{ZZ} interaction between any pairs of qubits from different subsets. We proceed by iterating this step, i.e., applying a $Z$ flip to half of the qubits of each subset and combining the obtained Hamiltonian with the previous one. After a total of $\log_2 n$ steps, we obtain a Hamiltonian $H$ that projects all interactions into their corresponding \textit{ZZ} term. The final expression for $H$ is a sum of $n$ terms of local transformations of $H_{\textit{XYZ}}$, Eq.~\eqref{eq:VHV:ZZ}, where $\gamma_k = 1/m$ and the set of $m$ unitary operations is given by\footnote{Here \unexpanded{$\langle \left\{ a, b, c \right\} \rangle$} is the whole group generated by $a, b$ and $c$. Note that in this case the generators $W_k$ are commuting and $W^2_k = \id$.}
\begin{equation*}
    \left\{ V_k \right\} = \left\langle \left\{ W_k \right\} \right\rangle.
\end{equation*}

Note that on a lattice with a finite interaction range one can implement the same procedure with a fixed number of unitary operations that do not grow with $m$. For example, a square lattice with n.n. interactions can be split into two sub-lattices of non-interacting qubits by grouping qubits laying on even (odd) diagonals in $\Omega_0$ ($\Omega_1$). Then $H_{\textit{XYZ}}$ does not involve interactions between qubits belonging to the same group. Hence, already after the first step in Eq.~\eqref{eq:H'first}, the Hamiltonian $H'$ takes the desired \textit{ZZ}-coupling form.

For the most general two-qubit interaction pattern 
\begin{equation*}
    H = \! \sum_{1\leq i<j \leq m} \sum_{\substack{a,b \\ \in \{x,y,z\} }} \! f^{(ij)}_{ab} \, \sigma^{(i)}_a \, \sigma^{(j)}_b + \sum_{i=1}^m \sum_{\substack{a \\ \in \{x,y,z\} }} \! g^{(i)}_a \, \sigma^{(i)}_a,
\end{equation*}
with $\big(\sigma^{(i)}_x,\sigma^{(i)}_y,\sigma^{(i)}_{z}\big) = \big( X^{(i)}, Y^{(i)}, Z^{(i)}\big)$, the procedure is very similar. One can define two unitary transformations $W_1 = \bigotimes_{i \in \Omega_0} Z^{(i)}$ and $W_1' = \bigotimes_{i \in \Omega_1} Z^{(i)}$, such that 
\begin{equation*}
    H'= \frac{1}{4} \left(H + W_1 H W_1 + W_2 H W_2 + W_1 W_2 H W_1 W_2 \right)
\end{equation*}
only contains the interaction term $f^{(ij)}_{zz} Z^{(i)} Z^{(j)}$ for all pairs of qubits $(ij)$ belonging to different groups. Repeating the procedure recursively as described above, brings the Hamiltonian to the desired \textit{ZZ}-interaction form plus a local term $\sum_i g_z^{(i)} Z^{(i)}$, that can also be compensated with local control.

\subsection*{General interaction type}
\subsection{Simulating \textit{XYZ} interactions}
\label{sec:general:interaction:type}

We have discussed how our methods can be implemented with many-body systems where the physical interaction is of a general two-qubit type. Now we will show how general interaction beyond the \textit{ZZ}-coupling can be simulated on the logical level. We start by showing how to simulate any \textit{XYZ} model using standard techniques from Hamiltonian simulation.

The Hamiltonian describing the interaction between the $N$ logical qubits correspond to logical two-body \textit{ZZ} interaction, Eq.~\eqref{eq:Ham2}. From this available Hamiltonian, we can generate an arbitrary target Hamiltonian by applying standard Hamiltonian simulation techniques \cite{dodd2002universal,jane2002simulation,bennett2002optimal,dur2008quantum}. In a similar way, as we did in Sec.~\ref{sec:general:physical:interactions}, we can implement the evolution generated by any logical Hamiltonian of the form of Eq.~\eqref{eq:sumH} by alternating the evolution of local transformations of the original logical Hamiltonian. Any unitary local transformation of the two-body \textit{ZZ} Hamiltonian is of the form
\begin{equation*}
    V H_0 V^\dagger = \sum_{1 \leq i<j \leq N} \lambda_{ij} \left( \boldsymbol{n}_i \cdot \boldsymbol{\sigma} \right)_i \left( \boldsymbol{n}_j \cdot \boldsymbol{\sigma} \right)_j,
\end{equation*}
where $V$ is a local unitary and $(\boldsymbol{n} \cdot \boldsymbol{\sigma} )_i = \big(n_x X_i + n_y Y_i + n_z Z_i\big)$ with $\boldsymbol{n} = (n_x, n_y, n_z)$ being a real normalized vector. In this section, we only consider operations acting on the logical qubits, and hence we avoid super-index $L$ to simplify notation. 

Therefore, an \textit{XYZ}-type interaction (e.g. Heisenberg interaction)
\begin{equation*}
    \tilde{H}_{\textit{XYZ}} = \! \sum_{1\leq i<j \leq N} \! \alpha_{ij} \, X_i X_j + \beta_{ij} \, Y_i Y_j + \lambda_{ij} \, Z_i Z_j,
\end{equation*}
can be reproduced by fast alternating between
\begin{equation*}
\begin{aligned}
    H_0 & = \sum_{ 1\leq i<j \leq N } \lambda_{ij} \, Z_i Z_j \\
    H_1 & = \sum_{ 1\leq i<j \leq N } \alpha_{ij} \, X_i X_j \\
    H_2 & = \sum_{ 1\leq i<j \leq N } \beta_{ij} \, Y_i Y_j
\end{aligned}
\end{equation*}
where we can set any interaction pattern for $\{ \alpha_{ij} \}$, $\{ \beta_{ij} \}$ and $\{ \lambda_{ij} \}$ as shown in Sec.~\ref{sec:int:control}.

\subsection{Simulating many-body interactions}

Another possible direction of generalization that we now address, is to simulate many-body interactions.

A direct way of generating a multiple qubit $Z$ interaction between all qubits, i.e., simulating the evolution generated by $H = \omega Z_1 \cdots Z_N$, is to first apply a maximally entangling operation between the qubits, i.e.,
\begin{equation*}
    U \, e^{ -\ti \, \omega X_1 } \, U^\dagger = e^{ -\ti \, \omega A_1 Z_2 \cdots Z_N }
\end{equation*}
where
\begin{equation*}
    U = e^{ -\ti \sum_{j = 2}^N Z_1 Z_j \pi/4 }
\end{equation*}
and
\begin{equation}\label{eq:A}
    A = \left( \frac{\ti}{2} \right)^{N-1} [ \cdots [ [ X, Z], Z ], \cdots Z ],
\end{equation}
where in Eq.~\eqref{eq:A} we concatenate $N-1$ commutators and hence $A \in \{\pm X, \pm Y\}$. $U$ is implemented by establishing an interaction pattern that couples all qubits to $S_1$ with the same strength $\lambda$ and then letting the system evolve for a time $t = \pi/(4\lambda)$. Obviously, this method allows one to generate an effective many-body $Z$ interaction on any selected set of qubits. It is however quite costly, as to produce even a weak interaction requires the implementation of $U$ which is a ``maximally entangling operation'' (it can generate a $N$-qubit GHZ-state from a product state). Of course, there are also different ways to combine local and two-qubit gates in order to simulate a many-qubit interaction, e.g. $e^{- \ti \lambda Z_1 Z_2 Z_3 } = e^{-\ti \pi X_1 Z_2 /4 } e^{-\ti \lambda Y_1 Z_3} e^{\ti \pi X_1 Z_2 /4 }$.

Alternatively, multiple qubit interactions are accomplished by means of Hamiltonian simulation techniques. For instance, alternating the evolution generated by two Hamiltonians $H_A$ and $H_B$, one can approximate the evolution generated by the commutator $[H_A, \, H_B]$ for small times \cite{dur2008quantum}, as 
\begin{equation*}
    e^{ \ti H_A t } e^{ \ti H_B t } e^{ -\ti H_A t } e^{ -\ti H_B t } \approx e^{ -[H_A, H_B] t^2 }
\end{equation*}
We can use this technique to generate an effective three-qubit interaction by just manipulating one of the qubits. Setting the two interaction patterns
\begin{equation*}
\begin{gathered}
    H_A = \lambda_{12} X_1 Z_2 \\
    H_B = \mu_{13} Y_1 Z_3
\end{gathered}
\end{equation*}
one approximates the evolution given by the commutator
\begin{equation*}
    \left[ H_A, H_B \right] = 2 \, \ti \lambda_{12} \mu_{13} Z_1 Z_2 Z_3.
\end{equation*}
Unlike the first method, this one is approximate but can be faster to implement.

\subsection{Simulating \textit{d}-dimensional systems}

So far we were interested in simulating various interactions on an ensemble of logical qubits. Here, we extend our setting to arbitrary \textit{d}-dimensional logical systems interacting with each other.

In Sec.~\ref{sec:control:logical:system}, we showed that once an ensemble of $N$ logical qubits is established, we can implement any logical qubit gate and any interaction pattern can be reproduced. Therefore, a straightforward way to obtain logical qudits is to embed several logical qubits and treat them as a single system of higher dimensions. This way, assuming that implementing $N$ logical qubits requires $N(N-1)$ physical qubits, see Sec.~\ref{sec:int:control}, simulating $N'$ qudits of dimension $d = 2^k$ requires $k N'$ logical qubits. Therefore, the number of required physical qubits is given by $n = \log_2(d) N' (\log_2(d) N' - 1)$. Note that with this method when the number of physical qubits in each set increases quadratically, the dimension of the qudit scales exponentially. 

\begin{figure}
    \centering
    \includegraphics[width=0.9\columnwidth]{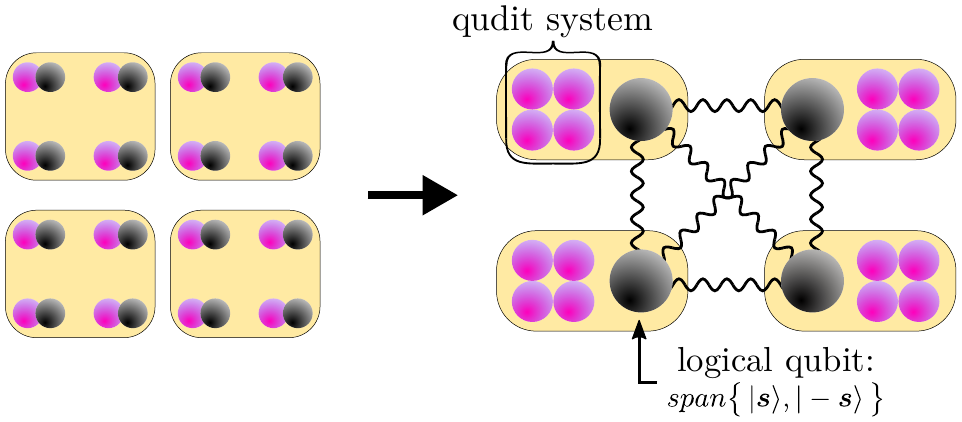}
    \caption{\label{fig:qudits} On the left, schematic representation of an ensemble of physical systems constituted of two two-dimensional subsystems (black and purple). Black subsystems couple to each other via a two-body \textit{ZZ} interaction. Purple subsystems are decoupled. On the right, is a schematic representation of the logical qudits (purple) and logical qubits (black) in each set. The logical qudits couple to each other through the logical qubits.}
\end{figure}

We also consider an alternative way of implementing logical qudits. In various setups, one has control over additional degrees of freedom of the physical qubits that do not couple to each other, e.g., mechanical degrees of freedom of trapped ions. Depending on the dimensional of these extra degrees of freedom, they can be used to implement logical qudits for various \textit{d}. Then, in each set, we implement a logical qubit and a qudit where the first one is used to control the second including the interactions with other logical systems. To see how this is achieved, consider a set $S_i$ of $n$ physical qubits where for each one we have control on an extra two-dimensional degree of freedom $e_i^{(k)}$, i.e., the Hilbert space of physical system $q_i^{(k)}$ is given by $\mathcal{H}_i^{(k)} = \mathcal{H}^{(k),s }_i \otimes \mathcal{H}^{(k),e}_i$. Qubit subsystems $\big\{ s_i^{(k)} \big\}$ interact with each other as described in Sec.~\ref{sec:setting}, while the extra qubit subsystems $\big\{ e_i^{(k)} \big\}$ are completely decoupled. Implementing a logical qubit in $\big\{ s_i^{(k)} \big\}$ as described in Sec.~\ref{sec:setting} and embedding all external degrees of freedom of the set, i.e., treating $\big\{ e_i^{(k)} \big\}$ as an indivisible $2^n$-dimensional system, we obtain a qubit and a qudit system in $S_i$, see Fig.~\ref{fig:qudits}. Full control of each physical system suffices to fully manipulate the qudit and its interactions. Once the logical qubit in $\big\{ s_i^{(k)} \big\}$ is established, any control operation between $e_i^{(k)}$ and $s_i^{(k)}$ corresponds to a control operation between the logical qubit and subsystem $e_i^{(k)}$. Therefore, through the logical qubit, we can perform any multiple qubit gate between $\big\{ e_i^{(k)} \big\}$, and thus, we get full control of the qudit, i.e., any gate $U_e \in \mathcal{H}_i^{(1),e} \otimes \cdots \otimes \mathcal{H}_i^{(n),e}$ can be decomposed as
\begin{equation*}
    U_e \otimes U_s = \prod_{j=1}^n U_j,
\end{equation*}
where $U_s$ is an arbitrary operation on the logical qubit system, and $U_j \in span\{ \ket{\pm \boldsymbol{s}_i} \} \otimes \mathcal{H}^{(j),e}_i$. In a similar way we can couple external subsystems of different sets through their respective logical qubits, and therefore implement any multiple qudit gate.

\section{Comparison with universal Hamiltonian simulation}
\label{sec:comparison}

As shown in \cite{dodd2002universal} given any intrinsic many-body entangling Hamiltonian, local control on the individual parties suffices to efficiently reproduce any other Hamiltonian. While different types of interactions are achieved by fast alternating between local transformations of the intrinsic Hamiltonian, different interaction patterns are generated by isolating two-body interactions and iterating them to mediate or cancel the interaction between other parties. Our setting can provide a significant enhancement of these techniques as it can lead to an increased coupling strength and a direct way of establishing different interaction patterns. These two features significantly reduce the required time in both processes at the price of a larger physical system.

To compare the performance of our approach with known schemes, we consider the particular task of simulating a triangular lattice with n.n. interactions from a physical square lattice with n.n. interactions, see Fig.~\ref{fig:scalable:patterns:b}. In this case, the aim is to reproduce a system with a larger number of interactions per qubit, as in the target pattern each qubit couples to six neighbours while in the original one each qubit only couples to four.

\subsection{Commuting interaction}

First, we consider the target interaction Hamiltonian corresponding to a two-body \textit{ZZ} interaction, i.e.,
\begin{equation*}
    \tilde{H}_{\textit{ZZ}} = \sum_{\langle i, j \rangle} \lambda \, Z_i Z_j,
\end{equation*}
where $\langle i, j \rangle$ refers to the n.n. pairs in a triangular lattice, see Fig.~\ref{fig:scalable:patterns:b}. This interaction is simple but allows one to efficiently prepare many qubit entangled states such as graph states, and can also be used as a resource to simulate more complex interactions as shown in the next section.

\subsubsection{Grouping method}
\label{sec:comparison:1}

As discussed in Sec.~\ref{sec:scalable:lattices}, our methods can be used to reproduce a triangular lattice with n.n. interactions by grouping the physical qubits in sets of eight as shown in Fig.~\ref{fig:scalable:patterns:b}. In this way we obtain a logical system of reduced size, but which has the desired interaction pattern $\tilde{H}_{\textit{ZZ}}$ with a coupling strength $\lambda = 2 J/\delta$. To compute the time required to reproduce the evolution given by the target Hamiltonian for a time $T$ we divide the process into two steps:

\begin{enumerate}
\item \textbf{System initialization.} As we explained in Sec.~\ref{sec:state:preparation:connected:sets}, in order to induce interactions between the logical qubit, each set has to be initialized in a GHZ state, i.e., in the state $\ket{+^L} = \big( \ket{0^L}+\ket{1^L} \big) / \sqrt{2}$. We implement a control-\textit{X} gate between n.n. physical qubits by letting them evolve under the intrinsic interaction for a time $\tau = \pi \delta/(4J)$, and some extra single qubit operations on the other qubits are performed to decouple them. In this particular setting, the sets are connected and some of the control-\textit{X} gates can be applied simultaneously, which allows us to reduce the implementation time as explained in Sec.~\ref{sec:state:preparation:connected:sets}. If in a group we label the physical qubits as shown in Fig.~\ref{fig:circuit4:a}, a way of preparing a GHZ state is given by the circuit shown in Fig~\ref{fig:circuit4:b}.
\begin{figure}
    \subfloat[\centering]{\includegraphics[width=0.4\columnwidth]{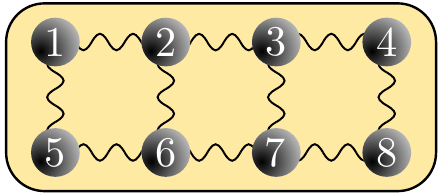} \label{fig:circuit4:a}} \hfill \\
    \subfloat[\centering]{\includegraphics[width=0.80\columnwidth]{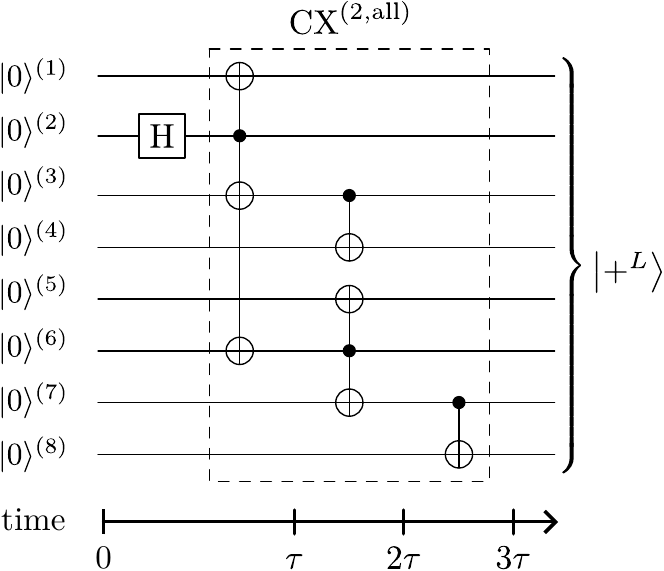} \label{fig:circuit4:b}}
    \caption{In (a), a set of eight physical qubits with n.n. interactions. In (b), a quantum circuit that initializes the set shown in (a) in the $\ket{+^L}$ state. A sequence of control-\textit{X} gates between n.n. to implement $\text{CX}^{(1,\text{all})}$ is also shown. Each control gate requires a time $\tau = \pi \delta/(4J)$ to be implemented. However, several quantum gates can be implemented in parallel.}
\end{figure}
Since this operation can be performed in all logical qubits simultaneously, the time required for the initialization step is given by 
\begin{equation*}
    \eta_0 \equiv \frac{3\pi \delta}{4J}.
\end{equation*}

\item \textbf{Hamiltonian evolution.} Since the interaction pattern for logical qubits is given by $\lambda_{ij} = 2J/\delta$, once these qubits are initialized, it suffices to let the system evolve under the intrinsic Hamiltonian for a time $T/2$. Therefore, the total time required for the simulation with our method is given by
\begin{equation*}
    t_g = \frac{T}{2} + \eta_0.
\end{equation*}
\end{enumerate}

\subsubsection{Standard methods}
\label{sec:comparison:2}

In standard Hamiltonian simulation methods, \cite{dodd2002universal,jane2002simulation,bennett2002optimal} one reproduces a triangular lattice by keeping the inherent interaction between n.n., and generating extra interactions along one diagonal of the square lattice. The process is divided into four steps. (1) The system evolves under the intrinsic Hamiltonian to implement the vertical and horizontal interactions. (2) The state of the qubits is rearranged in the lattice by vertically displacing $k-1$ positions the qubit states of the $k$-column. (3) The system evolves under the intrinsic Hamiltonian but with cancelled interactions along the vertical edges (after the lattice rearrangement the horizontal edges correspond to diagonal interaction on the original lattice). (4) Finally, the rearrangement of the lattice is undone. The time required is given by the evolution time plus the time required to rearrange the lattice, i.e.,
\begin{equation*}
    t_s = 2 T + 2 \zeta(n),
\end{equation*}
since steps (1) and (3) take a time $T$ each, as the coupling strength is not changed. The time to rearrange the lattice in (2) and (3), $\zeta(n)$, depends on the used method, see for instance \cite{Bapat2021quantumroutingfast}. In \cite{bapat2022nearly}, it is shown that the time to perform state reverse in a 1d $n$-qubit chain by means of a \textit{ZZ} n.n. interactions is lower bounded by $t^* \geq n \delta/ (\zeta J)$ where $\zeta \approx 1.912$. This bound is obtained by considering the entanglement generated between each half of the qubit chain. From the entanglement generation point of view, this situation is analogous to the permutation we perform to the $k = \ell/2$ column of the lattice. Therefore, this bound also applies in our case, and hence the time to rearrange the lattice must at least linearly increase with the length of the columns $\ell$ (what is given by $\ell = \sqrt{n}$ in a $\ell \times \ell$ lattice) as $\zeta(n) \geq \ell \delta/ (\zeta J)$. For instance, in a SWAP-based implementation of the rearrangement $\zeta(n) =  (\ell-1) 3\pi \delta/(4J)$, see Appendix~\ref{app:SAWP} for details.

\subsection{Non-commuting interactions}

Now we consider the task of reproducing a target Hamiltonian with non-commuting terms, e.g., a two-body \textit{ZZ} interaction with a logical $X$-field or a Heisenberg interaction.

In the previous section, we showed how from the original square lattice we can obtain the interaction pattern corresponding to a triangular lattice with n.n. interactions, i.e., $\tilde{H}_{\textit{ZZ}}$. Therefore, now we only need to establish the desired interaction type by fast alternating between different unitary transformations of the original Hamiltonian or local operations of the qubits, as explained in Sec.~\ref{sec:general:interaction:type}.

Concretely, we consider the task to reproduce the evolution generated by the two-body \textit{ZZ} interaction plus a local $X$-field for a time $T$, i.e., the target Hamiltonian is given by
\begin{equation*}
    \tilde{H} = \tilde{H}_{\textit{ZZ}} + \sum_{j=1}^N X_j,
\end{equation*}
one needs to fast alternate between the two non-commuting terms of $\tilde{H}$ to approximate its evolution, i.e.,
\begin{equation}\label{eq:interaction:xfield}
    e^{- \ti \tilde{H} T} = \left( e^{- \ti \tilde{H}_{\textit{ZZ}} T/k} e^{- \ti \sum_{j=1}^N X_j T/k} \right)^k + \mathcal{O}\left( \frac{T^2}{k^2} \right),
\end{equation}
where $k$ is chosen to be large enough to neglect the second and higher-order terms.

\subsubsection{Grouping method}

For our method, we showed that once the logical qubits have been initialized the time required to implement $e^{- \ti \tilde{H}_\textit{ZZ} T/k}$ is given by $T/(2k)$. On the other hand, since we consider the time consumed to implement single physical qubit gates as negligible, implementing
\begin{equation*}
    e^{- \ti \sum_{i=1}^N X^L_i t} = \Bigg( \bigotimes_{i=1}^N \text{H}^L_i \Bigg) e^{- \ti \sum_{j=1}^N Z^L_j t} \Bigg( \bigotimes_{k=1}^N \text{H}^L_k \Bigg)
\end{equation*}
requires a time $2\eta_{\text{H}}$, where $\eta_{\text{H}}$ is the time required to implement a logical Hadamard gate on each of the logical qubits. In Sec.~\ref{sec:LU:control}, we showed that a logical Hadamard gate is given by Eq.~\eqref{eq:UL}. Hence, if we apply $\text{CX}^{(1,\text{all})}$ as sown in Fig.~\ref{fig:circuit4:b}, the time required to implement the Hadamard gate is given by
\begin{equation*}
    \eta_\text{H} \equiv \frac{3\pi \delta}{2J}.
\end{equation*}
Therefore the total time required to simulate the evolution in Eq.~\eqref{eq:interaction:xfield} with our method is given by
\begin{equation*}
    t_g = \frac{T}{2} + 2k \, \eta_{\text{H}} + \eta_0.
\end{equation*}
Note that $t_g$ is independent of the size of the lattice and it scales linearly with the number of alternations $k$ between the two terms.

\subsubsection{Standard methods}

With the standard methods, one implements $e^{-\ti X_j t}$ with single-qubit operations, we will thus neglect the time consumed by this part of the evolution. Therefore, we just have to consider the time needed to implement $e^{-\ti \tilde{H}_{\textit{ZZ}} t}$ between each application of $e^{-\ti X_j t}$. In Sec.~\ref{sec:comparison:2}, we showed that this consumes a time $t_s$, and hence, the total time is given by
\begin{equation*}
    t_s = 2T + 2k \, \zeta(n).
\end{equation*}

\subsection{Comparison}

Comparing the two approaches, one notes a significant enhancement in efficiency offered by our method. While $t_g$ is independent of the total number of qubits $n$, $t_s$ increases linearly with $\ell$. The main advantage of our setting based on grouping physical qubits in logical sets is that it gives access to interactions not present on the original lattice. One can then simulate various interactions without the need to actively rearrange the states of the physical qubits, which is increasingly time-consuming when the size of the system increases. However, this enhancement comes at the price of a higher degree of complexity in the control of logical systems than of physical ones. Therefore, we expect our setting to provide a significant advantage in preparing specific entangled states, or in simulations involving mainly commuting interactions where only a few intermediate logical single qubit operations are required. 

In Sec.~\ref{sec:state:preparation:disconnected} we have also shown how one can directly generate arbitrary interaction patterns by delocalizing the logical sets. In this case, initializing the logical qubits requires a rearrangement of the lattice. But in contrast to standard Hamiltonian simulation techniques, this has to be done only once.

\section{Summary and outlook}
\label{sec:summary}

In this paper, we have introduced a quantum simulator based on logical systems. In a physical many-body system with inherent distance-dependent two-body interactions, we group the qubits in logical sets that are treated as an effective two-level system. Even though we assume no control over the physical qubit-qubit interactions, we can establish different interaction patterns between the logical systems by properly controlling the internal state of the logical qubits, which can be accomplished only by means of single physical qubit operations. We also show how single physical qubit control suffices to obtain full control of the logical systems.

We showed the performance of our setting in particular examples given by different groupings of the physical qubits and different physical interaction ranges. In these examples, our approach can be used to increase the interaction strength, change the interaction range, or just the possibility of turning off and on interactions at our own will. In general, we saw the size of the logical sets has to increase linearly with the number of tunable interactions between logical qubits. We found particular solutions to establish finite-range interaction patterns for arbitrary large systems with a common underlying physical system.

Even though our setting assumed an intrinsic \textit{ZZ} interaction type between the physical systems, we showed how it can be implemented from arbitrary physical two-body interactions by utilizing standard techniques from Hamiltonian simulation. The same techniques allow us to simulate an arbitrary interaction type between the logical systems including many-body interactions. In a similar way, we can extend our setting to implement logical qudits systems. This can be done by joining several logical qubits or by using external degrees of freedom of the physical systems.

Finally, we compared our quantum simulator with known techniques to simulate a triangular lattice in a square lattice with n.n. interactions. The comparison shows how our model provides an enhancement on the implementation of multiple qubit interactions at the cost of higher complexity on single-qubit operations.

We want to point out that noise and error treatment is beyond the scope of this paper and will be treated in future works \cite{ferran_noise}. However, observe that the encoding used for the logical systems allows us to correct $X$ errors. This is due to the fact that the encoding we use for simulation constitutes at the same time a bit flip error correction code \cite{nielsenchuang2010}. For logical systems carried by $n$ physical qubits, up to $\lfloor n/2 \rfloor$ bit-flip errors can be corrected. Also, in \cite{riera2022remotely}, we already showed how one can mitigate the effect of thermal noise on the trapped physical qubits by enlarging the logical systems.

Our results provide an alternative or complement to the standard Hamiltonian simulation techniques for simulating arbitrary many body systems. The novel feature that we exploit is to use logical subspaces of multiple physical systems together with by local control in order to implement a programmable simulator. We have presented several examples, which illustrate the flexibility of our approach and show that it compares advantageously with standard Hamiltonian simulation methods.

\section*{Acknowledgements}

This work was supported by the Austrian Science Fund (FWF) through projects No. P30937-N27, No. P36009-N and No. P36010-N. Finanziert von der Europ\"aischen Union - NextGenerationEU.

\bibliography{Quantum_Simulator.bib}

\onecolumngrid
\appendix

\section{Effective spin values: Explicit example}
\label{app:sec:flipping}

\begin{figure}
    \includegraphics[width=\columnwidth]{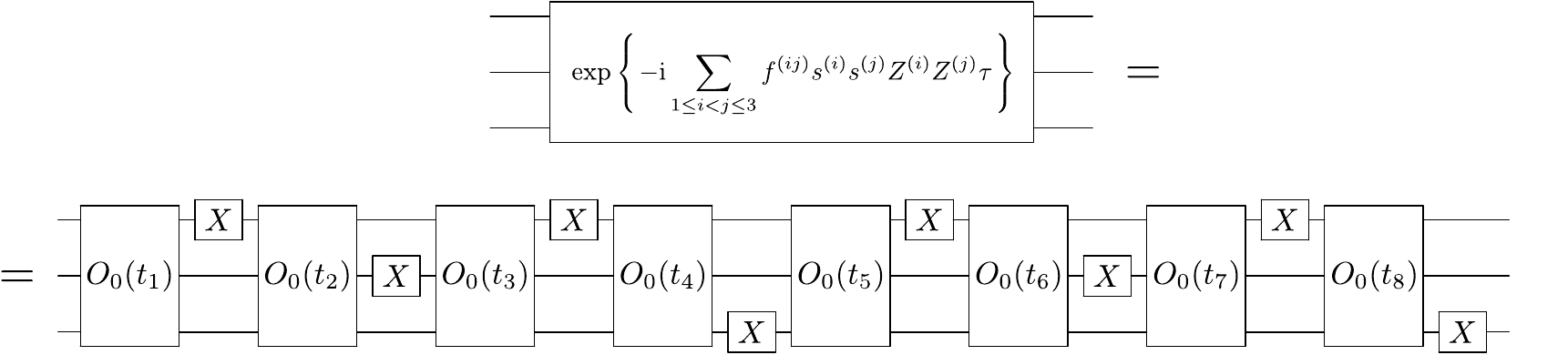}
    \caption{\label{fig:fliping} We show a circuit that implements a \textit{ZZ} interaction between the three qubits with effective spin values $s^{(1)}$, $s^{(2)}$ and $s^{(3)}$ for an arbitrary time $\tau$. The circuit concatenates eight \textit{ZZ} evolutions, i.e., of the form $O_0(t)$ (see Eq.~\eqref{eq:app:O0} and Eq.~\eqref{eq:app:ti}), with single qubit $X$ gates in between}
\end{figure}

Here we compute an explicit example to establish an effective spin value for three qubits interacting with a two-body \textit{ZZ} interaction, i.e., if the inherent interaction is given by
\begin{equation}\label{eq:app:O0}
    O_0(t) = e^{-\ti \sum_{1\leq i<j\leq 3} f^{(ij)} Z^{(i)} Z^{(j)} t},
\end{equation}
and we aim to obtain $O_3(\tau)$ what is given by
\begin{equation*}
    O_3(\tau) = e^{-\ti \sum_{1\leq i<j\leq 3} f^{(ij)} s^{(i)} s^{(j)} Z^{(i)} Z^{(j)} \tau}.
\end{equation*}
As we explained in Sec.~\ref{sec:eff:spin}, that is achieved by flipping each qubit at specific times of the evolution as given in Eq.~\eqref{eq:Oj}. By following the procedure introduced in Sec.~\ref{sec:eff:spin} we need to divide $\tau$ into eight intervals where after each interval we flip one of the qubits, see Fig.~\ref{fig:fliping}. We flip each qubit at times:
\begin{equation*}
\begin{aligned}
    \text{qubit-1:}& \; \{ t_1, \, t_2 + t_3, \, t_4 + t_5, \, t_6 + t_7 \} \\
    \text{qubit-2:}& \; \{ t_1 + t_2, \, t_3 + t_4 + t_5 + t_6 \} \\
    \text{qubit-3:}& \; \{ t_1 + t_2 + t_3 + t_4, \, t_5 + t_6 + t_7 + t_8 \}
\end{aligned}
\end{equation*}
where
\begin{equation}\label{eq:app:ti}
\begin{aligned}
    t_1 & = \frac{\tau}{8} \left(1+s^{(3)}\right) \left(1+s^{(2)}\right) \left(1+s^{(1)}\right) \quad & \quad
    t_5 & = \frac{\tau}{8} \left(1-s^{(3)}\right) \left(1-s^{(2)}\right) \left(1-s^{(1)}\right) \\
    t_2 & = \frac{\tau}{8} \left(1+s^{(3)}\right) \left(1+s^{(2)}\right) \left(1-s^{(1)}\right) \quad & \quad
    t_6 & = \frac{\tau}{8} \left(1-s^{(3)}\right) \left(1-s^{(2)}\right) \left(1+s^{(1)}\right) \\
    t_3 & = \frac{\tau}{8} \left(1+s^{(3)}\right) \left(1-s^{(2)}\right) \left(1-s^{(1)}\right) \quad & \quad 
    t_7 & = \frac{\tau}{8} \left(1-s^{(3)}\right) \left(1+s^{(2)}\right) \left(1+s^{(1)}\right) \\
    t_4 & = \frac{\tau}{8} \left(1+s^{(3)}\right) \left(1-s^{(2)}\right) \left(1+s^{(1)}\right) \quad & \quad
    t_8 & = \frac{\tau}{8} \left(1-s^{(3)}\right) \left(1+s^{(2)}\right) \left(1-s^{(1)}\right).
\end{aligned}
\end{equation}

\section{Effective spin values: Logical qubits}
\label{app:sec:flipping2}

In Sec.~\ref{sec:eff:spin}, we showed that in an ensemble of $m$ physical qubits, to establish an effective spin value for each qubit we need to perform $\chi \leq 2^m$ flips. Here, we show that the number of flips can be considerably reduced if we adapt the procedure to each case. First, we consider the scenario where the interaction between physical qubits is of finite range. In this case, some flips can be performed in ``parallel'' which allows us to reduce the number of flips. We illustrate this with a particular example.

Consider $m$ physical qubits arranged in a square lattice with n.n. interactions. Note that the interaction graph of the $m$ qubits is 2-colourable, and hence, we can distinguish between red qubits, $\{r^{(i)}\}_{i=1}^{m_r}$, and blue qubits $\{b^{(i)}\}_{i=1}^{m_b}$ (where $m_r+m_b=m$ and $m_r\leq m_b$), in such a way that qubits of the same colour do not directly interact with each other. This distinction allows us to write the evolution of the qubits under the intrinsic interactions as
\begin{equation*}
    O_0(t) = e^{-\ti \sum_{i=1}^{m_r} \sum_{j=1}^{m_b} f_{rb}^{(ij)} Z^{(i)}_r Z^{(j)}_b t},
\end{equation*}
where $Z_r^{(i)}$ acts on qubit $r^{(i)}$, $Z_b^{(i)}$ acts on qubit $b^{(i)}$ and $f_{rb}^{(ij)}$ is the coupling function between qubits $r^{(i)}$ and $b^{(j)}$. Considering the evolution under $O_0(\tau)$, we obtain a non-integer spin value for the red qubits by flipping each of them twice. In particular, for a fixed but arbitrary time $\tau$ we flip qubit $r^{(i)}$ at $t_i = \big[ 1+s_r^{(i)} \big]\tau/2$ and at $\tau$ for $i=1,\dots,m_r$, and we obtain
\begin{equation}\label{eq:Ur}
\begin{gathered}
    O_r(\tau) = X_r^{(m_r)} \cdots X_r^{(1)} \, O_0(\tau-t_{m_r}) X^{(m_r)} \cdots U_0(t_3-t_2) \, X^{(2)}_r \, O_0(t_2-t_1) \, X^{(1)}_r \, O_0(t_1) 
    \\ = e^{-\ti \sum_{i=1}^{m_r} \sum_{j=1}^{m_b} f_{rb}^{(ij)} s_r^{(i)} Z^{(i)}_b Z^{(i)}_r \tau},
\end{gathered}
\end{equation}
where without loss of generality we assumed $t_1<t_2<\dots<t_{m_r}$. We say the red qubits are flipped ``in parallel'' as they are flipped in the same iteration. This can be done because the red qubits do not couple with each other. Note that to implement $O_r(\tau)$, Eq.~\eqref{eq:Ur}, we perform $2 m_r$ flips. 

$O_r(\tau)$ corresponds to an interaction where the spin value of qubits $r^{(i)}$ is given by $s^{(i)}_r$. Next, we perform the same step for the blue qubits, i.e., considering the evolution described by $O_r(t)$ we flip qubit $b^{(j)}$ at $t'_j = \big[ 1+s_b^{(i)} \big]\tau/2$, obtaining at time $\tau$
\begin{equation}\label{eq:Urb}
\begin{gathered}
    O_{rb}(\tau) = X_b^{(m_b)} \cdots X_b^{(1)} \, O_r(\tau-t'_{m_b}) \, X_b^{(m_b)} \cdots O_r(t'_3-t'_2) \, X^{(2)}_b \, \widetilde{O}_r(t'_2-t'_1) \, X^{(1)}_b \, O_r(t'_1) \\
    = e^{-\ti \sum_{i=1}^{m_r} \sum_{j=1}^{m_b} f_{rb}^{(ij)} s_r^{(i)} s_b^{(i)} Z^{(i)}_b Z^{(i)}_r \tau},
\end{gathered}
\end{equation}
where
\begin{equation*}
    \widetilde{O}_r(\tau) = O_0(t_1) \, X^{(1)}_r \, O_0(t_2-t_1) \, X^{(2)}_r \, O_0(t_2-t_1) \cdots X^{(m_r)} \, O_0(\tau-t_{m_r}) \, X_r^{(1)} \cdots X_r^{(m_r)}
\end{equation*}
and without loss of generality we assumed $t_1<t_2<\dots<t_{m_b}$. Note that $\widetilde{O}(t) = O(t)$, however, just as in Eq.~\eqref{eq:Oj}, alternating between $\widetilde{O}(t)$ and $O(t)$ makes a difference in the resulting gate sequence, as it appears some terms of the form $\big(X_r^{(i)}\big)^2 = \id$ that we can ignore. Note that $O_{rb}(\tau)$ corresponds to the intrinsic interaction with an effective spin value for the qubits. 

To implement $O_{rb}(\tau)$, Eq.~\eqref{eq:Urb}, we perform $2m_b$ flips on blue qubits, $2m_r$ flips on red qubits for each $O_r(t)$, and $2m_r$ for each $\widetilde{O}_r(t)$. However, we have to take into account that in Eq.~\eqref{eq:Urb} appear $\lceil m_b/2 \rceil$ terms of the form $\widetilde{O}(t)X^{(i)}_b O(t')$, wherein each of these terms, $2m_r$ flips are repeated and hence cancelled. Therefore, in total, the number of flips performed is given by
\begin{equation*}
    \chi_{\text{n.n.}} = 2m_b + 2 m_r (m_b + 1) - 2m_r \left\lceil\frac{m_b}{2}\right\rceil .
\end{equation*}
Note, in the case of $m_r=m_b=m/2$
\begin{equation*}
    \chi_{\text{n.n.}} \leq \left(\frac{m}{2}\right)^2 + 2m
\end{equation*}
where the inequality saturates for $m/4 \in \mathbbm{N}$. This is a substantial reduction of the number of flips performed with respect to the general method, as now the number of flips is just polynomial in $m$ instead of exponential.

Another relevant example in this paper is given when the physical qubits in the lattice also contain diagonal interactions. In this case, the interaction graph of the physical qubits is 4-colourable. Here, we can obtain an effective spin value by following the same procedure. Now, we divide the qubits into four colours and perform four iterations instead of two, where in each iteration the qubits of one colour are flipped in parallel. If we have $m/4$ qubits of each colour, iterating Eq.~\eqref{eq:Urb} twice more we obtain that the number of flips is given by
\begin{equation*}
    \chi_{\text{diag.}} \leq \left(\frac{m}{4}\right)^4 + 5 \left(\frac{m}{4}\right)^3 + 9\left(\frac{m}{4}\right)^2 + 2 m,
\end{equation*}
where the inequality saturates for $m/8 \in \mathbbm{N}$.

For general $m$ physical qubits with a $\kappa$-colourable interaction graph, where there are $m_\gamma$ qubits of color $\gamma \in \{1,\dots,\kappa\}$, the number of flips can be computed as
\begin{equation}\label{eq:generalchi}
    \chi_{\kappa} = y_\kappa,
\end{equation}
where
\begin{equation*}
\begin{aligned}
    y_1 & = 2 m_1 \\
    y_j & = 2 m_j + y_{j-1} (m_j+1) - 2 m_{j-1} \left\lceil \frac{m_j}{2} \right\rceil ,
\end{aligned}
\end{equation*}
and we assume $m_{\gamma}\leq m_{\gamma+1}$. From Eq.~\eqref{eq:generalchi} we obtain that the number of flips for a system of $m$ physical qubits described by a $\kappa$-colourable interaction graph is given by $\mathcal{O}\big[(m/\kappa)^\kappa\big]$.

In the case of full-range interaction between the physical qubits, we can perform a similar analysis once the physical qubits are grouped in the logical qubit. The physical qubits are insensitive to the interactions with the physical qubits of the same logical set. Therefore, in this case, we can repeat the same analysis by flipping in parallel the physical qubits of the same set instead of the qubits of the same colour. In this way, if we group the qubits in $N$ logical qubits, we can use Eq.~\eqref{eq:generalchi}, where now $m_{\gamma} = n_i$ and $\kappa=N$. This results that we need to perform $N$ iterations and hence, if $n_i=n$, the number of flips required is given by $\mathcal{O}\left(n^N\right)$.

\section{Control of interactions: Explicit example}
\label{app:sec:eqsystem.example}

In order to illustrate how algorithm-1 introduced in Sec.~\ref{sec:int:control} works, we compute in detail its steps for a particular example. We consider the setting shown in Fig.~\ref{fig:toy:modelg1}. Four sets of four physical qubits each located in a square lattice of qubit-qubit distance $\delta=1$, and the coupling strength between two physical qubits is given by $f_{ij} = \left| \boldsymbol{r}_i - \boldsymbol{r}_j \right|^{-1}$. The considered target pattern is the one shown in Fig.~\ref{fig:toy:model3}, i.e., $\lambda_{12} = \lambda_{13} = \lambda_{14}$ and $\lambda_{23} = \lambda_{24} = \lambda_{34} = 0$.

We want to find a set of vectors $\{\boldsymbol{s}_{i}\}$ that are a solution of the system of equations
\begin{equation*}
    \left\{ \boldsymbol{x}^T_i \boldsymbol{F}_{ij} \boldsymbol{x}_j = \lambda_{ij} \right\}_{1\leq i < j\leq N}.
\end{equation*}
The first step consists in choosing a random vector for set 1:
\begin{equation*}
    \boldsymbol{s}_1 = \big( 1, \, 1, \, 1, \, 1 \big)^T.
\end{equation*}
Then, we consider the equation corresponding to the interaction between the $S_1$-$S_2$ pair:
\begin{equation*}
    \boldsymbol{s}_1^T \boldsymbol{F}_{12} \boldsymbol{x}_2 = \lambda_{12}.
\end{equation*}
We initially set $\lambda_{12} = 4$, and as $\boldsymbol{s}_1$ is fixed, the equation is given by:
\begin{equation*}
    2.654 x_2^{(1)} + 1.597 x_2^{(2)} + 2.654 x_2^{(3)} + 1.597 x_2^{(4)} = 4.
\end{equation*}
This is a linear equation with an infinite number of solutions. For $S_2$, we choose a random solution from the solution set of the equation
\begin{equation*}
    \boldsymbol{s}_2 = \big( \, 0.5, \, -9.3\, 3.2\, 5.65 \, \big)^T .
\end{equation*}
Then we consider the equations corresponding to interactions between $S_1$-$S_3$ and $S_2$-$S_3$:
\begin{equation*}
    \begin{gathered}
    \begin{pmatrix}
       \boldsymbol{s}_1^T \boldsymbol{F}_{13} \\ \boldsymbol{s}_2^T \boldsymbol{F}_{23}
    \end{pmatrix} 
    \boldsymbol{x}_3
    =
    \begin{pmatrix}
      \lambda_{13} \\ \lambda_{23}
    \end{pmatrix}
    \end{gathered}
\end{equation*}

As $\boldsymbol{s}_1$ and $\boldsymbol{s}_2$ are already fixed the equations can be written as
\begin{equation*} \left\{
    \begin{matrix}
        2.654 x_3^{(1)} + 2.654 x_3^{(2)} + 1.597 x_3^{(3)} + 1.597 x_3^{(4)} = 4 \\ 
        0.817 x_3^{(1)} + 1.727 x_3^{(2)} + 0.646 x_3^{(3)} + 1.009 x_3^{(4)} = 0
    \end{matrix} \right. ,
\end{equation*}
what is a linear system of equations with a two-dimensional solution set as vectors $\left\{ \boldsymbol{s}_1^T \boldsymbol{F}_{13}, \boldsymbol{s}_2^T \boldsymbol{F}_{23} \right\}$ are linearly independent. For $S_3$ we choose a random solution from the solution set of the system
\begin{equation*}
    \boldsymbol{s}_{3} = \big( -2.4, \, -0.5, \, 12.595, \, -5.269 \, \big)^T
\end{equation*}
Then we consider equations corresponding to interactions between $S_1$-$S_4$, $S_2$-$S_4$ and $S_3$-$S_4$:
\begin{equation*}
    \begin{pmatrix}
      \boldsymbol{s}_1^T \boldsymbol{F}_{14} \\ \boldsymbol{s}_2^T \boldsymbol{F}_{24} \\ \boldsymbol{s}_3^T \boldsymbol{F}_{34}
    \end{pmatrix} 
    \boldsymbol{x}_4
    =
    \begin{pmatrix}
      \lambda_{14} \\ \lambda_{24} \\ \lambda_{34}
    \end{pmatrix}
    .
\end{equation*}
As $\boldsymbol{s}_1$, $\boldsymbol{s}_2$ and $\boldsymbol{s}_3$ are already fixed the set of equations can be written as
\begin{equation*} \left\{
    \begin{matrix}
      1.955 x_4^{(1)} + 1.394 x_4^{(2)} + 1.394 x_4^{(3)} + 1.144 x_4^{(4)} = 4 \\
      3.289 x_4^{(1)} + 3.491 x_4^{(2)} + 1.355 x_4^{(3)} + 1.316 x_4^{(4)} = 0 \\
      0.207 x_4^{(1)} + 0.576 x_4^{(2)} - 0.399 x_4^{(3)} + 0.581 x_4^{(4)} = 0
    \end{matrix} \right. ,
\end{equation*}
what is a linear system of four equations with a one-dimensional set of solutions, as vectors $\{ \boldsymbol{s}_1^T \boldsymbol{F}_{14}, \boldsymbol{s}_2^T \boldsymbol{F}_{24}, \boldsymbol{s}_3^T \boldsymbol{F}_{34} \}$ are linearly independent. For $S_4$, we choose a random solution from the solution set of the system,
\begin{equation*}
    \boldsymbol{s}_4 = \big( -2.4, \, -0.638, \, 3.617, \, 3.967 \big)^T
\end{equation*}
Finally, the vectors have to be divided by $\max_{i,k}\big|s_i^{(k)}\big| = s_3^{(2)}$ to ensure that $s_i^{(k)} \in [-1, 1]$ $\forall \, i, k$
\begin{equation*}
\begin{aligned}
    \boldsymbol{s}_1 & = \big( 0.079, \, 0.079, \, 0.079, \,0.079 \big)^T \\
    \boldsymbol{s}_2 & = \big( 0.040, \, -0.738, \, 0.254, \, 0.449 \big)^T \\ 
    \boldsymbol{s}_3 & = \big( -0.191, \,-0.040, \, 1, \, -0.418 \big)^T \\
    \boldsymbol{s}_4 & = \big( -0.191, \,-0.051, \, 0.287, \, 0.315 \big)^T,
\end{aligned}
\end{equation*}
this scales the interaction pattern by $\left( s_3^{(3)} \right)^{-2} = 0.0063$, and therefore it is given by
\begin{equation*}
    \begin{gathered}
        \lambda_{12} = \lambda_{13} = \lambda_{14} = 0.025 \\
        \lambda_{23} = \lambda_{24} = \lambda_{34} = 0
    \end{gathered}
\end{equation*}
Note that the solution obtained is 3 orders of magnitude smaller than the optimal one that reaches $\lambda = 2.04$.

In Tables~\ref{tab:g1:algorithm} and~\ref{tab:g2:algorithm} we show the effective coupling strength obtained with our method to generate the patterns shown in Fig.~\ref{fig:toy:model}.

\section{Full range interaction pattern from an n.n. square lattice: Polynomial scaling}
\label{app:nn:scale}

In Sec.~\ref{Sec:nn:1}, we have shown how to implement four and five logical qubits in a square lattice with n.n. interactions of 16 and 25 physical qubits respectively. We group the qubits by imposing four physic qubit-qubit interactions between any pair of logical sets. One can try to find a similar grouping to implement $N$ logical qubits in a $N \times N$ square lattice. However, as we cannot prove that this is always possible, we show an alternative way of implementing $N$ logical qubits with a polynomial scaling in the number of physical qubits. For that, we consider several $4 \times 4$ sub-lattices $\{ L_i \}$ with n.n. interactions. In each sub-lattice, we make interact with 4 logical qubits in a way that each pair of sets interact at least in one of the sub-lattices. Notice in this way, the qubits of each set are distributed in several sub-lattices.

For instance, to implement eight logical qubits we need six sub-lattices. One possible way of distributing the grouping is given by
\begin{equation*}
\begin{aligned}
    L_1 : (S_1, S_2, S_3, S_4) \quad & \quad L_4 : (S_3, S_4, S_7, S_8) \\
    L_2 : (S_5, S_6, S_7, S_8) \quad & \quad L_5 : (S_1, S_2, S_7, S_8) \\
    L_3 : (S_1, S_2, S_5, S_6) \quad & \quad L_6 : (S_3, S_4, S_5, S_6),
\end{aligned}
\end{equation*}

\begin{figure}
    \centering
    \includegraphics[width=0.8\columnwidth]{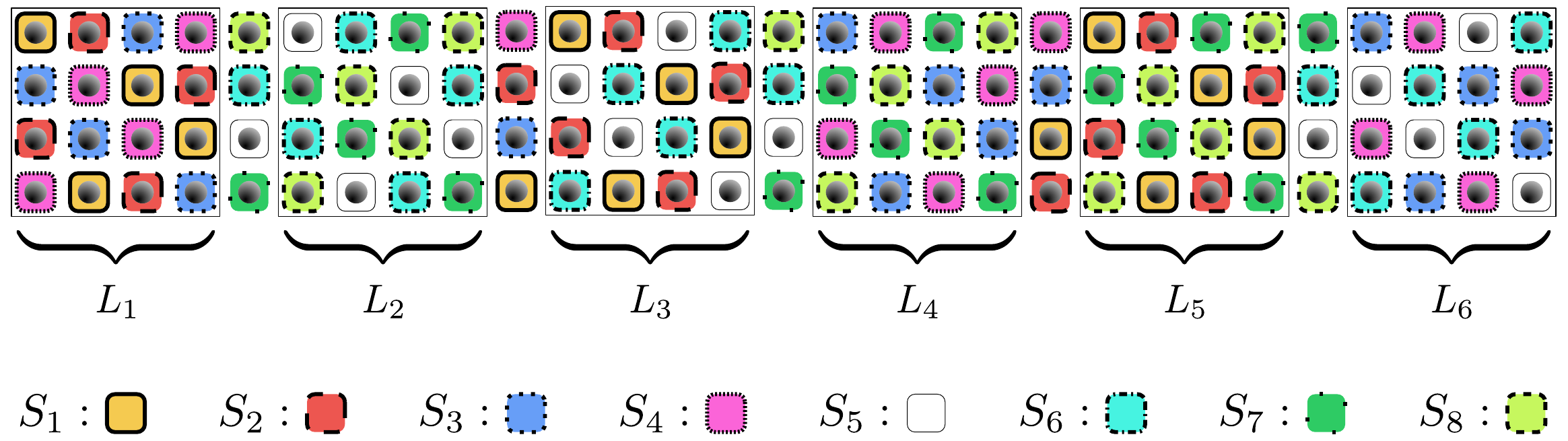}
    \caption{\label{app:fig:lattices} Representation of the grouping of eight logical qubits in a rectangular lattice of $4 \times 29$ qubits with n.n. interactions. The whole lattice is divided in six sub-lattices }
\end{figure}

Then as we do not consider isolated lattices, we placed them together with an extra row of four physical qubits between which we assign each physical qubit to one of the sets. With this method, we achieve to implement the eight logical qubits in a lattice of 116 physical qubits, where we can generate any interaction pattern, see Fig.~\ref{app:fig:lattices}.

For implementing $N = 2^k$ logical ensembles, we can iterate the procedure described for $N=8$. Like with the $N=8$ case, we define 6 sets of $N/2$ elements each. For each of these sets, we iterate this step $\log_2 (N) - 2$ times. In this way we obtain $a = 6^{\log_2 (N) - 2}$ sets of $4$ elements each. To each set corresponds a sublattice of 16 physical qubits each where we make interact with the corresponding logical qubits. If we take into account that between lattices we need to leave a column of 4 qubits to isolate them, we obtain that to implement obtain $N$ logical qubits the number of physical qubits required is given by
\begin{equation*}
    n = 16 a + 4 (a-1) = \frac{5}{9} N^{\log_2 6} - 4 \approx \frac{5}{9} N^{2.58} - 4.
\end{equation*}

\begin{figure}
    \centering
    \subfloat[\centering]{\includegraphics[width=0.15\columnwidth]{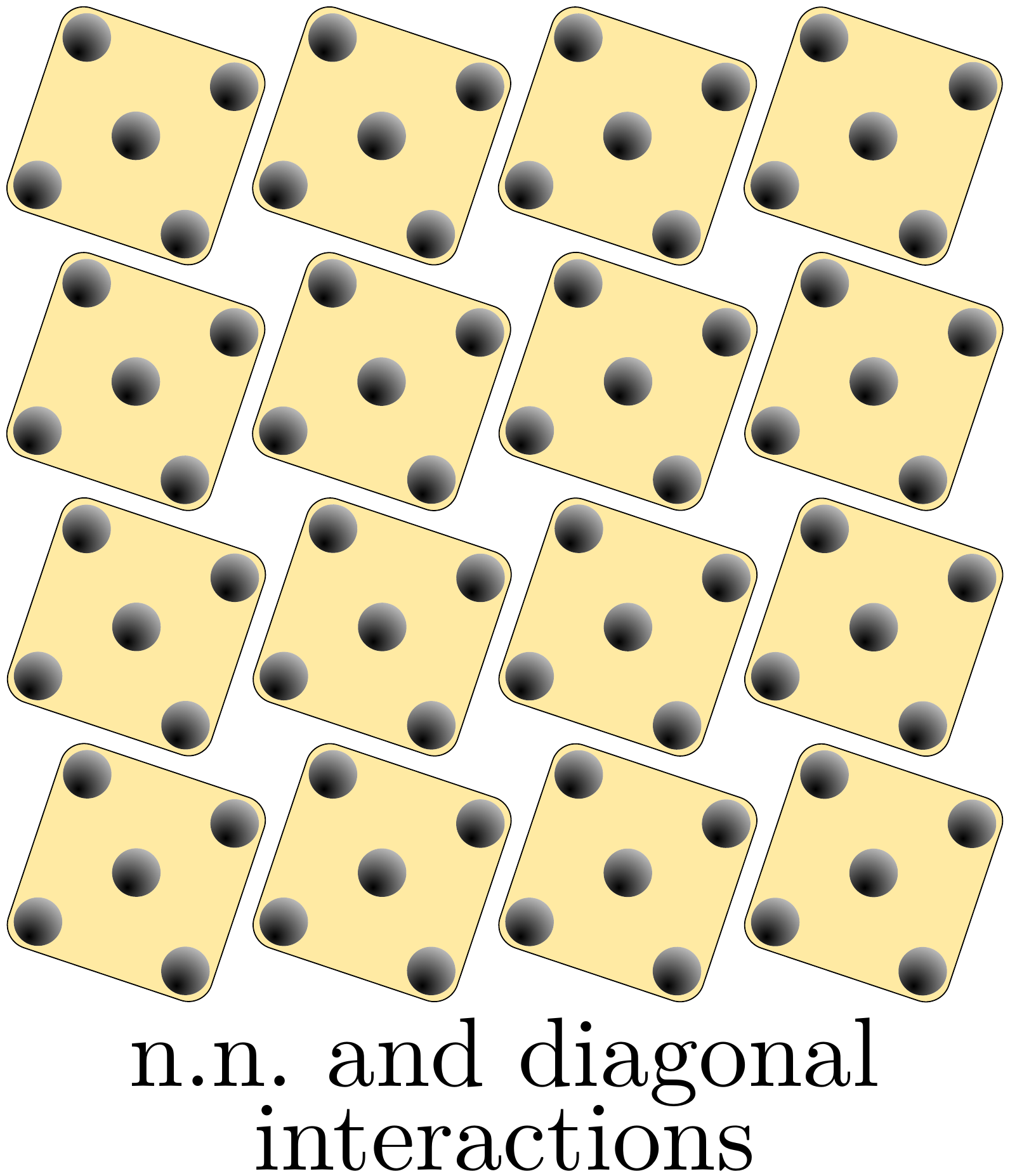} \label{fig:prog:lattices:a} } \hspace{0.4in}
    \subfloat[\centering]{\includegraphics[width=0.15\columnwidth]{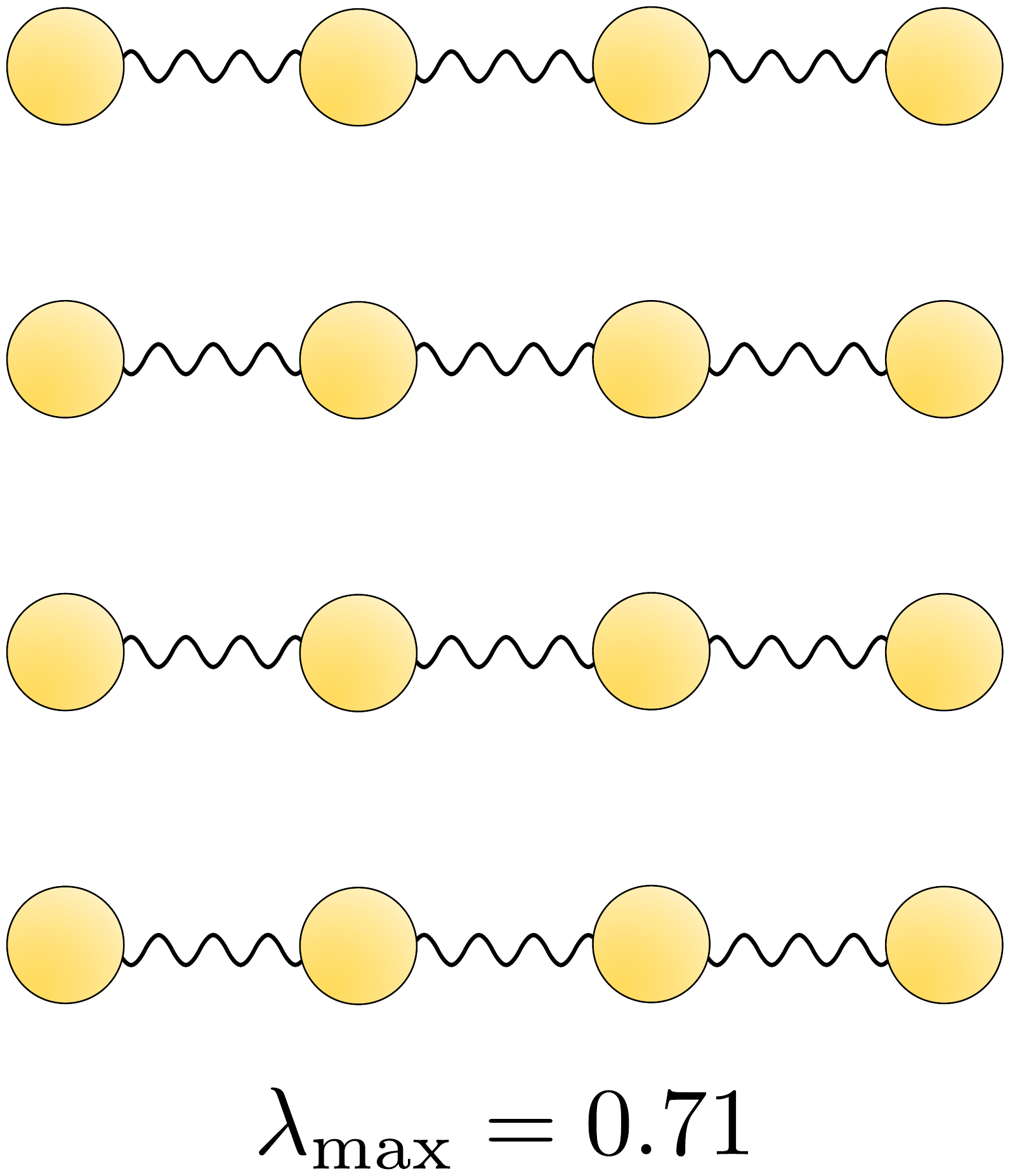} \label{fig:prog:lattices:b} } \hspace{0.4in}
    \subfloat[\centering]{\includegraphics[width=0.15\columnwidth]{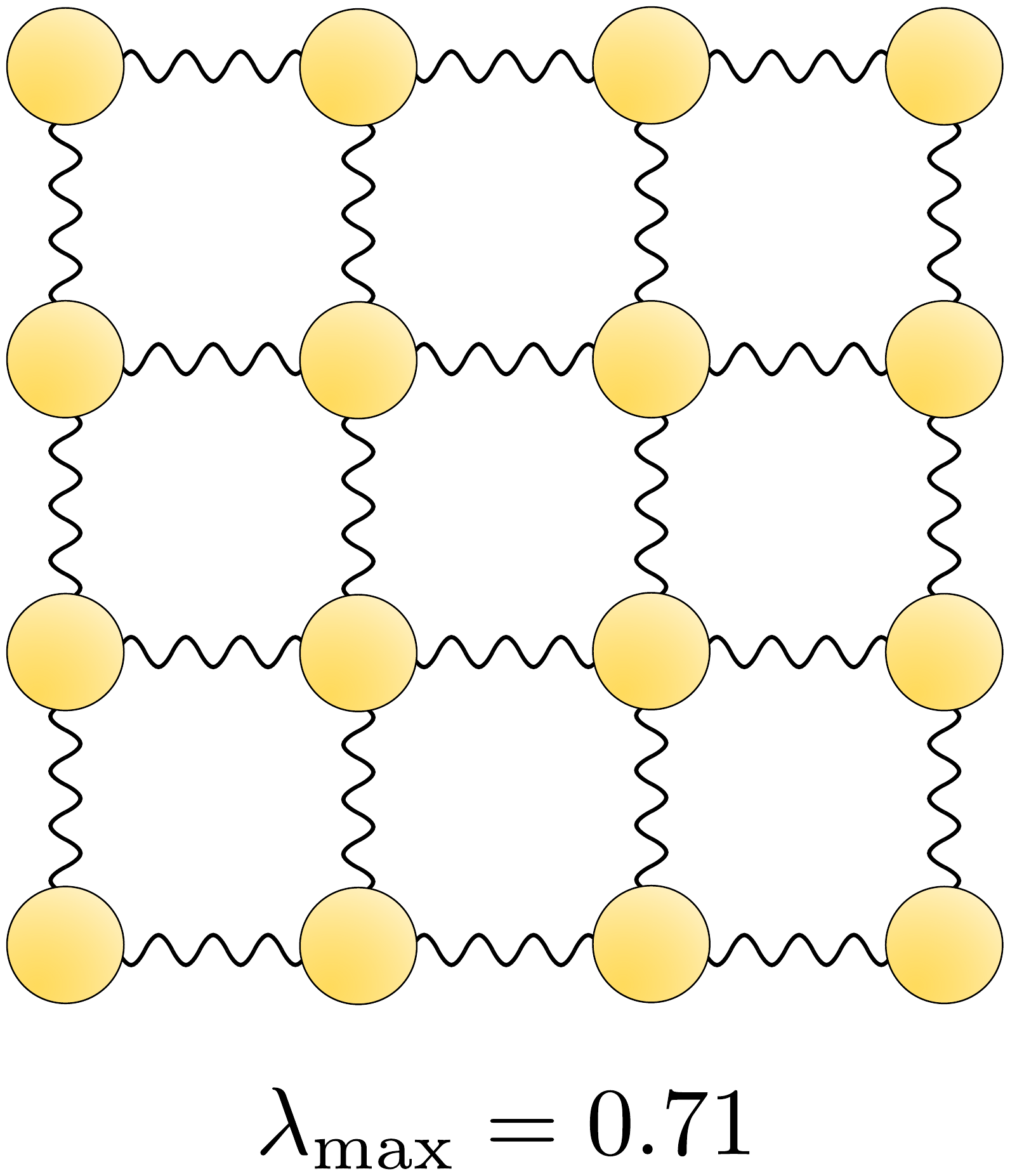}
    \label{fig:prog:lattices:c} } \hspace{0.4in}
    \subfloat[\centering]{\includegraphics[width=0.15\columnwidth]{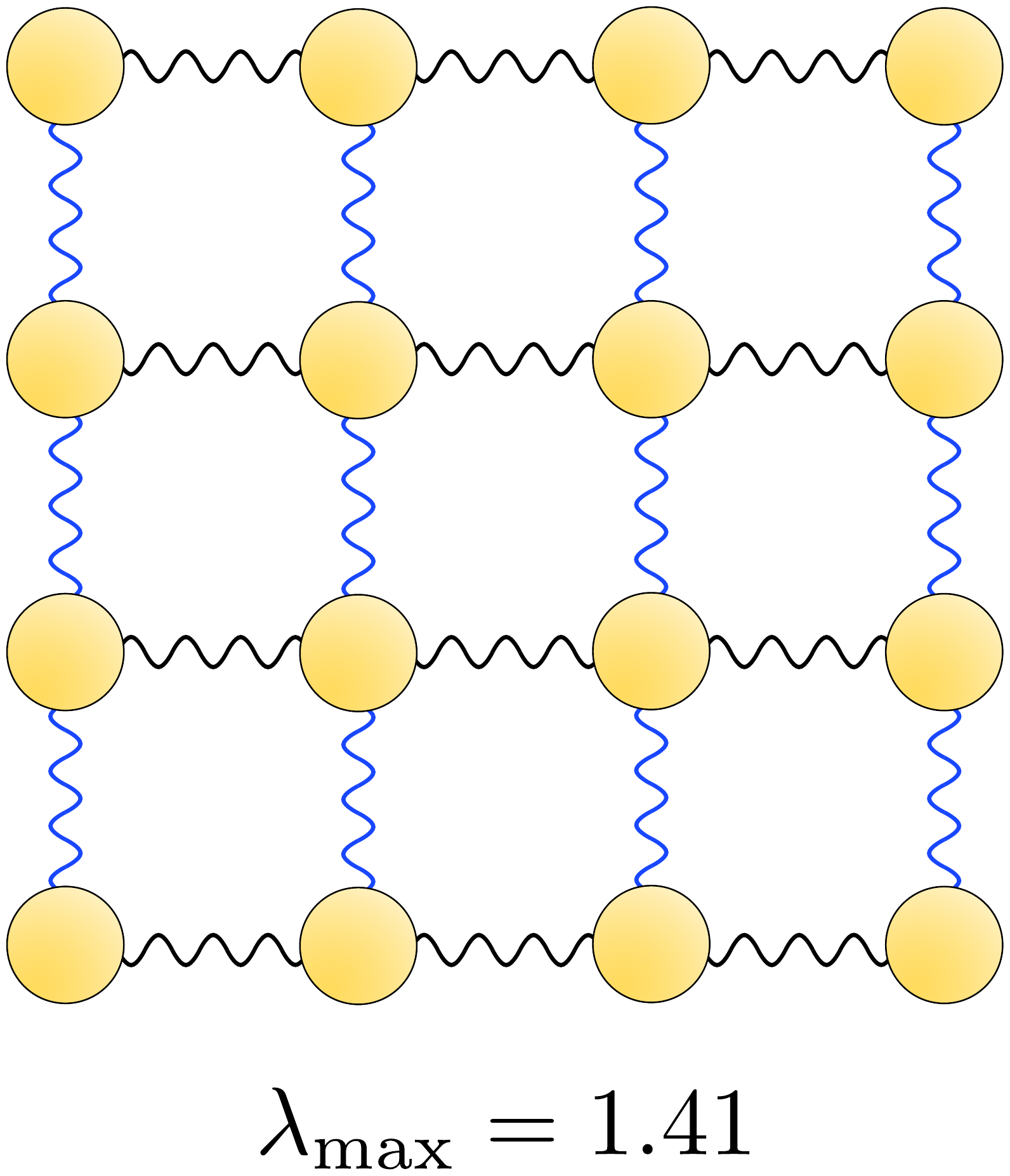}
    \label{fig:prog:lattices:d} } \hspace{0.4in} \\
    \subfloat[\centering]{\includegraphics[width=0.15\columnwidth]{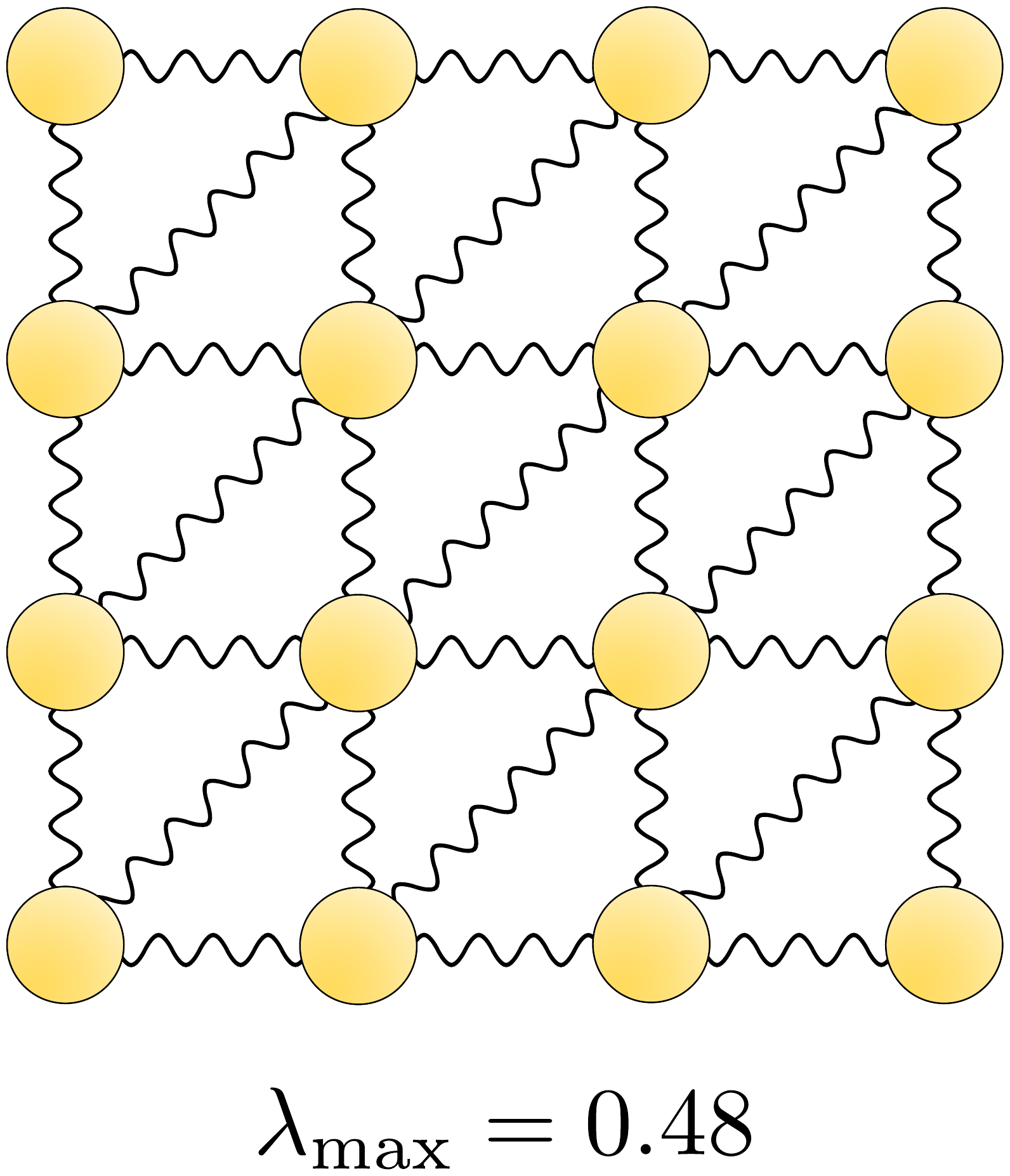}
    \label{fig:prog:lattices:e} } \hspace{0.4in}
    \subfloat[\centering]{\includegraphics[width=0.15\columnwidth]{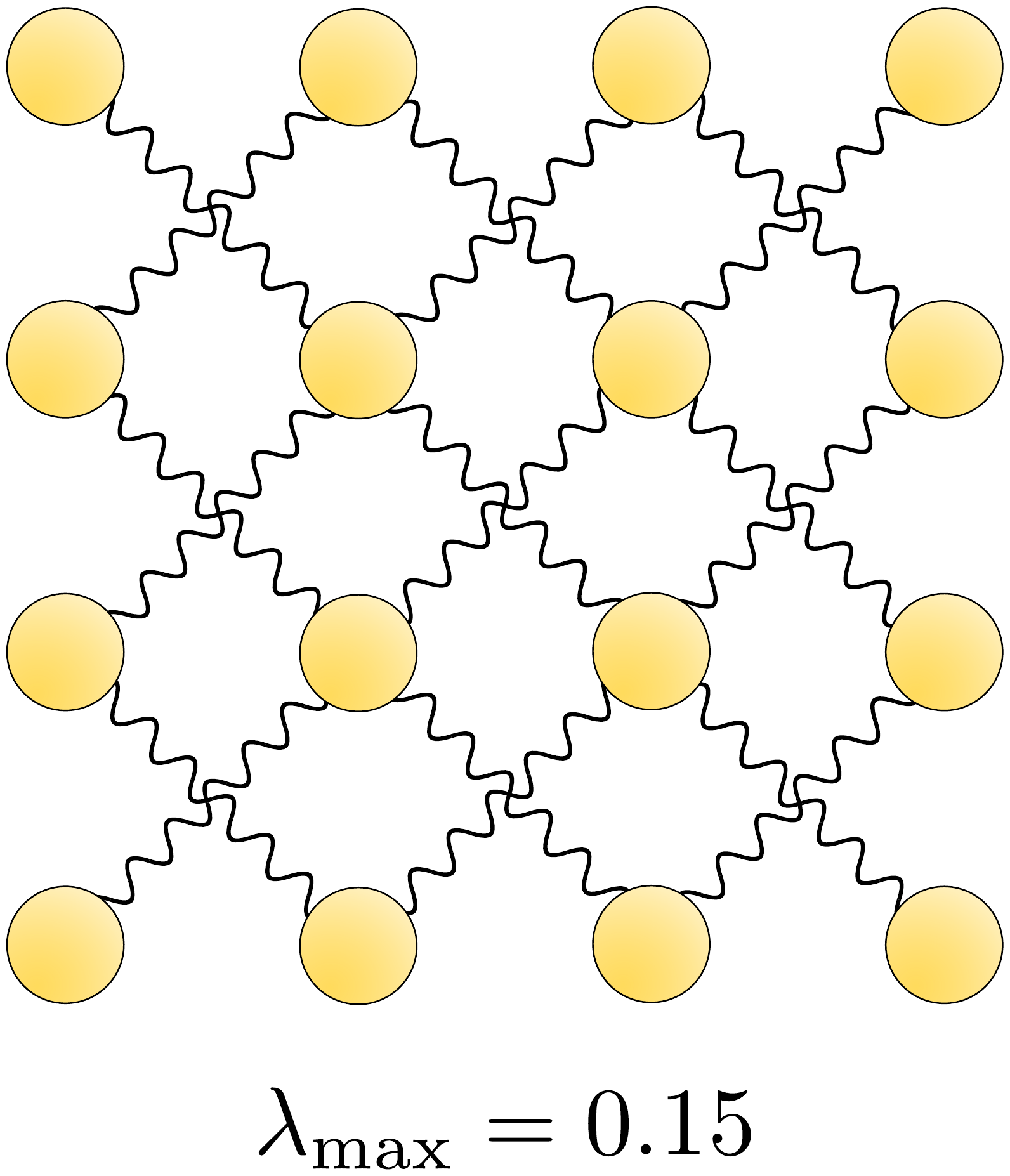}
    \label{fig:prog:lattices:f} } \hspace{0.4in}
    \subfloat[\centering]{\includegraphics[width=0.15\columnwidth]{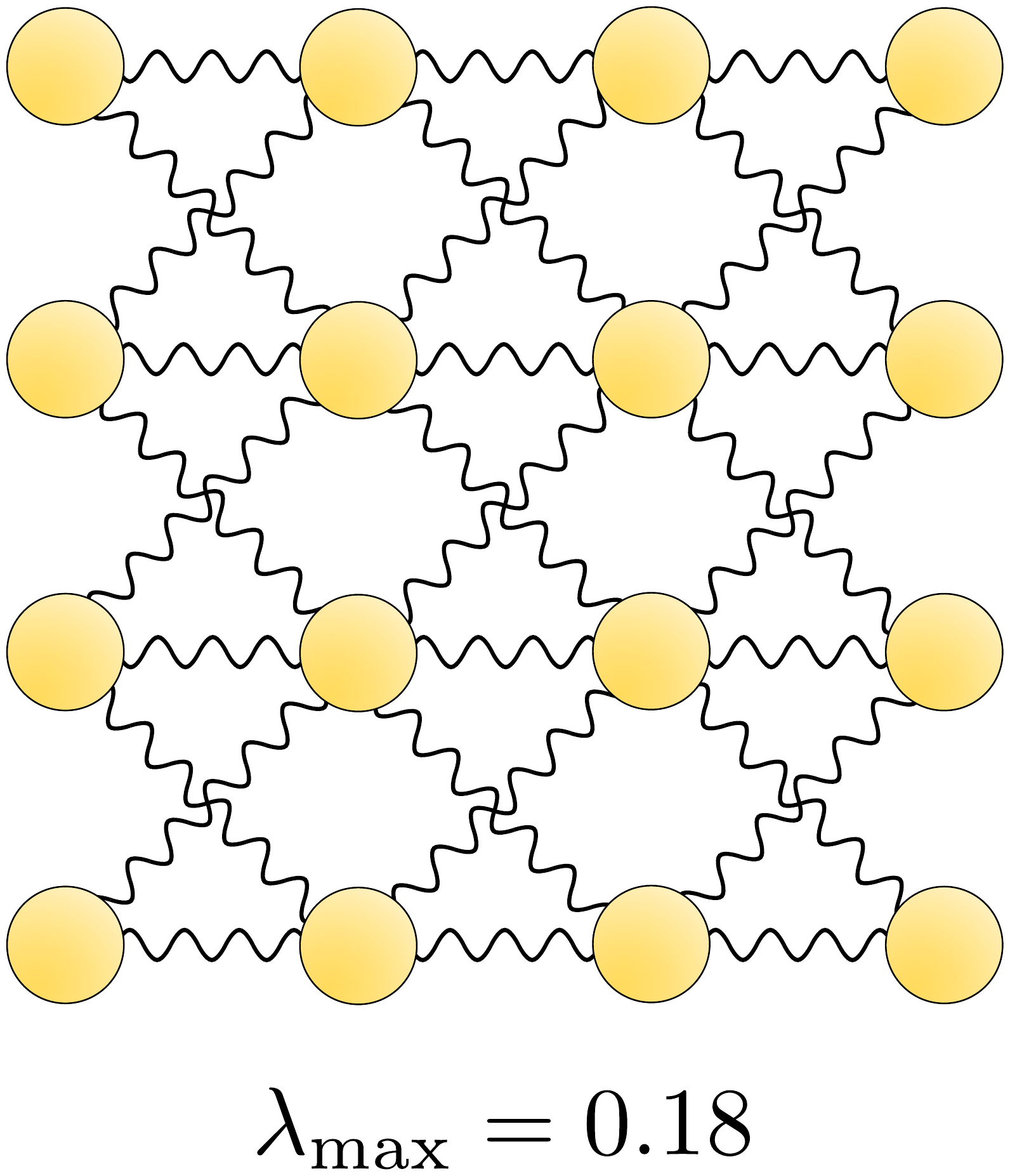}
    \label{fig:prog:lattices:g} } \hspace{0.4in}
    \subfloat[\centering]{\includegraphics[width=0.15\columnwidth]{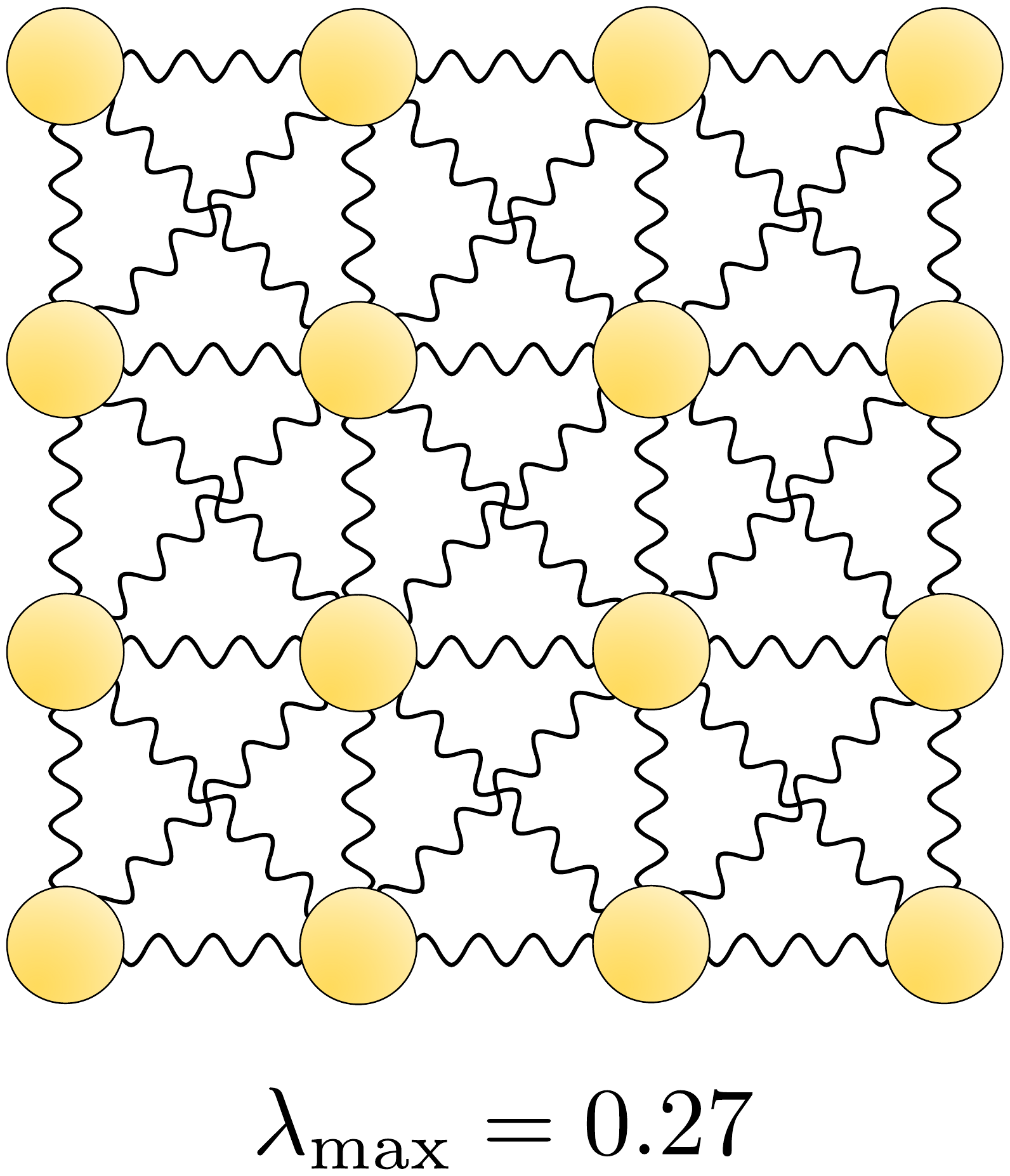} 
    \label{fig:prog:lattices:h} }
    \caption{\label{fig:prog:lattices} In (a) we show a square lattice of physical qubits with n.n. and diagonal interactions. We group the qubits in groups of five qubits each. In (b)-(h) we show different interaction patterns that we generate setting all logical qubits in the same logical subspace. The interaction strength between two linked qubits is given by $\lambda_{ij} = \lambda_{\max} J/\delta$. A blue wavy line links two logical qubits if the interaction strength is given by $\lambda_{\text{max}} / 2$. Find in Table \ref{table:fig12} the logical subspaces to generate the interaction patterns.}
\end{figure}
\begin{figure}
    \centering
    \subfloat[\centering]{\includegraphics[width=0.30\columnwidth]{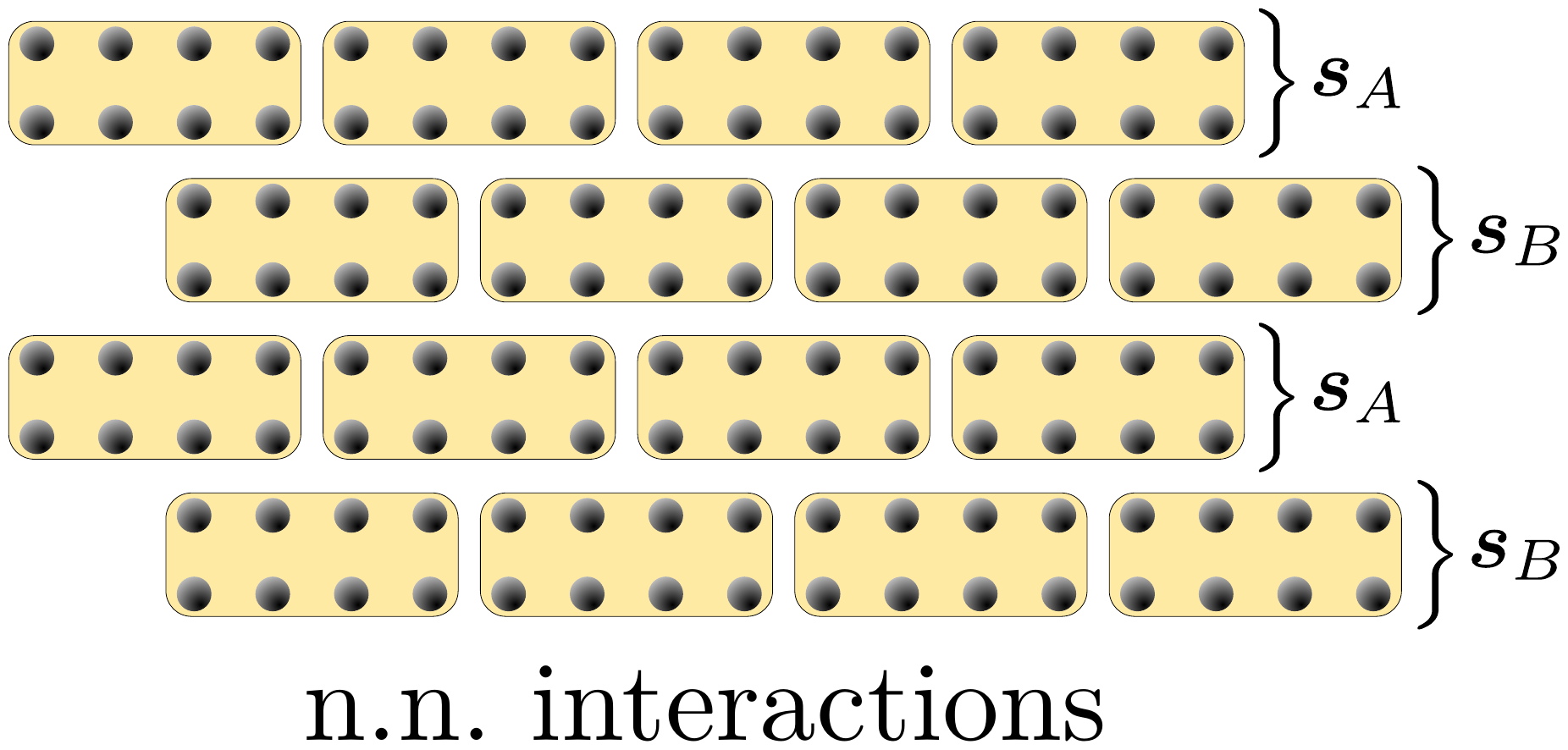} \label{fig:prog:lattices2:a} } \hspace{0.3in}
    \subfloat[\centering]{\includegraphics[width=0.15\columnwidth]{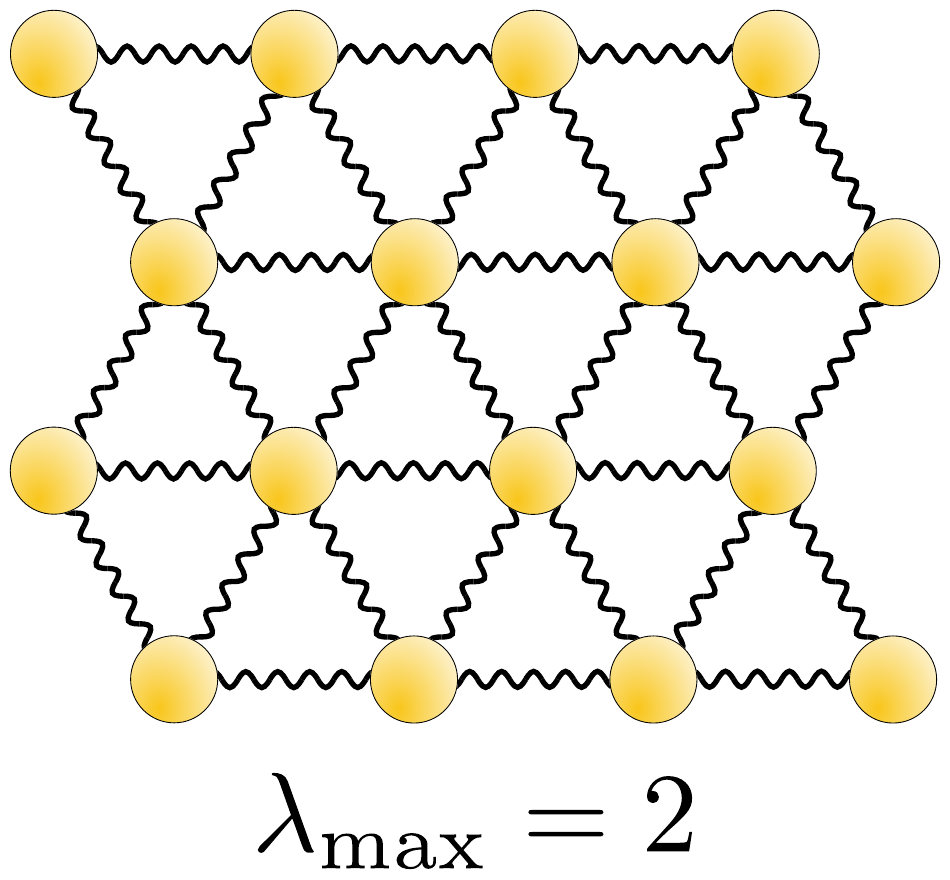} \label{fig:prog:lattices2:b} } \hspace{0.3in}
    \subfloat[\centering]{\includegraphics[width=0.15\columnwidth]{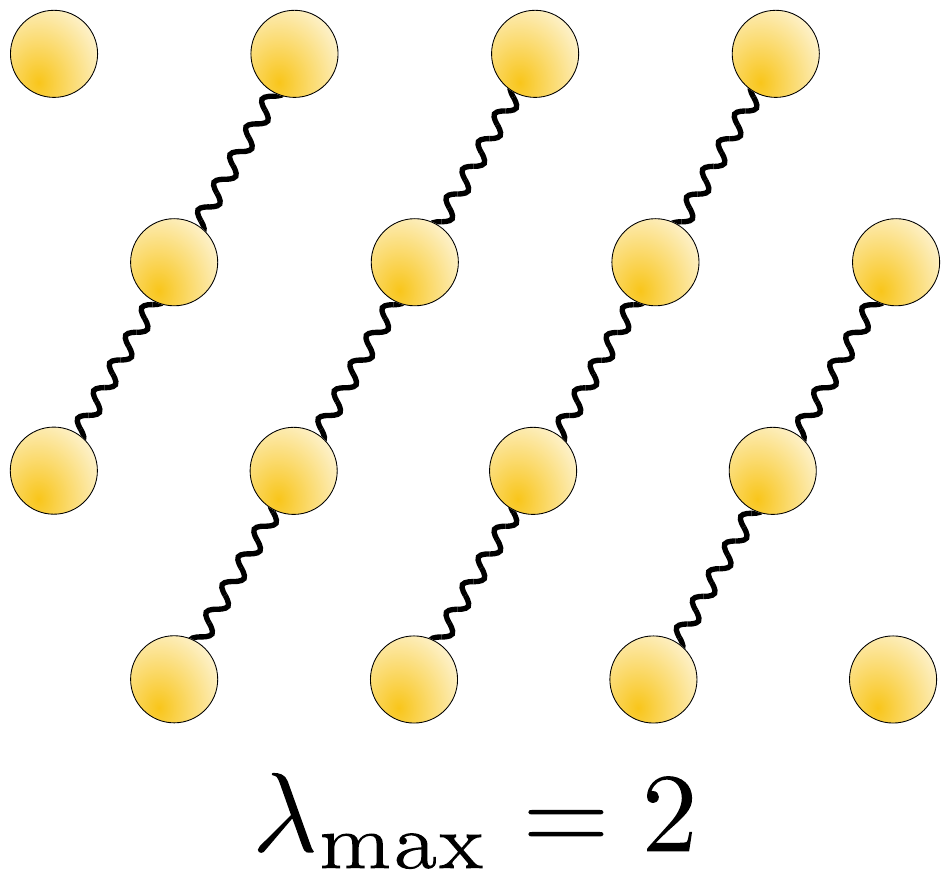}
    \label{fig:prog:lattices2:c} } \hspace{0.3in}
    \subfloat[\centering]{\includegraphics[width=0.15\columnwidth]{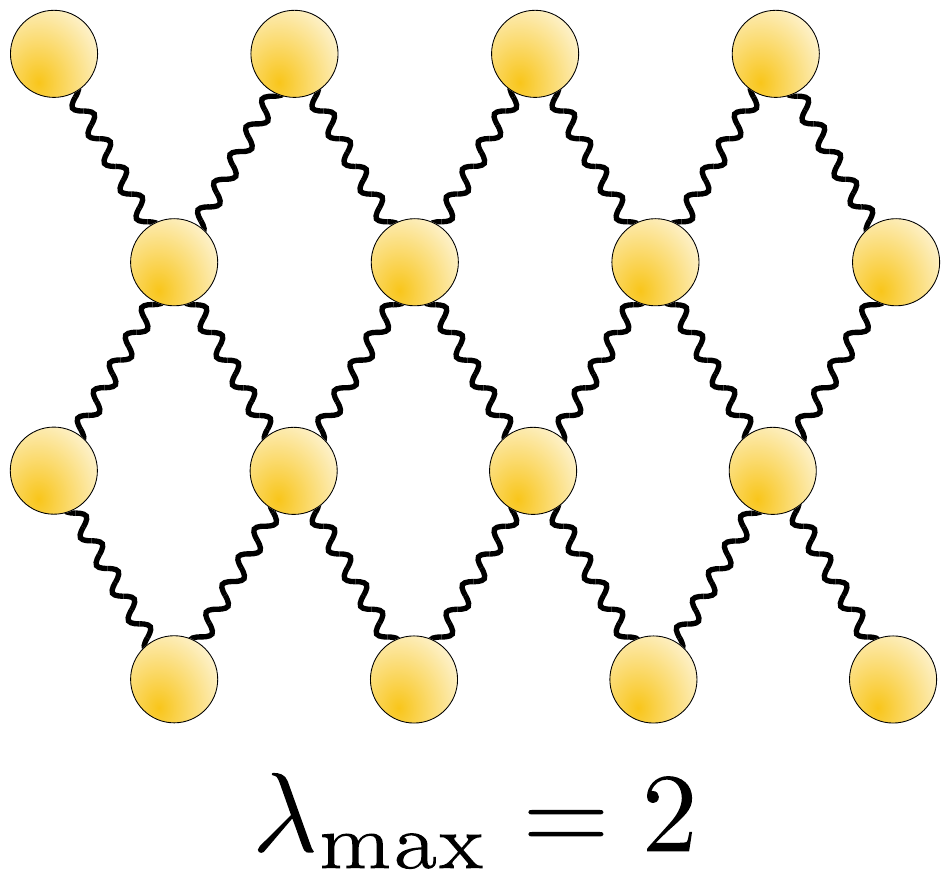}
    \label{fig:prog:lattices2:d} } \hspace{0.3in} \\
    \subfloat[\centering]{\includegraphics[width=0.15\columnwidth]{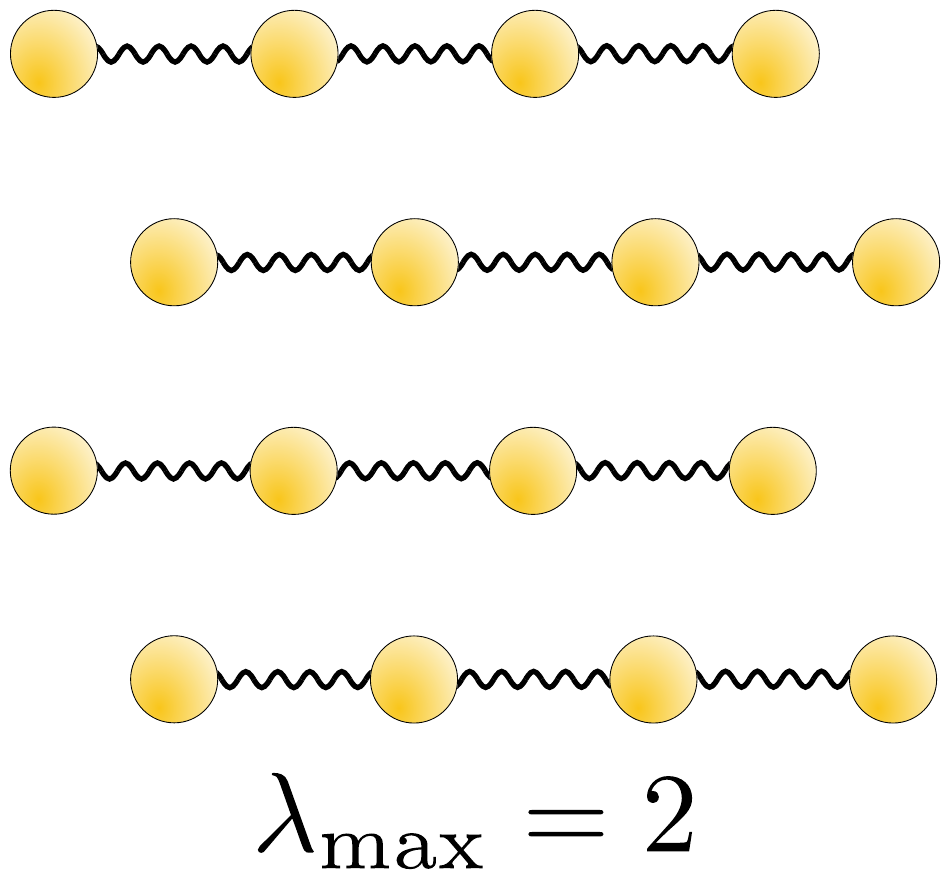}
    \label{fig:prog:lattices2:e} } \hspace{0.3in}
    \subfloat[\centering]{\includegraphics[width=0.15\columnwidth]{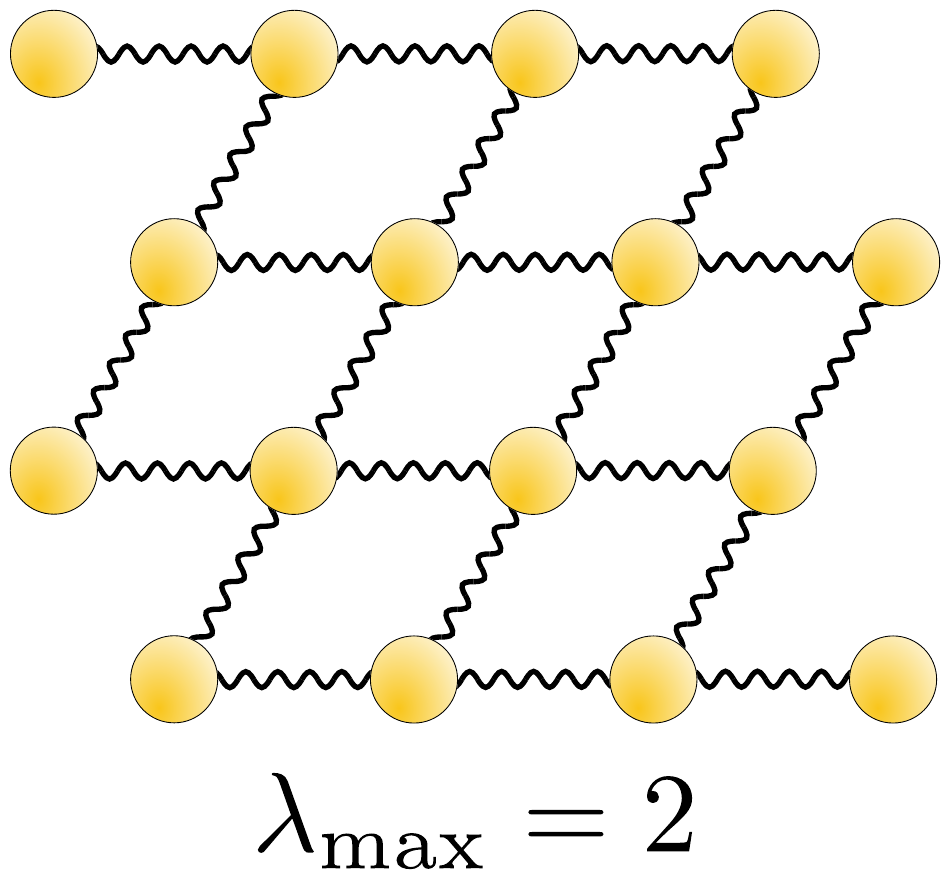}
    \label{fig:prog:lattices2:f} } \hspace{0.3in}
    \subfloat[\centering]{\includegraphics[width=0.15\columnwidth]{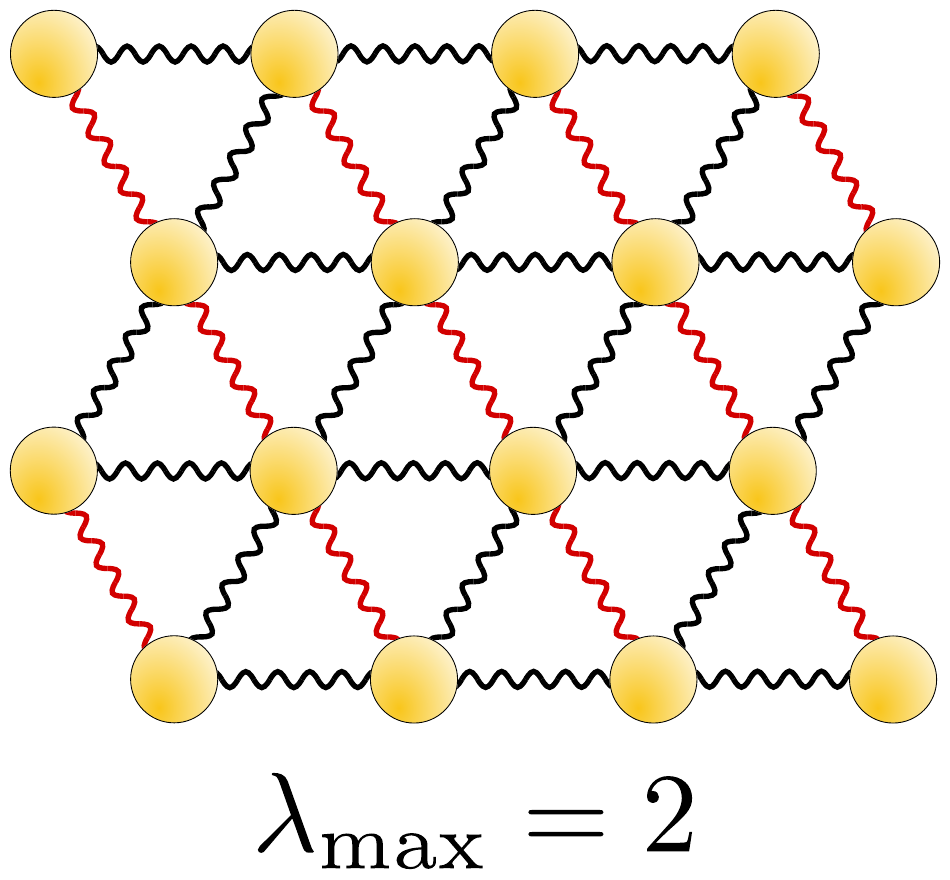}
    \label{fig:prog:lattices2:g} } \hspace{0.3in}
    \subfloat[\centering]{\includegraphics[width=0.15\columnwidth]{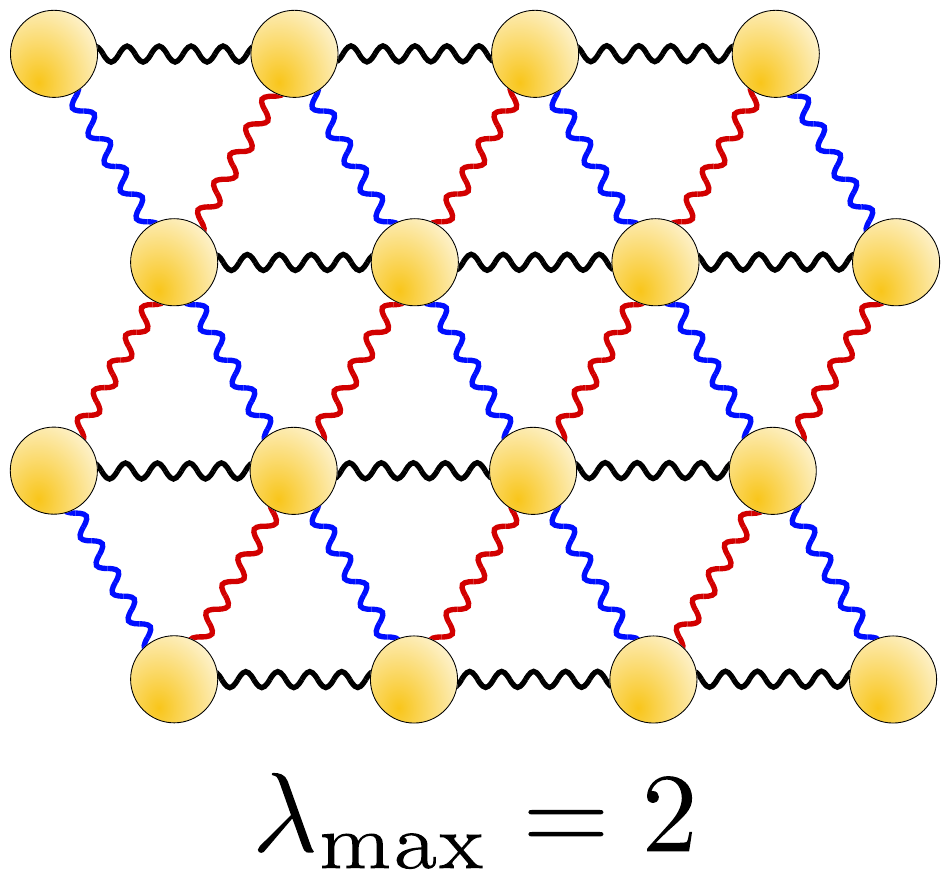}
    \label{fig:prog:lattices2:h} }
    \caption{\label{fig:prog:lattices2} In (a) we show a square lattice of physical qubits with n.n. interactions. We group the qubits in groups of eight qubits each. In (b)-(h) we show different interaction patterns that we generate by setting all qubits in the same row in the same logical subspace and we alternate between two logical subspaces in each diagonal. The interaction strength between two linked qubits is given by $\lambda_{ij} = \lambda_{\max} J/\delta$. In (g) red wavy lines correspond to a coupling strength of $\lambda_{ij} = - \lambda_{\max} J/\delta$. In (h) blue and red wavy lines correspond to a coupling strength of $\lambda_{ij} = (\lambda_{\max}/3) (J/\delta)$ and $\lambda_{ij} = (\lambda_{\max}/2) (J/\delta)$ respectively. Find in Table \ref{table:fig13} the logical subspaces to generate the interaction patterns.}
\end{figure}

\section{Rearranging a square lattice}
\label{app:SAWP}

Consider the task of rearranging the state of the qubits in a $n\times n$ square lattice by applying SWAP gates between n.n. For that one has to vertically displace the $k$-column $k-1$ positions. The first column is left untouched and no operations are required. For the second column, the last qubits have to be permuted with all the others to reach the first position, and for what we need to perform $n-1$ SWAP operations. For the third column, the states of the last two qubits have to be permuted with the other $n-2$ qubits to reach the two first positions of the column. Generalizing this step we obtain that in column $k$, the last $(k-1)$ qubits have to be permuted $n-k+1$ positions, and therefore the total number of SWAP gates required to rearrange the lattice is given by 
\begin{equation*}
    \zeta(n) = \sum_{k=1}^n (k-1)(n-k+1) = \frac{1}{6}(n^3-n).
\end{equation*}
Given the fact that to perform a single SWAP gate takes a time $\tau$ we can compute how much time $\tau'$ we need to rearrange the lattice. However, that is not a tribal task if we consider that SWAP gates that do not involve the same qubits can be applied simultaneously. To compute $\tau'$, we define a \textit{permutation shot} which is a reordering of the qubit states that is achieved by permuting the state of several pairs of n.n. qubits in a way that two permutations cannot involve the same qubits. As the columns permutations are independent, we just need to consider the number of permutation shots necessary to rearrange the qubit states of each column which is given by $n-1$. Therefore, the time to rearrange the lattice is given by $\tau' = (n-1) \tau$.

\section{Logical subspaces}
\label{app:sec:data}

Find here the logical subspaces to generate interaction patterns shown in Figures~\ref{fig:toy:model},~\ref{fig:5},~\ref{fig:NN:FR}, \ref{fig:prog:lattices} and \ref{fig:prog:lattices2}.

\renewcommand{\arraystretch}{1.0}
\begin{table*}[h!]
\begin{tabular}{|c|c|}
    \hline
    $\;$ Fig.~\ref{fig:prog:lattices} $\;$ & $\boldsymbol{s}$ \\ \hline
    b & $(1, 0, 0, 1, 0)$                 \\
    c & $(1, 1, 0.41, 0, 0)$              \\
    d & $(1, 0, 1, 0.83, 0)$              \\
    e & $(-0.74, 1, 1, 0.68, 0)$          \\
    f & $(-1, 0.65, 0.42, 0.34, 0.52)$    \\
    g & $(1, -0.41, -0.83,-0.65, -0.32)$  \\
    h & $(1, -0.62, -0.62, -0.62, -0.62)$ \\ \hline
\end{tabular}

\caption{\label{table:fig12}The logical subspace of the logical qubits to generate the different interaction patterns of Fig.~\ref{fig:prog:lattices}.}
\end{table*}

\renewcommand{\arraystretch}{1.0}
\begin{table*}[h!]
\begin{tabular}{|c|c|c|}
    \hline
    $\;$ Fig.~\ref{fig:prog:lattices2} $\;$ & $\boldsymbol{s}_A$ & $\boldsymbol{s}_B$ \\ \hline
    b & $(1,1,1,1,1,1,1,1)$           & $(1,1,1,1,1,1,1,1)$           \\
    c & $(1,1,0,0,0,0,1,1)$           & $(1,1,0,0,0,0,1,1)$           \\
    d & $(1,-1,1,-1,1,1,1,1)$         & $(1,1,1,1,1,-1,1,1)$          \\
    e & $(1,0,0,1,1,0,0,1)$           & $(1,0,0,1,1,0,0,1)$           \\
    f & $(1,1,0,1,1,0,1,1)$           & $(1,1,0,1,1,0,1,1)$           \\
    g & $(1,1,-1,1,1,-1,1,1)$         & $(1,1,-1,1,1,-1,1,1)$         \\
    h & $(1,0.5,0.33,1,1,0.33,0.5,1)$ & $(1,0.5,0.33,1,1,0.33,0.5,1)$ \\ \hline
\end{tabular}
\caption{\label{table:fig13}The logical subspace of the logical qubits to generate the different interaction patterns of Fig.~\ref{fig:prog:lattices2}.}
\end{table*}

\begin{table*}
\begin{tabular}{|c|c|c|c|c|c|}
    \hline
    $\;$ Fig.~\ref{fig:toy:model} $[G_1]\;$ & $\quad \lambda_{\text{max}}$ \quad & $\boldsymbol{s}_1$ & $\boldsymbol{s}_2$ & $\boldsymbol{s}_3$ & $\boldsymbol{s}_4$ \\ \hline
    c & $0.85$ & $(1,1,1,1)$         & $(0,0,0,0)$         & $(1,1,1,1)$       & $(0,0,0,0)$             \\
    d & $2.04$ & $(1,1,1, 1)$        & $(1,1,-1,0.28)$     & $(0.97,1,-1,-1)$  & $(0.97,-0.7,0.77,0.05)$ \\
    e & $1.08$ & $(1,1,-1,-0.88)$    & $(0.61,-1,0.67,-1)$ & $(-1,-1,-0.05,1)$ & $ (1,-1,-0.75,-1)$      \\
    f & $1.01$ & $(1,-0.25,-0.25,1)$ & $(1,1,-0.9,1)$      & $(1,-0.9,1,1)$    & $(1,-0.25,-0.25,1)$     \\
    g & $0.73$ & $(-1,-1,1,0.6)$     & $(-1,-1,0.71,0.9)$  & $(1,0.07,1,1)$    & $(1,1,-1,-0.12)$        \\
    h & $0.43$ & $(1,-0.41,-0.9,1)$  & $(1,1,-1,0.07)$     & $(1,-1,1,1)$      & $(1,1,-0.5,-0.9)$       \\ \hline
\end{tabular}
\caption{\label{tab:g1}Logical subspace for each logical qubit to generate the different interaction patterns of Fig.~\ref{fig:toy:model} with grouping $G_1$.}
\end{table*}

\begin{table*}
\begin{tabular}{|c|c|c|c|c|c|}
    \hline
    $\;$ Fig.~\ref{fig:toy:model} $[G_1]\;$ & $\lambda$ \quad & $\boldsymbol{s}_1$ & $\boldsymbol{s}_2$ & $\boldsymbol{s}_3$ & $\boldsymbol{s}_4$ \\ \hline
    d & $0.025$ & $(0.08,0.08,0.08,0.08)$ & $(0.04,-0.74,0.25,0.45)$ & $(-0.19,-0.04,1,-0.42)$    & $(-0.19,-0.051,0.29,0.32)$ \\
    e & $0.046$ & $(0.11,0.11,0.11,0.11)$ & $(0.05,-1,0.34,0.34)$    & $(-0.26,-0.054,0.34,0.45)$ & $(-0.26,0.83,0.24,-0.87)$  \\
    f & $0.04$  & $(0.09,0.09,0.09,0.09)$ & $(0.05,-0.89,0.31,0.54)$ & $(-0.23,-0.05,0.15,0.55)$  & $(-0.23,0.06,-0.28,1)$     \\
    g & $0.02$  & $(0.08,0.08,0.08,0.08)$ & $(0.04,-0.74,0.25,0.45)$ & $(-0.19,-0.04,1,-0.42)$    & $(-0.19,0.25,-0.18,0.24)$  \\
    h & $0.01$  & $(0.04,0.04,0.04,0.04)$ & $(0.02,-0.35,0.12,0.12)$ & $(-0.09,-0.02,-0.82,1)$    & $(-0.09,0.14,-0.06,0.18)$  \\ \hline
\end{tabular}
\caption{\label{tab:g1:algorithm}Logical subspaces found by using algorithm 1 for interaction patterns in Fig.~\ref{fig:toy:model} with grouping $G_1$.}
\end{table*}

\begin{table*}
\begin{tabular}{|c|c|c|c|c|c|}
    \hline
    $\;$ Fig.~\ref{fig:toy:model} $[G_2]\;$ & $\lambda$ \quad & $\boldsymbol{s}_1$ & $\boldsymbol{s}_2$ & $\boldsymbol{s}_3$ & $\boldsymbol{s}_4$ \\ \hline
    d & $0.030$ & $(0.09,0.09,0.09,0.09)$ & $(0.04,-0.8,0.28,0.52)$  & $(-0.21,-0.04,1,-0.48)$    & $(-0.21,0.27,0.092,0.13)$ \\
    e & $0.019$ & $(0.07,0.07,0.07,0.07)$ & $(0.03,-0.63,0.22,0.28)$ & $(-0.16,-0.03,1,-0.59)$    & $ (-0.16,0.03,0.11,0.11)$ \\
    f & $0.050$ & $(0.11,0.11,0.11,0.11)$ & $(0.05,-1,0.35,0.67)$    & $(-0.26,-0.055,1,-0.34)$   & $(-0.26,0.27,0.13,0.22)$  \\
    g & $0.057$ & $(0.12,0.12,0.12,0.12)$ & $(-0.29,-0.06,-0.22,1)$  & $(-0.29,-0.06,0.83,-0.11)$ & $(-0.29,0.8,0.13,-0.69)$  \\
    h & $0.018$ & $(0.07,0.07,0.07,0.07)$ & $(0.03,-0.63,0.22,0.28)$ & $(-0.16,-0.03,1.,-0.59)$   & $(-0.16,0.03,0.11,0.11)$  \\ \hline
\end{tabular}
\caption{\label{tab:g2:algorithm} Logical subspaces found by using algorithm 1 for interaction patterns in Fig.~\ref{fig:toy:model} with grouping $G_2$.}
\end{table*}

\begin{table*}[h!]
\begin{tabular}{|c|c|c|}
    \hline
    Fig.~\ref{fig:cube} & nearest neighbor interactions $\lambda_{\text{max}} = 0.31$            & full range $\lambda_{\text{max}} = 0.57$  \\ \hline
    $\boldsymbol{s}_1$ & $(-1, -1, -1, 0.81, -0.53, 1, 0.57, 0.12)$  & $(1, -0.39, 1, -0.45, 1, -1, -0.7, 1)$         \\
    $\boldsymbol{s}_2$ & $(1, -0.53, -1, -1, 0.09, -0.49, 1, 0.77)$  & $(1, 0.74, -0.29, 1, 0.22, -1, -1, 1)$         \\
    $\boldsymbol{s}_3$ & $(-1, -0.04, 1, 1, 0.02, 1, -0.92, -0.73)$  & $(0.31, 1, 1, -0.97, 1, -1, -0.97, 1)$         \\
    $\boldsymbol{s}_4$ & $(1, 1, 1, 1, 1, 1, -1, 0.36)$              & $(1, 1, 0.46, -0.88, 0.45, -1, -0.76, 1)$      \\
    $\boldsymbol{s}_5$ & $(1, 1, -1, -1, 1, -0.18, -1, -1)$          & $(0.76, 0.18, 1, -0.24, 0.2, -1, -0.09, 0.46)$ \\
    $\boldsymbol{s}_6$ & $(0.86, -1, -0.7, 1, 0.3, -0.63, 1, -0.78)$ & $(0.94, 0.02, 1, 0.04, 1, -1, -1, 0.44)$       \\
    $\boldsymbol{s}_7$ & $(1, 0, 0.61, -1, -0.77, -1, 0.01, 1)$      & $(1, 0.47, 1, 0.75, 1, -1, -1, -0.66)$         \\
    $\boldsymbol{s}_8$ & $(0.27, -0.89, 0.81, -1, 1, 1, -0.51, -1)$  & $(1, -1, -1, 1, 1, 0.36, -0.4, 1)$             \\ \hline
\end{tabular}
\caption{\label{tab:fig:5a} Logical subspace for each logical qubit to generate the different interaction patterns of Fig.~\ref{fig:cube}.}
\end{table*}

\begin{table*}
\begin{tabular}{|c|c|c|}
    \hline
    Fig.~\ref{fig:FR-NN} & nearest neighbor interactions $\lambda_{\text{max}} = 0.64$ & nearest neighbor and diagonal interactions $\lambda_{\text{max}} = 0.43$ \\ \hline
    $\boldsymbol{s}_1$ & $(-1,-0.36,-1,0.85,1,1,-0.6,1,-1)$           & $(-1,-1,0.19,0.06,0.2,1,1,-1,0.22)$         \\
    $\boldsymbol{s}_2$ & $(-1,-0.97,1,1,-0.49,1,-1,1,-0.59)$          & $(1,0.01,0.57,1,1,-0.73,-0.93,-1,0.44)$     \\
    $\boldsymbol{s}_3$ & $(0.3,-0.64,-1,0.66,1,-1,0.08, 1,-0.67)$     & $(1,1,-0.41,-1,-1,-1,0.38,0.76,0.21)$       \\
    $\boldsymbol{s}_4$ & $(-0.15,1,-1,1,1,0.54,0.07,-0.84,-1)$        & $(1,-1,-0.58,1,0.84,-1,-0.78,1,0.56)$       \\
    $\boldsymbol{s}_5$ & $(-1,1,0.02,-1,-0.91,1,-1,-1,1)$             & $(1,1,1,1,1,1,0.87,1,0.67)$                 \\
    $\boldsymbol{s}_6$ & $(1,-0.74,-1,1,1,0.74,-0.02,-1,-1)$          & $(0.38,1,1,-0.39,0.11,1,-0.49,-1,-1)$       \\
    $\boldsymbol{s}_7$ & $(0.04,-1,-0.78,1,1,1,-0.36,-1,0.87)$        & $(-0.26,0.91,0.46,-1,-1,-1,0.99,0.29,0.69)$ \\
    $\boldsymbol{s}_8$ & $(-0.16,-0.42,0.41,1,-1,0.88,0.63,-1,-0.74)$ & $(1,-0.09,-0.4,-1,0.57,-1,1,1,-1)$          \\ 
    $\boldsymbol{s}_9$ & $(0.58,-1,-0.6,0.84,0.4,1,-1,-1,1)$          & $(1,-1,-0.65,-1,0.94,1,-0.58,1,-0.66)$      \\ \hline
\end{tabular}
\caption{\label{tab:fig:5b}Logical subspace for each logical qubit to generate the different interaction patterns of Fig.~\ref{fig:FR-NN}.}
\end{table*}
\end{document}